\documentclass[final,3p,times]{elsarticle}
\usepackage{amssymb}
\usepackage{amsmath}
\usepackage{bm}
\usepackage{color}
\usepackage[colorlinks=true, allcolors=blue]{hyperref}
\usepackage{setspace}
\usepackage{float}
\usepackage{multirow}
\usepackage{ulem,xpatch}

\newcommand{\commentLZ}[1]{\textcolor{cyan}{[LZ: #1]}}
\journal{Acta Materialia}

\begin{document}

\begin{frontmatter}
	
	
	
	\title{Efficiency, Accuracy, and Transferability of Machine Learning Potentials: \\ Application to Dislocations and Cracks in Iron}

	
	\author{Lei Zhang$^{a*}$, Gábor Csányi$^b$, Erik van der Giessen$^c$, Francesco Maresca$^{a*}$}
	
	\affiliation{organization={Engineering and Technology Institute, 
			Faculty of Science and Engineering,
			University of Groningen},%Department and Organization
		addressline={ Nijenborgh 4}, 
		city={Groningen},
		postcode={9747 AG}, 
		state={Groningen},
		country={The Netherlands}}
	\affiliation{organization={Engineering Laboratory, University of Cambridge},%Department and Organization
		%        	addressline={ Nijenborgh 4}, 
		%        	city={Groningen},
		postcode={CB2 1PZ}, 
		state={Cambridge},
		country={United Kingdom}}
	
	\affiliation{organization={Zernike Institute for Advanced Materials, 
			University of Groningen},%Department and Organization
		addressline={ Nijenborgh 4}, 
		%        	city={Groningen},
		postcode={9747 AG}, 
		state={Groningen},
		country={The Netherlands}}
	
	\begin{abstract}
	Machine learning interatomic potentials (ML-IAPs) enable quantum-accurate, classical molecular dynamics simulations of large systems, beyond reach of density functional theory (DFT). 
	Yet, their efficiency and ability to predict systems larger than DFT supercells are not fully explored, posing a question regarding transferability to large-scale simulations with defects (e.g. dislocations, cracks). 
	Here, we apply a three-step validation approach to body-centered-cubic iron. 
	First, accuracy and efficiency are assessed by optimizing ML-IAPs based on four state-of-the-art ML packages. 
	The Pareto front of computational speed versus testing root-mean-square-error (RMSE) is computed. 
	Second, benchmark properties relevant to plasticity and fracture are evaluated. 
	Their average relative error $Q$ with respect to DFT is found to correlate with RMSE. 
	Third, transferability of ML-IAPs to dislocations and cracks is investigated by using per-atom model uncertainty quantification. 
	The core structures and Peierls barriers of screw, M111 and three edge dislocations are compared with DFT. Traction-separation curve and critical stress intensity factor ($K_{\rm Ic}$) are also predicted. 
	Cleavage on the pre-existing crack plane is found to be the zero-temperature atomistic fracture mechanism of pure body-centered-cubic iron under mode-I loading, independent of ML package and training database. 
	Quantitative predictions of dislocation glide paths and $K_{\rm Ic}$ can be sensitive to database, ML package, cutoff radius, and are limited by DFT accuracy. 
	Our results highlight the importance of validating ML-IAPs by using indicators beyond RMSE. 
	Moreover, significant computational speed-ups can be achieved by using the most efficient ML-IAP package, yet the assessment of the accuracy and transferability should be performed with care.
		
	\end{abstract}
	
	
	%\item Pareto front: speed vs. accuracy;
	%\item Q-factor: RMSE vs. physical properties;
	%\item Dislocation mobility revealed by DFT-accurate ML-IAPs
	%\item Influence of cutoff and database
	
	
\end{frontmatter}

%\linenumbers

%% add a section  of abbreviations 
\section*{List of abbreviations}

\noindent 
\textbf{ACE}: Atomic cluster expansion \\
\textbf{BCC}: Body-centered cubic \\
\textbf{DFT}: Density-functional theory \\
\textbf{DOFs}: Degrees of freedom \\
\textbf{GAP}: Gaussian approximation potential \\
\textbf{LAE}: Local atomic environment \\
\textbf{MEAM}:  Modified embedded atom method\\
\textbf{ML-IAP}: Machine learning interatomic potential \\
\textbf{MTP}: Moment tensor potential \\
\textbf{MS/MD}: Molecular statics/dynamics \\
\textbf{NEB}: Nudged elastic band \\
\textbf{NN}: Neural network (NN) \\
\textbf{PES}: Potential energy surface \\
\textbf{qSNAP}: quadratic spectral neighbor analysis potential \\
\textbf{RMSE}: Root-mean-square error \\
\textbf{SNAP}: Spectral neighbor analysis potential \\
\textbf{SOAP}: Smooth Overlap of Atomic Positions\\

\setstretch{1.5}
\section{Introduction}
\label{chap1}
Plastic deformation and fracture in metals are controlled by the motion and interaction of extended defects, such as dislocations, grain boundaries, and cracks.
For example, plastic deformation in body-centered cubic (bcc) metals at moderate temperatures is primarily dictated by the mobility of screw dislocations, which is significantly slower than the edge dislocations as a consequence of the compact core structure \cite{dorn1963nucleation,proville2013prediction,itakura2012first,ventelon2013ab}. 
Additionally, the competition between crack propagation and dislocation emission from crack-tips controls the intrinsic ductility, which plays a role in determining the fracture toughness \cite{mak2021ductility,andric2017new}.
Therefore, accurate modelling of extended defects at the atomic scale is a fundamental step to understand and engineer the mechanical properties of metals.

Classical molecular statics/dynamics (MS/MD) simulations are routinely employed to investigate the structure and propagation mechanism of extended defects at the atomic scale.
However, in the case of bcc metals like iron, the empirical interatomic potentials (IAPs) struggle to accurately predict the compact screw dislocation core structure and atomic scale crack propagation mechanism \cite{mendelev2003development,gordon2011screw,MOLLER_2015197}. 
Recent advances in machine learning (ML) techniques have allowed the development of ML-IAPs that are able to predict extended defects with similar accuracy as density-functional theory (DFT) calculations ($\sim$1 meV/atom), but at a computational speed that is orders of magnitude faster.
For instance, ML-IAPs can faithfully replicate the compact core structure of screw dislocations in various bcc transition metals as predicted by DFT \cite{szlachta2014accuracy,Dragoni2018_PhysRevMaterials.2.013808,alam2021artificial_Molybdenum,wang2022classical_vanadium}.
More recently, an active learning scheme has been developed within the Gaussian Approximation Potential (GAP) framework, which enables the prediction of fracture mechanisms in bcc iron \cite{lei_2022}.
These successful applications demonstrate that ML-IAPs are promising tools for simulating mechanisms at the atomic scale.
Yet, ML-IAPs suffer from significant computational cost and require extensive validation to test transferability.

To improve both accuracy and computational efficiency, multiple ML-IAP frameworks have been developed, that differ by the local atomic environment (LAE) expansion approach and the regression method \cite{behler2007generalized,bartok2010gaussian,shapeev2016moment,ARTRITH2016135,drautz2019atomic}.
A benchmark study of moment tensor potential (MTP), GAP, spectral neighbour analysis potential (SNAP), and quadratic SNAP (qSNAP) for a wide range of materials (Li, Mo, Cu, Ni, Si, and Ge) was conducted in Ref. \cite{zuo2020performance}, showing that root-mean-square error (RMSE) of $\sim 1$ meV/atom can be achieved for all ML-IAPs at computational speeds spanning two orders of magnitude. 
The Pareto front of RMSE against computational speed showed that MTP occupies the frontier for all material systems.
Following the results in Ref. \cite{zuo2020performance}, atomic cluster expansion (ACE) potential has been added and has been shown to occupy the Pareto frontier for copper and silicon, being mildly faster and more accurate than MTP  \cite{lysogorskiy2021performant}.   
However, extended defects such as dislocations and cracks have not been investigated in Ref. \cite{zuo2020performance,lysogorskiy2021performant}. 
A recent summary showed that most ML-IAPs consider point defects, stacking faults, and free surfaces as training configurations \cite{freitas2022machine}. 
For dislocations, grain boundaries and cracks, only a few elements, such as Al, Si, Mo, and W have been explored \cite{alam2021artificial_Molybdenum,lin2022development,bartok2018machine,nishiyama2020application}. 
The dislocation core structures in bcc iron have been studied by several ML-IAPs \cite{ann_mori2020neural,mtp_wang2022machine,goryaeva2021efficient}, yet, the efficiency, accuracy and transferability of ML-IAPs in predicting extended crystal defects, i.e., dislocations and cracks, have not been fully investigated. 

\begin{figure}[H]
	\centering
	\includegraphics[trim=0 0 0 0, width=10cm]{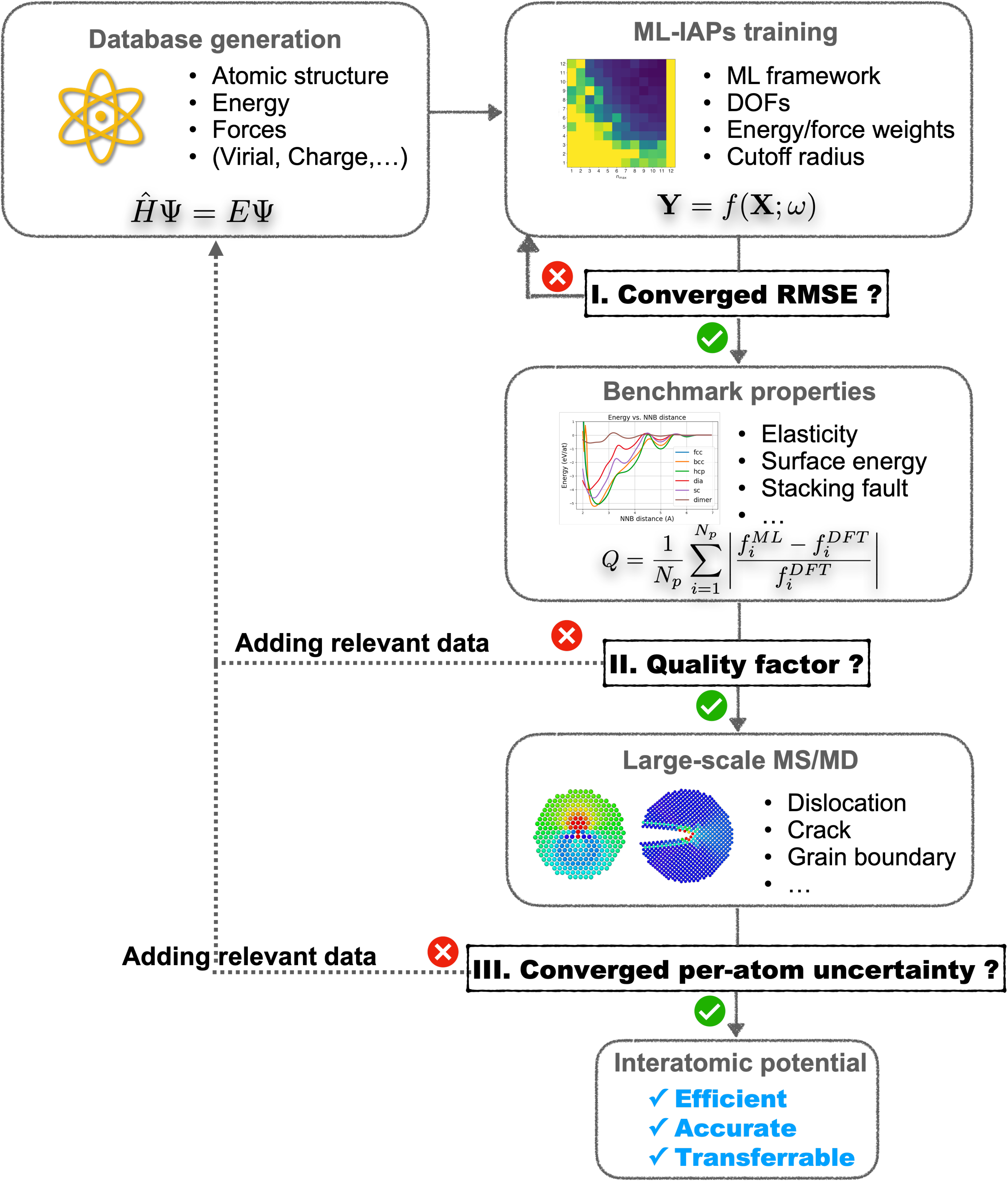} 
	\caption{Flowchart for ML-IAPs training and validation proposed in this work. Three criteria (i.e., \textit{I. Converged RMSE, II. Quality factor, III. Converged per-atom model uncertainty}) need to be met for an efficient, accurate and transferable ML-IAP.
	}
	\label{fig:validation}
\end{figure}
Here, we present a systematic training and validation procedure of an array of state-of-the-art ML-IAPs for ferromagnetic bcc iron, which is based on a three-step approach that is summarized in Fig. \ref{fig:validation}.
During ML-IAPs training, several fitting (hyper)parameters are optimized \cite{Dragoni2018_PhysRevMaterials.2.013808,lei_2022}. 
This hyperparameter optimization procedure is an essential step to achieve both accuracy and efficiency, since it affects the convergence behaviour of the ML-IAP to the training data as a function of the number of degrees of freedom (DOF). 
Therefore, the convergence of RMSE (typically below 5 meV/atom) with respect to the DOF constitutes the first criterion that is assessed to evaluate the accuracy (\textit{criterion I. Converged RMSE} in Fig. \ref{fig:validation}), and is a standard procedure for verifying the fitting accuracy and avoiding poor/overfitting issues. 
In general, an increased number of DOFs may be needed if the ML-IAP fails to pass criterion \textit{I.}
The second checkpoint of an accurate ML-IAP is the average error factor \textit{Q} (\textit{criterion II. Quality factor}), which is measured by considering a broad range of benchmark properties that are relevant for plasticity and fracture, such as elasticity, surface/stacking-fault energy \textit{etc}., and that can be computed with DFT supercells thus enabling direct comparison.
Passing \textit{criterion II} ensures the training of a general purpose potential that is capable of replicating DFT predictions for elementary properties. 
However, its transferability is not guaranteed to the simulation of extended defects encompassing features (e.g. crack-tips, kinks) that were not part of the training database.
Therefore, we propose here the assessment of the transferability by computing the per-atom uncertainty as a measure of the degree of extrapolation from the DFT database, for large-scale simulations of extended defects that cannot be computed with direct DFT supercells (\textit{criterion III. Converged per-atom uncertainty}).
A converged per-atom model uncertainty below 15 meV/atom (for GAP trained on current database) or below 3 ($\gamma$ parameter in ACE) indicates that the model is interpolating or extrapolating mildly the DFT data, hence the potential is suitable for simulations involving the tested extended defects.
The ML-IAPs are therefore accurate, efficient and transferable if all the three criteria are met.
Iterative/active learning can be applied to include more relevant DFT data, if the ML-IAPs fail to pass criteria \textit{II} and \textit{III}.

In this study, dislocation glide and crack propagation are investigated including model uncertainty quantification.
The core structure and Peierls barrier of five dislocations, including $a_0[100](010)$ edge, $a_0[100](011)$ edge, $a_0/2[\bar{1}\bar{1}1](1\bar{1}0)$ edge, $a_0/2[111](1\bar{1}0)$ $71^{\circ}$ mixed (also known as M111), and $a_0/2\langle 111\rangle$ screw
are evaluated by using the Nudged Elastic Band (NEB) approach \cite{henkelman2000climbing}. 
Furthermore, we compute the traction-separation (T-S) curves and critical stress intensity factors for \{100\} and \{110\} crack planes. 
In addition, we train an array of GAP models based on an independent, larger iron DFT database to study the influence of the database choice on the predictions \cite{ogata_nnp_meng2021general}. 
We demonstrate the consistent accuracy and transferability of GAP ML-IAPs, which can be obtained with the smaller database, as well as the efficiency of ACE ML-IAPs, which can achieve comparable accuracy with up to two orders of magnitude computational speed-up.

The paper is organized as follows. 
In \textit{section \ref{chap2}}, we present the approach used to train and optimize the different ML-IAPs, discussing the main steps of the fitting procedure and the training scripts. This section is intended to guide ML-IAPs users who need to train a new ML-IAP, or extend the database of an existing ML-IAP to enhance the transferability to materials science applications.
In \textit{section \ref{chap3}}, we predict a variety of properties that are related to plasticity and fracture based on the ML-IAPs from \textit{section \ref{chap2}}. 
In \textit{section \ref{chap4}}, we compute the core structures and Peierls barriers for five dislocation characters.
In \textit{section \ref{chap5}}, we perform fracture simulations for crack systems on \{100\} and \{110\} planes at T=0K and 100K.
We discuss critical issues regarding the validation of ML-IAPs, the choice of the ML framework, the cutoff radius, and the DFT database in \textit{section \ref{chap6}}. 
The main findings are summarized in \textit{section \ref{chap7}}. 

\section{Consistent training of ML-IAPs}
\label{chap2}
To construct a ML-IAP, a set of consistent quantum mechanical data (atomic positions, energy, force and virial stress) needs to be obtained by performing first principle calculations, typically DFT.  
The atomic coordinates of DFT configurations are represented by descriptors that incorporate permutational, translational and rotational invariance.
The machine learning algorithm provides the (unknown) potential energy surface (PES) as a function of the descriptors, by fitting to the DFT data.
The training of ML-IAPs involves a range of (hyper)parameters that are crucial for the performance of ML-IAPs, such as the DOFs, the cutoff radius, the weights of energy/force/virial, and the choice of hyperparameters for the regression task.
For MTP and ACE, the DOFs are set by the expansion level of the local atomic environment (LAE), i.e., the descriptor that encodes the local atomic position information.
Since the two databases considered in this work are relatively large, a sparse version of the Gaussian process is employed, in which the fitting parameters of GAP are determined by the number of representative sparse points ($M$). 
For the neural network (NN) potential, the DOFs depend on the number of layers and nodes of the network architecture.
Because the number of potentials to be trained grows exponentially with the number of parameters to vary, we focus on optimizing the DOFs and energy/force/virials weights.
In the subsequent subsections, we briefly list the main configurations included in the database and review the key ingredients behind each ML potential that are involved in the optimization of the (hyper)parameters. 
The training scripts used in this work are available at the url specified in the \textit{Data Availability} Section.

\subsection{DFT database}\label{chap2.1}
Two iron DFT databases are employed in the current study. 
The first database, referred to as DB-I, was originally developed for thermomechanics and defects, including dislocations \cite{Dragoni2018_PhysRevMaterials.2.013808,maresca2018screw}, and has been extended recently with an active-learned database to predict fracture in single crystal \cite{lei_2022}. 
DB-I has 14,476 configurations and 160,280 LAEs, including the deformed bcc, face-centered cubic (fcc) and hexagonal close-packed (hcp) primitive cells/supercells, point defects, low index surfaces, $\gamma$ surfaces, surface separation paths and small crack-tips.
The details of the database can be found in Refs. \cite{Dragoni2018_PhysRevMaterials.2.013808,lei_2022}.
We train five ML-IAPs on DB-I to assess their accuracy and compare their computational efficiency.

Another iron DFT database considered here is extracted from a Fe-H database that was used to train a NN potential, which was developed for modelling hydrogen embrittlement \cite{ogata_nnp_meng2021general}.
The pure iron configurations from the Fe-H database are selected, referred to as DB-II.
DB-II contains 9,622 configurations (768,986 LAEs) which include deformed supercells, point defects, low index surfaces, $\gamma$ surfaces, a few symmetric tilt grain boundaries, various dislocation core structures, and inherent structures of the liquid state. 
More details about DB-II and the DFT calculations can be found in Ref. \cite{ogata_nnp_meng2021general}. 

Comparing the size of the two databases, DB-II encompasses a number of atomic environments that is $\sim 5$ times larger than DB-I. 
The difference between DB-I and DB-II in terms of configuration types is that DB-I includes hcp primitive cells, surface separation paths and crack tips, while DB-II has Bain path, diffusion pathways, symmetric tilt grain boundaries, dislocation structures, and inherent structures of the liquid phase. 
To investigate the influence of the choice of the database on the prediction of dislocation and fracture properties, we train GAP on both DB-I and DB-II, and we refer to the two potentials as GAP-DB-I and GAP-DB-II.

\subsection{Training of ML-IAPs}\label{chap2.2}

Here, we use DB-I to train ML-IAPs based on four state-of-the-art packages. 
All the packages have been previously applied to metallic systems, are open-source, well-documented, ready-to-use, and provide various examples, including basic training and validation, prediction, and active learning if applicable.
Furthermore, these codes are interfaced with the Large-scale Atomic/Molecular Massively Parallel Simulator (LAMMPS), which enables the efficient parallelization of large-scale atomistic simulations (e.g., plasticity and fracture). 
Although the training is not needed once the potential is obtained, the training efficiency should be considered as one of the performance indices of ML-IAPs, especially when active/iterative learning is used, or in the context of applications where extensive training is required. 
Note that a fair comparison of the training efficiency is not possible among the employed packages and is not conducted in this study. 
A brief summary of the main features of the packages employed in the current study is provided in \textit{Supplementary Material S1.1}. 

The basic assumption of all these ML-IAPs is the localization of the total energy, whereby the total energy of an atomic system can be written as the sum of per-atom contributions
	\begin{equation}
		E_{\rm tot} =\sum_i E_i,
	\end{equation}
where $i$ runs over all atoms of the system.
A central task of ML-IAPs is to approximate the local energy $E_i$, the expression (expansion) of which differs in each ML framework.
In the following, we summarize the LAE expansion approach and the regression method of each ML framework. 
The training and convergence analysis of RMSE is also presented. 
The database is randomly split into training and testing sets with a 9:1 ratio for all training realizations. 
The cutoff radius is set to 6.5 Å if not specified otherwise.

\subsubsection{GAP}
GAP computes the local energy of atom $i$ based on the Smooth Overlap of Atomic Positions (SOAP) descriptor $\bm{q_i}$ by \cite{bartok2010gaussian}
	\begin{equation}
		E_i=\sum_s \alpha_s K(\bm{q}_s,\bm{q_i})
	\end{equation}
where $\bm{q}_s$ is the descriptor of atom $s$ from the database. 
The array $\bm{\alpha}$ of weights $\alpha_s$ is determined via \cite{szlachta2014accuracy}
	\begin{equation}\label{eq.5.2.21}
		\bm{\alpha}=(\bm{K}_{MM}+\bm{K}_{MN}\bm{L}\bm{\Lambda}^{-1}\bm{L}^T\bm{K}_{NM})^{-1}_{MM} \bm{K}_{MN}\bm{L}\bm\Lambda^{-1}\bm{t}, 
	\end{equation}
where $\bm{\Lambda}=\sigma_v^2\bm{I}$ is a regularization noise and $\sigma_v$ is the tolerance (expected error) in fitting the DFT data. 
Here, we set the default expected energy ($\sigma_v^{\rm energy}$), force ($\sigma_v^{\rm force}$) and virial ($\sigma_v^{\rm virial}$) errors to be 5 meV/atom, 0.2 eV/Å, and 0.01 eV/atom. 
In practice, it is found that the accuracy of the potential can be improved by choosing different values of the expected errors for different types of input data \cite{szlachta2014accuracy,Dragoni2018_PhysRevMaterials.2.013808}.
Therefore, we apply different $\sigma_v^{\rm energy}, \sigma_v^{\rm force}$, and $\sigma_v^{\rm virial}$ to subsets of the DFT data according to their expected accuracy (see GAP training scripts in \textit{Data availability}). 
$\bm{t}$ is the DFT observable, i.e. total energies, forces and virials.
$\bm{L}$ is a linear operator that converts the DFT total energy, force and virial into the unknown localized atomic energies. 
Note that this is a sparse version of the Gaussian process (GP) coefficients, where $N$ is the total number of LAE from the database and $M$ is a representative subset of $N$\footnote{
The $M$ ($M<<N$) representative data points are selected using a leverage-score CUR algorithm that is implemented in the \textit{QUIP} package \cite{bartok2010gaussian,Csanyi2007-py,Kermode2020-wu}.
Thus, the sparse GP filters out part of the redundant data and reduces the computational cost of both the training and the evaluation steps.}. 
At the evaluation step, the computational cost reduces from $\mathcal{O}(N)$ to $\mathcal{O}(M)$, enabling the simulation of large-scale systems based on a large DFT database.
The elements of $\bm{K}$ ($K_{ij}$) are the covariances between $\bm{q}_i$ and $\bm{q}_j$, which is also the kernel that measures their similarity. 
The dot product kernel is applied
	\begin{equation}
		K_{ij}=|\bm{q}_i \cdot \bm{q}_j|^\zeta\ , 
	\end{equation}
where $\zeta$ is a parameter that is used to sharpen the sensitivity towards changes of the atomic positions. 
Here, $\zeta=4$ is used \cite{szlachta2014accuracy,Dragoni2018_PhysRevMaterials.2.013808}.

\begin{figure}[H]
	\centering
	% Answer: [trim={left bottom right top},clip]
	\includegraphics[trim=0 80 0 0, width=14cm]{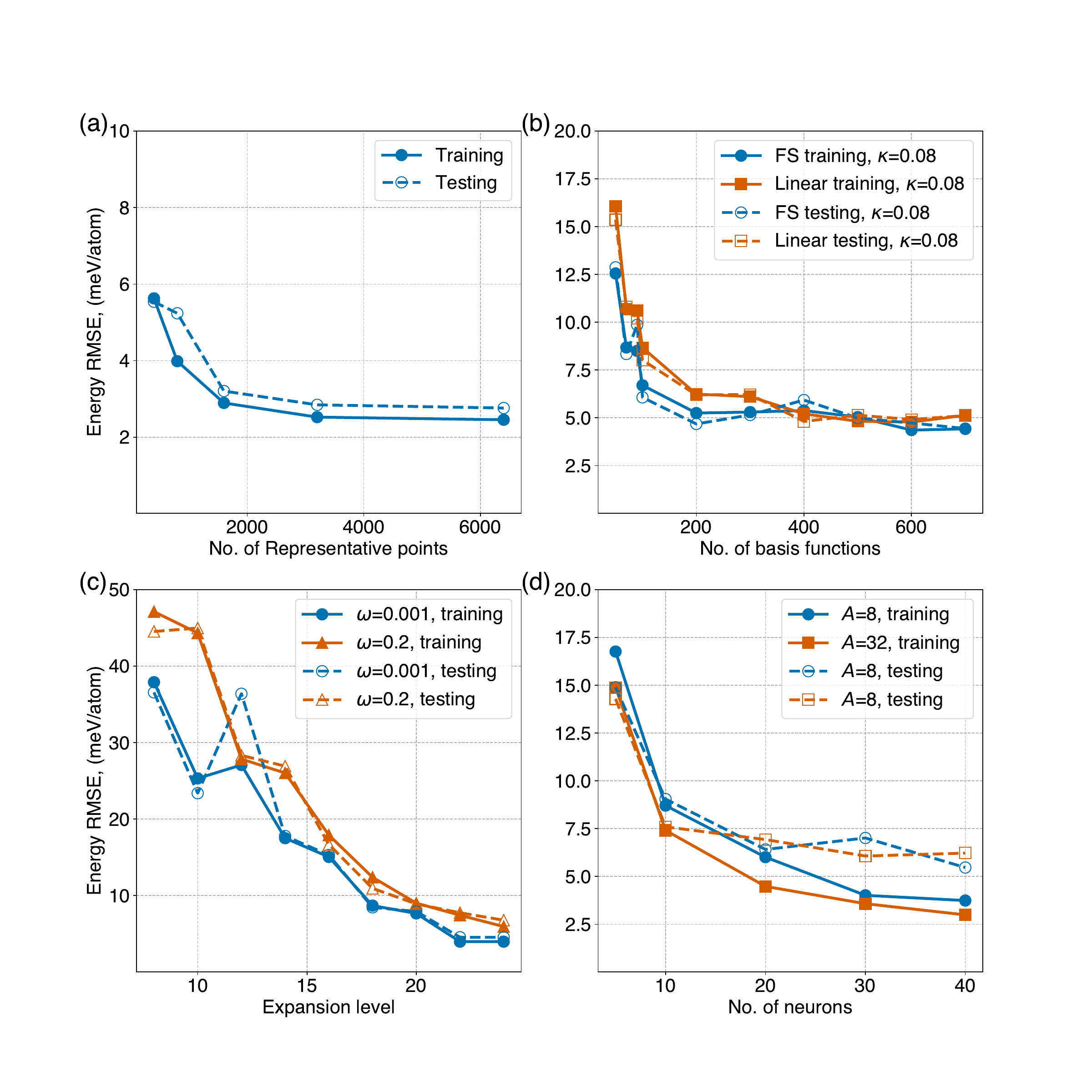} 
	\caption{
    Training and testing RMSE on energy for \textbf{(a)} GAP, \textbf{(b)} PACE, \textbf{(c)} MTP, and \textbf{(d)} NNP.
		Solid and dashed lines indicate the training and testing errors, respectively.		
		The number of representative LAE in GAP equals the number of sparse points ($M$).
        The energy/force weight ratio is defined as $1/\kappa-1$ in PACE-FS and PACE-Linear. 
		$\omega$ is the energy/force weight ratio of MTP.
		$A$ is the number of angular basis functions of NNP. 
        For clarity, only representative hyperparameter choices are shown here. The complete hyperparameter optimization is reported in Fig. S2 of the \textit{Supplementary Material S1.3}.
	}
	\label{fig:convergence_panel}
\end{figure}

We use the open-source package\textit{ QUantum mechanics and Interatomic Potentials} (\textit{QUIP}) to fit GAP \cite{bartok2010gaussian,Csanyi2007-py,Kermode2020-wu}.
A distance-based 2-body descriptor and two TurboSOAP descriptors are employed \cite{bartok2013representing,turboSOAP}, i.e., one “inner” TurboSOAP with $r_{\rm cut}=3Å$ and another “outer” TurboSOAP with $r_{\rm cut}=6.5Å$.
To ensure convergence of the sparse GP, we investigate the influence of $M$ on RMSE by fixing the LAE expansion ($n_{\rm max}=8,\ l_{\rm max}=8$). 
As shown in Fig. \ref{fig:convergence_panel}a, the training and testing RMSE on energies converges at increasing $M$ ($>3200$).
Therefore, we fix $M=4800$ for the rest of the GAP potentials trained on DB-I.
Note that the saturated $M$ is database-dependent. 
Moreover, the accuracy of GAP depends also on the LAE expansion level \cite{patrick_2020_gap}, which is determined by the radial ($n_{\rm max}$) and angular ($l_{\rm max}$) expansion degrees. 
In order to find the optimal combination of $n_{\rm max}$ and $l_{\rm max}$, we conduct a grid search. 
Our findings highlight a fundamental difference between TurboSOAP and original SOAP descriptors (see \textit{Supplementary Material S1.2.1}).
The optimal combination of hyperparameters that yields the lowest energy and force RMSE are listed in Table S1 of \textit{Supplementary Material S1.2}. 
In addition, we provide the convergence analysis of GAP trained on DB-II in \textit{Supplementary Material S1.4}.

\subsubsection{ACE}
In the general ACE formalism, the local energy is expressed as a function of atomic properties \cite{drautz2019atomic,drautz2020atomic}
	\begin{equation}\label{eq:E_ace}
		E_i=\mathcal{F}(\varphi_i^{(1)},\cdots,\varphi_i^{(P)})\ ,
	\end{equation}
where $\varphi_i^{(p)}$  $(p=1,...,P)$ can be expanded as 
	\begin{equation}
		\varphi_i^{(p)}=\sum^{\nu_{\rm max}}_{\nu=1}c_{\nu}^{(p)} B_{i\nu}.
	\end{equation}
$c_\nu^{(p)}$ are the expansion coefficients to be fitted by the regression algorithm.
$B_{i\nu}$ is the basis constructed by the atomic cluster expansion, where the permutation, reflection, and rotation invariants are incorporated \cite{drautz2019atomic}. 
The $\bm{B}$ basis is efficiently reconstructed by the multiplication between the generalized Clebsch-Gordan coefficients and the permutation-invariant $\bm{A}$ basis functions (see details in Ref. \cite{drautz2019atomic,lysogorskiy2021performant}).
Two functional forms of $\mathcal{F}$ (Eq. \ref{eq:E_ace}) have been applied recently \cite{lysogorskiy2021performant,bochkarev2022efficient}.
The simplest model expands the local energy as a linear function of the density 
	\begin{equation}\label{eq:linear-ace}
		E_i=\varphi_i^{(1)}
	\end{equation}
while a nonlinear Finnis-Sinclair-type model is formulated as
	\begin{equation}\label{eq:fs-ace}
		E_i=\varphi_i^{(1)}+\sqrt{\varphi_i^{(2)}}. 
	\end{equation}

We use the \textit{PACEMAKER} package \cite{lysogorskiy2021performant,bochkarev2022efficient} to train these two versions of ACE, i.e., the linear and Finnis-Sinclair models (referred to as PACE-L and PACE-FS, respectively). 
We train PACE-L and PACE-FS with correlation orders $N_{\rm c}=3-6$ and energy/force weight coefficients ratio $\kappa = 0.08, 0.16, 0.32$ (weights are defined as $1/\kappa-1$). 
The results are shown in Fig. \ref{fig:convergence_panel}b.
We find that the correlation order and the weight coefficients ratio have no significant influence on RMSE.
The training and testing RMSE on energy for both PACE IAPs are saturated with the increased number of basis functions. 
PACE-FS converges faster than PACE-L with the increasing number of basis functions, and can reach a smaller RMSE when the number of basis functions is large. 
The optimal combinations of hyperparameters that yield the lowest energy and force RMSE are given in Table S2 of \textit{Supplementary Material S1.2}. 
The full convergence plots for the PACE potentials are shown in Fig. S2b of the \textit{Supplementary Material S1.3}.

\subsubsection{MTP}

MTP represents the local energy of atom $i$ as a linear combination of the complete basis functions set $B_\alpha$, which is constructed from the moment tensor descriptor \cite{shapeev2016moment}
	\begin{equation}
		E_i = \sum_\alpha \zeta_\alpha B_\alpha(\mathbf{r}),
	\end{equation}
where $\zeta_\alpha$ are the coefficients to be fitted. 
The moment descriptor is designed to be invariant with respect to atomic permutations, rotations, and reflections, and it consists of a radial part and an angular part
	\begin{equation}
		M_{\mu,\nu}(\mathbf{r})=\sum_{j\in S_i^{R_c}} 
		\overbrace{f_\mu (|r_{ij}|)}^{\rm\scriptsize radial} \overbrace{\underbrace{\mathbf{r}_{ij}\otimes\cdots\otimes\mathbf{r}_{ij}}
			_{{\scriptstyle\nu \rm \ times}}}^{\rm\scriptsize angular},
	\end{equation}
where $j$ runs over all atoms inside the cutoff radius $R_c$, $\mathbf{r}_{ij}$ is the vector of the relative coordinates between atoms $i$ and $j$, and $|r_{ij}|$ is the distance between atoms $i$ and $j$.
The radial part is expanded inside $R_c$ as
	\begin{equation}
		f_\mu(|r_{ij}|)=\sum_{\beta=0}^{\beta_{\rm max}} c_{\mu,\beta}
		\varphi_{\beta}(|r_{ij}|)(R_c-|r_{ij}|)^2,
	\end{equation}
where $\varphi_\beta$ are Chebyshev polynomials and $c_{\mu,\beta}$ are the radial expansion fitting coefficients.
The angular part is defined as $\nu$-times the tensor product of atomic position vectors, encoding the angular information of the atomic environment $S_i\in{R_c}$.
The expansion degree of MTP is determined by $\mu$, $\nu$ and $\beta_{\rm max}$.
The developers of MTP \cite{gubaev2019accelerating,novikov2020mlip} define the \textit{level} of moments as
	\begin{equation}
		{\rm lev} M_{\mu,\nu} = 2+4\mu+\nu.
	\end{equation}

This is an optimal relation obtained after a number of tests \cite{gubaev2019accelerating}, which is implemented into the \textit{MLIP-2} package \cite{novikov2020mlip}. 
The detailed relation between basis function $B_\alpha$ and expansion level $M_{\mu,\nu}$ is coded into the \textit{MLIP-2} package and can be found in Ref. \cite{gubaev2019accelerating,novikov2020mlip}.
The number of basis functions is determined by the \textit{level} of moments, which determines how many times the contractions of the moments are performed.  
We train MTP with \textit{levels} ranging from 8 to 24.
The force weights ($\omega$) are optimized by fixing the energy weights to 1 and virial weights to 0.05, as plotted in Fig. \ref{fig:convergence_panel}c.
Both energy and force RMSE are converged with the increased level of expansion, reaching the accuracy limit at $lev=22$.
$\omega=0.001$ has lower energy RMSE than $\omega=0.2$, which is expected because of the reduction of the force weights in the loss function.
The optimal combination of hyperparameters that yields the lowest energy and force RMSE are given in Table S3 of \textit{Supplementary Material S1.2}. 

\subsubsection{NNP}
NN potentials were first introduced by Behler and Parrinello to describe the atomistic PES \cite{behler2007generalized}.
Here, we apply a 2-layers NN, in which the atomic energy of atom $i$ is expressed as
	\begin{equation}\label{eq:nn_arch}
		E_i=f_1^{(3)}
		\left\{b^{(3)}_1+\sum_{k=1}^{k_{\rm max}} w_{k1}^{(23)}\cdot f_k^{(2)}
		\left[b_k^{(2)}+\sum_{j=1}^{j_{\rm max}}w_{jk}^{(12)}\cdot f_j^{(1)}
		(b_j^{(1)}+\sum_{i=1}^{{\mu}} w_{ij}^{(01)}\cdot G_i^{\mu})
		\right]
		\right\},
	\end{equation}
where $j_{\rm max}$ and $k_{\rm max}$ are the number of nodes in the first and second hidden layer, $w$ and $b$ are the weights and the bias to be fitted, $f$ are the activation functions (which can be different for each node), and $G_i^\mu$ is the atomic symmetry function (ASF) that describes the LAE.

We use the \textit{n2p2} package to train NNPs \cite{behler2007generalized,morawietz2016van,singraber2019parallel}. 
\textit{n2p2} implements multiple atomic symmetry functions, activation functions, and cutoff functions.
The radial and angular symmetry functions are chosen to be 
	\begin{equation}\label{eq:acsf1}
		G^2_i = \sum_{j \neq i} \mathrm{e}^{-\eta(r_{ij} - r_\mathrm{s})^2}
		f_c(r_{ij})
	\end{equation}
and
	\begin{equation}\label{eq:acsf2}
		\begin{split}
			G^9_i = 2^{1-\zeta} \sum_{\substack{j,k\neq i \\ j < k}}
			\left( 1 + \lambda \cos \theta_{ijk} \right)^\zeta
			\mathrm{e}^{-\eta( (r_{ij}-r_s)^2 + (r_{ik}-r_s)^2 ) }
			f_c(r_{ij}) f_c(r_{ik}),
		\end{split}
	\end{equation}
respectively \cite{behler2007generalized,behler2011atom}, where $r_{ij}$ is the distance between atoms $i$ and $j$, $\theta_{ijk}$ is the angle formed by atom triplets centered on atom $i$, $r_s$ and $\eta$ are Gaussian center and width parameters, $\zeta\in\{1,6\}$ and $\lambda\in\{-1,1\}$ are two hyperparameters. 
A logistic activation function $f_a(x) = 1 / (1 + \mathrm{e}^{-x})$ is applied for two hidden layers and the identity function $f_a(x) = x$ is used for the output layer. 
The cutoff function is set to $f_c(x) = ((15 - 6x)x - 10) x^3 + 1$.

The influence of the angular expansion level and the number of neurons is explored by fixing the radial expansion to 10 and the force weight ratio (relative to energy) to 0.1.
Fig. \ref{fig:convergence_panel}d plots the energy RMSE as a function of neuron numbers for different numbers of angular basis functions. 
The test RMSE is saturated with 20 neurons.
The increase of $A$ does not reduce the RMSE significantly, which indicates that $A=8$ is enough to encode the 3-body interactions. 
The optimal combinations of hyperparameters that yield the lowest energy and force RMSE are given in Table S4 of \textit{Supplementary Material S1.2}. 

\subsection{Model uncertainty quantification}\label{chap2.3}
In this work, model uncertainty quantification is proposed as the final step of the ML-IAP validation procedure (Fig. \ref{fig:validation}).
In \textit{section \ref{chap4}}, we make extensive use of the per-atom uncertainty of GAP and PACE in order to assess the degree of extrapolation of the LAE from the configurations that are present in the training database. 
This approach can be applied within the context of benchmark simulations of extended defects, and it enables to verify the transferability of the ML-IAPs to large scale simulations involving these defects.
The per-atom uncertainty of GAP, which is the square root of the Gaussian process variance and has units of energy, has been widely used in iterative/active learning \cite{lei_2022,Jinnouchi_2019_PhysRevLett.122.225701,Jinnouchi_2020}. 
We use \textit{QUIP} package to evaluate the GAP variance.
\textit{PACEMAKER} package implements the per-atom uncertainty based on the extrapolation degree $\gamma$ of the local atomic environment.
This extrapolation degree reflects a geometric distance between the predicted atomic environment and the atomic configurations in the database, and is computed based on D-optimality \cite{lysogorskiy2023active}. 
The evaluation of $\gamma$ can be directly invoked via LAMMPS commands. 

\subsection{ML-IAP performance and cost assessment }

\begin{figure}[H]
	\centering
	\includegraphics[trim=0 30 0 20, width=14cm]{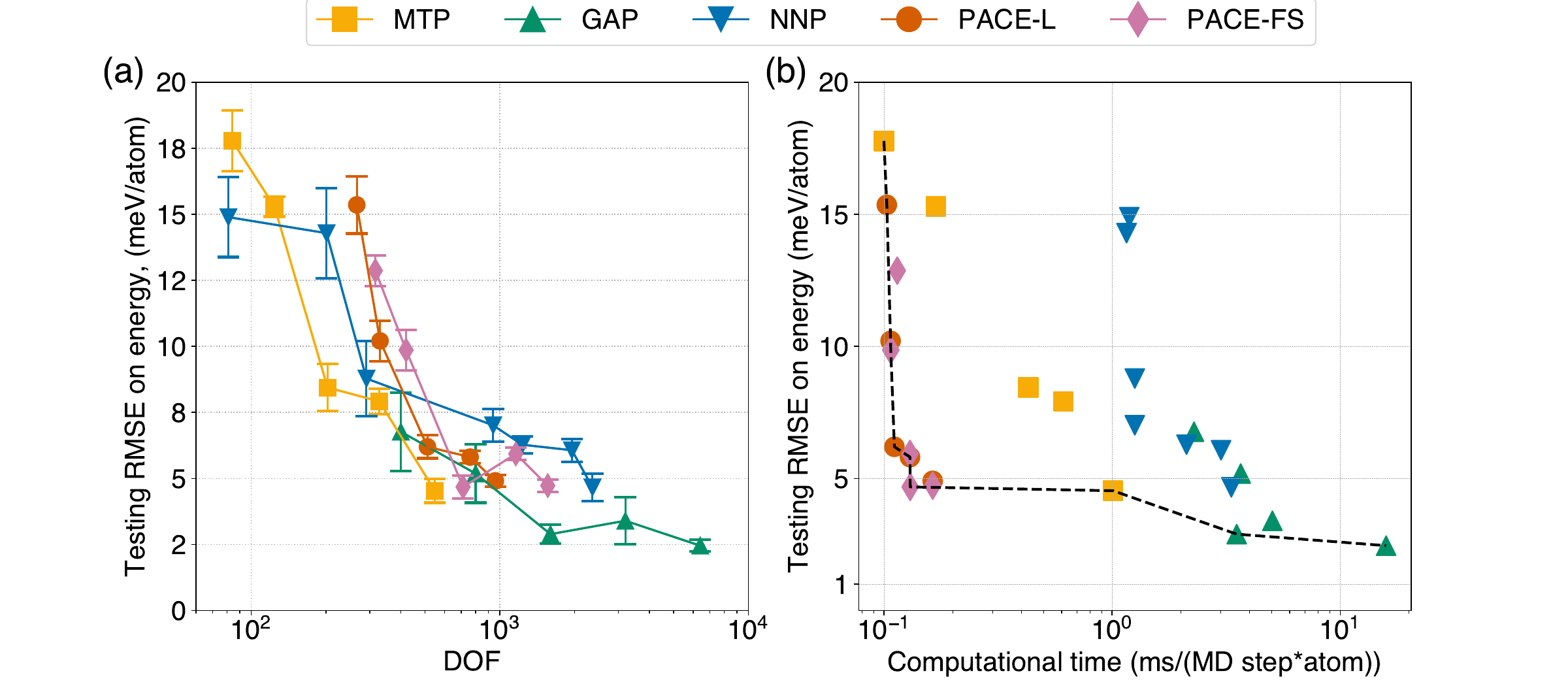}
	\caption{
    \textbf{(a)} Testing energy RMSE as a function of the number of DOFs for the five ML-IAPs considered in the current study. 
	\textbf{(b)} Testing energy RMSE versus computational cost for the different ML-IAPs. 
	The black dashed line indicates an approximate Pareto frontier formed by the convex hull of points lying on the bottom left of the chart, which approximates an optimal trade-off between accuracy and computational cost.
	The computational cost is assessed via LAMMPS calculations of parallelized simulations using 32 cores on a AMD Rome CPU, 7H12 (2x), 64 Cores/Socket, 2.6GHz, 280W.
	}
	\label{fig:pareto_front}
\end{figure}

Fig. \ref{fig:pareto_front}a shows the testing energy RMSE of the optimized ML-IAPs as a function of the number of DOFs (see Fig. S4 of the \textit{Supplementary Material S1.3} for the force RMSE counterpart).
The error bars indicate the standard deviation, which is computed based on five independent trainings. 
All ML-IAPs are able to reach a testing RMSE lower than 5 meV/atom, indicating that a good fitting accuracy can be achieved by increasing the number of DOFs.
NNP shows the largest error bar compared with the other ML-IAPs, reflecting the stochastic nature of neural networks. 
In particular, GAP shows the minimum test RMSE ($\sim$2 meV/atom) and MTP can reach 5 meV/atom accuracy with the minimum number of DOFs, compared with the other ML-IAPs.
Since all ML-IAPs can reach comparable testing RMSE, albeit at different DOFs and hence model complexity, efficiency is expected to be an important benchmark property. 
We calculate the computing cost of all ML-IAPs by performing MD in the NVE ensemble for 16,000 atoms using 32 cores.
The performance of parallelization is involved in this benchmark study, since our interest is in large-scale MD simulations where parallelization is unavoidable. 
The test energy RMSE against computational cost is plotted in Fig. \ref{fig:pareto_front}b. 
The force RMSE versus computational cost is reported in Fig. S4b of the \textit{Supplementary Material S1.3} and shows a similar trend.
For each ML-IAP, energy and force RMSEs decrease with the increase of the computational time. 
The energy RMSE of $\sim 5$ meV/atom can be achieved for all ML-IAPs.
In particular, GAP is able to reach an accuracy of $\sim 2$ meV/atom. 
All force RMSEs converge to $\sim 0.04$ eV/Å except for NNP ($\sim$ 0.06 eV/Å). 
PACE-L and PACE-FS are close to each other, occupying the Pareto frontier of the energy and force RMSE.
GAP is also on the Pareto front of the energy RMSE, yet at much longer computational timescales. 
The speed of MTP ranges from being the fastest to being one order of magnitude slower, indicating the rapid growth of the computational cost upon increasing the expansion level. 
GAP has the lowest energy RMSE but is up to two orders of magnitude slower than PACE-FS. 
It is worth noting that TurboSOAP descriptor is applied in this study, which is already two orders of magnitude faster than the ordinary SOAP \cite{turboSOAP}.

We have optimized an array of ML-IAPs, that achieve close-to-DFT accuracy (measured in terms of test RMSE) at different degrees of computational efficiency. 
In the following, we investigate the ability of these ML-IAPs to predict benchmark properties that are relevant to applications, i.e. dislocations and cracks.

\section{Accuracy of the prediction of benchmark properties}
\label{chap3}
The purpose of ML-IAPs is to predict physical/chemical properties/processes in extended systems (that include defects such as dislocations, cracks, grain boundaries) with quantum accuracy. 
The atomistic structure and the mobility of these extended defects depend on a number of elementary properties, which should be predicted with sufficient accuracy by ML-IAPs.
Here, we predict the lattice parameter ($a_0$), the vacancy formation energy ($E_{\rm v}$), the elastic constants ($C_{ij}$), the surface energies of low-index planes ($\gamma_{\rm s}$) and the unstable stacking fault energies associated with active slip planes in bcc crystals ($\gamma_{\rm us}$).
These physical properties are related to either dislocation (see Hirth and Lothe \cite{hirth1983theory}]) or fracture properties (see Griffith \cite{Griffith} and Rice \cite{RICE1992239} theories). 

\begin{figure}[H]
 	\centering
 	\includegraphics[trim=100 60 0 0,  width=18cm]{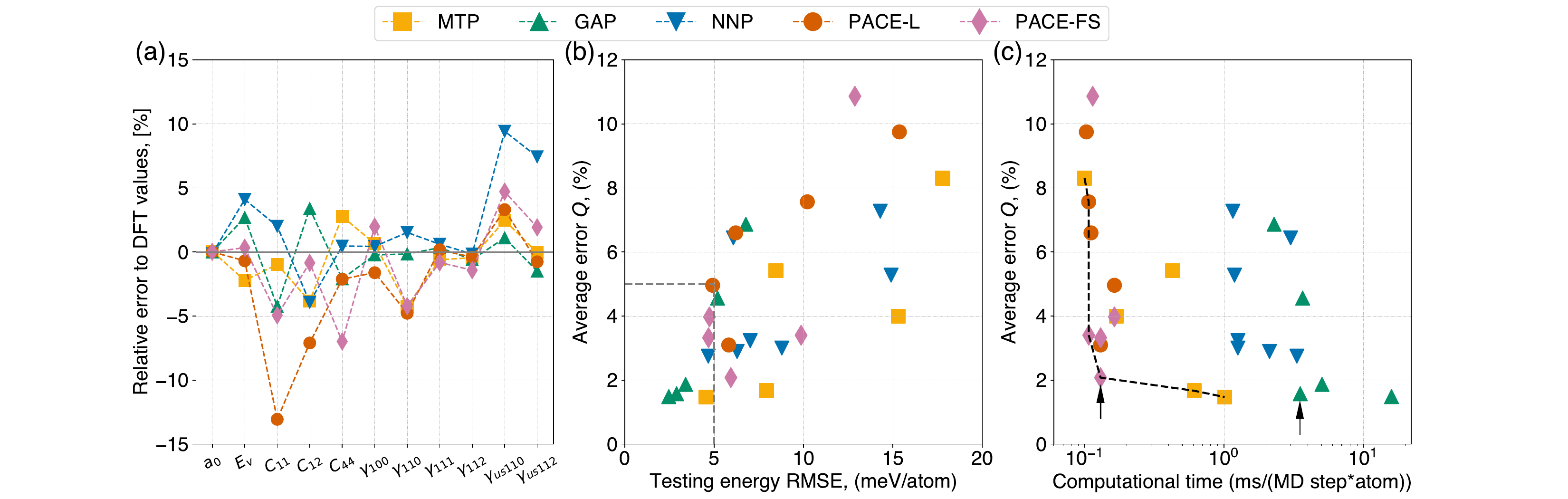}
 	\caption{
 		\textbf{(a)} Relative errors with respect to DFT for various material properties predicted by five ML-IAPs with the lowest $Q$.		
        In particular, lattice parameter ($a_0$), vacancy formation energy ($E_v$), elastic constants ($C_{11},C_{12},C_{14}$), surface energies ($\gamma_{100},\gamma_{110},\gamma_{111},\gamma_{112}$) and maximum (unstable) stacking fault energies on \{110\} and \{112\} planes are considered.
		\textbf{(b)} Error index $Q$ as a function of the test  energy RMSE. 
        \textbf{(c)} Error index $Q$ as a function of the computational speed. 
        The measurement of the computational speed is detailed in Fig. \ref{fig:pareto_front} and related main text.
        The potentials highlighted by the arrows are used for further validation in the following sections.
	}
 	\label{fig:qfactor}
 \end{figure}

To allow for a quick comparison of the quality of these ML-IAPs, we introduce the average error or ``quality factor'' $Q$
\begin{equation}\label{eq:q_factor}
		Q=\frac{1}{N_p}\sum_{i=1}^{N_p}\left|\frac{{f^{ML}_i-f^{DFT}_i}}{f^{DFT}_i}\right|,
\end{equation}
where $f_i$ is the value of the property ($a_0$, $C_{ij}$, $E_{\rm v}$, $\gamma_{\rm s}$, and $\gamma_{\rm us}$) computed with either the ML-IAP or DFT (see Fig. S6 of the \textit{Supplementary Material S2.1} for generalized stacking fault energy profile).
$N_p$ is the total number of properties.
All the ML-IAP calculations are performed in LAMMPS \cite{plimpton1995fast} using the workflow made publicly available at the url provided in the \textit{Data Availability} section.
The benchmark properties predicted by DFT are listed in Ref. \cite{Dragoni2018_PhysRevMaterials.2.013808}, and are consistent with the database that has been used for the training of the five ML-IAPs (DB-I).
Fig. \ref{fig:qfactor}a shows the benchmark properties predicted by the ML-IAPs with the smallest $Q$.
With only a few exceptions, all ML-IAPs are capable of reproducing the benchmark properties with a relative error equal or smaller than 5\%, highlighting the accuracy of the ML-IAPs.
PACE-L underestimates $C_{11}$ by 13\% and NNP overestimates $\gamma_{\rm us}$ (\{110\} plane) by 9.8\%. 
The PACE-L and NNP performance can be improved by increasing the training weights, in the loss function, of the elastically-deformed and $\gamma$ surface structures, respectively.
The results however suggest that a converged RMSE does not indicate the ability to predict all the desired physical properties at high accuracy. 

Fig. \ref{fig:qfactor}b shows the correlation between the test energy RMSE and the accuracy of the physical property predictions (see Fig. S7 of the \textit{Supplementary Material S2.2} for the force RMSE counterpart). 
The overall trend is that a smaller RMSE correlates to lower $Q$, which holds when the energy RMSE is larger than 5 meV/atom. 
$Q$ is not significantly improved with a further reduction of the energy RMSE, therefore confirming the observation that low RMSE alone cannot be used as an indicator for highly accurate physical property predictions.
For example, there is no apparent relationship between $Q$ and RMSE in the case of PACE-FS, when RMSE is smaller than 8 meV/atom. 
Non-linear models, such as NNP and PACE-FS, predict $Q<5\%$ for RMSE  8-10 meV/atom, indicating that non-linear models with significant RMSE can still predict physical properties accurately. 

Moreover, $Q$ is plotted as a function of the computational speed, as shown in Fig. \ref{fig:qfactor}c. 
The Pareto front is occupied by PACE-L, PACE-FS and MTP, and is slightly different from the Pareto front based on RMSE in Fig. \ref{fig:pareto_front}b. 
This result further indicates the limitation of using RMSE as the only measure of the ML-IAP quality. 
All potentials can reach an average error $Q$ lower than 3\%, and the smallest $Q$, which is predicted by GAP and MTP, is less than 1.5\%. 
Since PACE-FS and GAP occupy the Pareto front (Fig. \ref{fig:pareto_front}b and Fig. \ref{fig:qfactor}c) as the most efficient and accurate ML-IAPs respectively, they are selected to perform further validation based on benchmark MD simulations of extended defects, i.e., dislocations and cracks.
Additionally, since the per-atom model uncertainty quantification is implemented in both GAP and PACE frameworks, this will be used in order to benchmark the ML-IAP performance for the prediction of extended defects.
As highlighted by the arrows in Fig. \ref{fig:pareto_front}c, GAP ($l_{\rm max}=8, \ n_{\rm max}=8$) and PACE-FS ($B=400,\ \kappa=0.08$) with the lowest $Q$ are selected for further testing. 
GAP is trained on two databases (DB-I and DB-II) in order to assess the impact of the database selection on the prediction of dislocation properties and fracture mechanism.

\section{Dislocation core structures and Peierls barriers}
\label{chap4}

In this section, we focus on the prediction of dislocation properties and fracture mechanisms using three potentials, i.e., GAP-DB-I , GAP-DB-II ($l_{\rm max}=8, \ n_{\rm max}=8$) and PACE-FS (trained on DB-I).
Screw, edge and mixed dislocation characters are investigated. 
The screw dislocation is known to control the low-temperature plasticity in bcc iron \cite{dorn1963nucleation,ventelon2013ab}.
$a_0/2[111](1\bar{1}0)$ $70.5^{\circ}$ mixed dislocation (referred to as M111 dislocation) has been shown to be an important character during dislocation loop expansion in bcc Ta at low temperature \cite{kang2012singular}, yet the importance for bcc iron remains to be established.
Since edge dislocations in bcc iron are known to be extremely mobile, they are of less interest than screw characters. 
However, edge dislocations can also be formed via dislocation interactions, which may control the intersection node mobility that plays an important role in dislocation network evolution \cite{bertin2022enhanced}.
Therefore, the core structures and Peierls barriers of screw, $a_0[100](010)$ edge, $a_0[100](011)$ edge, $a_0/2[\bar{1}\bar{1}1](1\bar{1}0)$ edge, and M111 dislocations are computed at T=0K under zero applied stress. 

\begin{figure}[H]
	\centering
	\includegraphics[width=14cm]{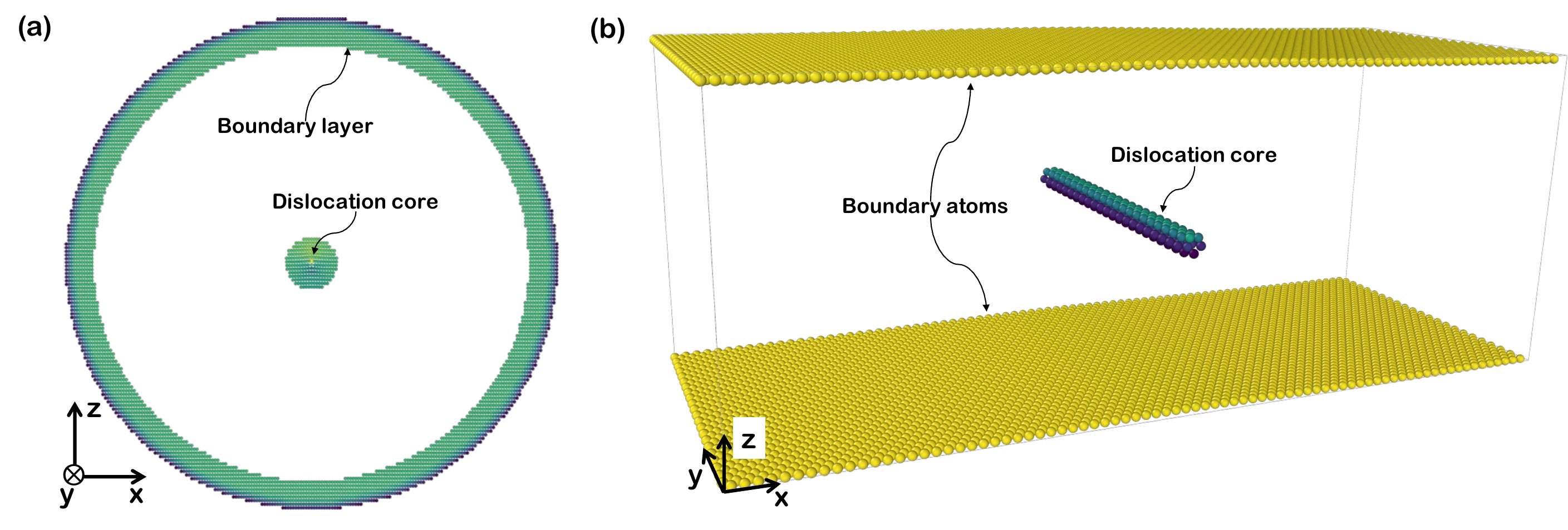}
	\caption{\textbf{(a)} Rigid boundary (RB) configuration and
		\textbf{(b)} Periodic array of dislocation (PAD) configuration. 
        Both configurations are used to find the equilibrium structure of the dislocation.
        PAD configuration is further used for the NEB calculations.
	}
	\label{fig:disl_model}
\end{figure}

%Two simulation configurations with different boundary conditions (BCs) are employed, i.e., the rigid boundary (RB) configuration and the periodic array of dislocations (PAD) configuration \cite{BACON20091}, as illustrated in Fig. \ref{fig:disl_model}a and Fig. \ref{fig:disl_model}b respectively.
The rigid body (RB) configuration is employed to find the equilibrium core structure of the dislocation (Fig. \ref{fig:disl_model}a). 
In the RB configuration, a single dislocation is put at the center of a cylinder ($r\approx 160 Å$) by applying the anisotropic linear elastic displacement field to all atoms \cite{BACON20091}. 
The open source software \textit{Atomsk} is used to generate the initial configurations with the linear elastic displacement field \cite{hirel2015atomsk}. 
The atoms in the inner cylinder ($r\approx 140 Å$) are allowed to relax while the atoms in the outer rim of the cylinder are kept fixed \cite{dftCore_fellinger2018geometries}. 
Periodic boundary conditions (PBC) are applied along the dislocation line direction ($y$). 
A convergence test is conducted to ensure that the predicted dislocation core structure is converged with respect to the configuration radius. 
In order to compare with the DFT predictions \cite{dftCore_fellinger2018geometries}, we apply the RB configuration and relax the dislocation structures using the conjugate gradient (CG) algorithm with a force tolerance of 10$^{-12}$ eV/Å. 

Next, periodic array of dislocations (PAD) configuration is used to compute the Peierls barrier \cite{BACON20091} (Fig. \ref{fig:disl_model}b).
The sample orientations used for modelling different dislocation characters are listed in Table \ref{tab:model_orient}.
PBCs are applied along the dislocation line ($y$) and the slip ($x$) directions.  
We apply the climbing-image nudged elastic band (CI-NEB) method to calculate the Peierls barrier for all dislocation characters \cite{henkelman2000climbing}.
CI-NEB requires the initial and final configurations of the dislocation glide process. 
The initial and final PAD configurations are constructed by following the procedure described in section 2.3 of Ref. \cite{BACON20091}.
All models are generated in LAMMPS and the corresponding scripts are provided at the url reported in the \textit{Data Availability} section.  
At the start of the NEB calculations, replicas are created by linear interpolation of the atomic positions between the initial and final states. 
We use 32 replicas and the FIRE algorithm with a force tolerance of 10$^{-3}$ eV/Å \cite{bitzek2006structural}. 
The predicted atomic positions around the dislocation cores are compared quantitatively with published DFT calculations \cite{dftCore_fellinger2018geometries}.

\begin{table}[H]
	\begin{center}
		\caption{\label{tab:model_orient}Crystallographic orientations of the dislocation configurations.}
		\begin{tabular}{ c|c|c|c|c } 
			\hline
			Dislocation & Character & $x$ & $y$ & $z$ \\
			\hline 
			$a_0/2[111]$ & screw &  $[1\bar{2}1]$ & $[111]$ & $[\bar{1}01$] \\ 
			$a_0[100](010)$ & edge &  $[100]$ & $[010]$ & $[001]$ \\ 
			$a_0[100](011)$ & edge & $[100]$ & $[01\bar{1}]$ & $[011]$ \\\
			$a_0/2[\bar{1}\bar{1}1](1\bar{1}0)$  & edge & $[11\bar{1}]$ & $[\bar{1}21]$  & [101] \\
			$a_0/2[111](1\bar{1}0)$ &  $70.5^{\circ}$ mixed & $[12\bar{1}]$  & [$\bar{1}$11] & [101] \\
			\hline
		\end{tabular}
	\end{center}
\end{table}

\subsection{Screw dislocation} \label{chap4.1}

\begin{figure}[H]
	\centering
	\includegraphics[trim=100 80 0 0,  width=18cm]{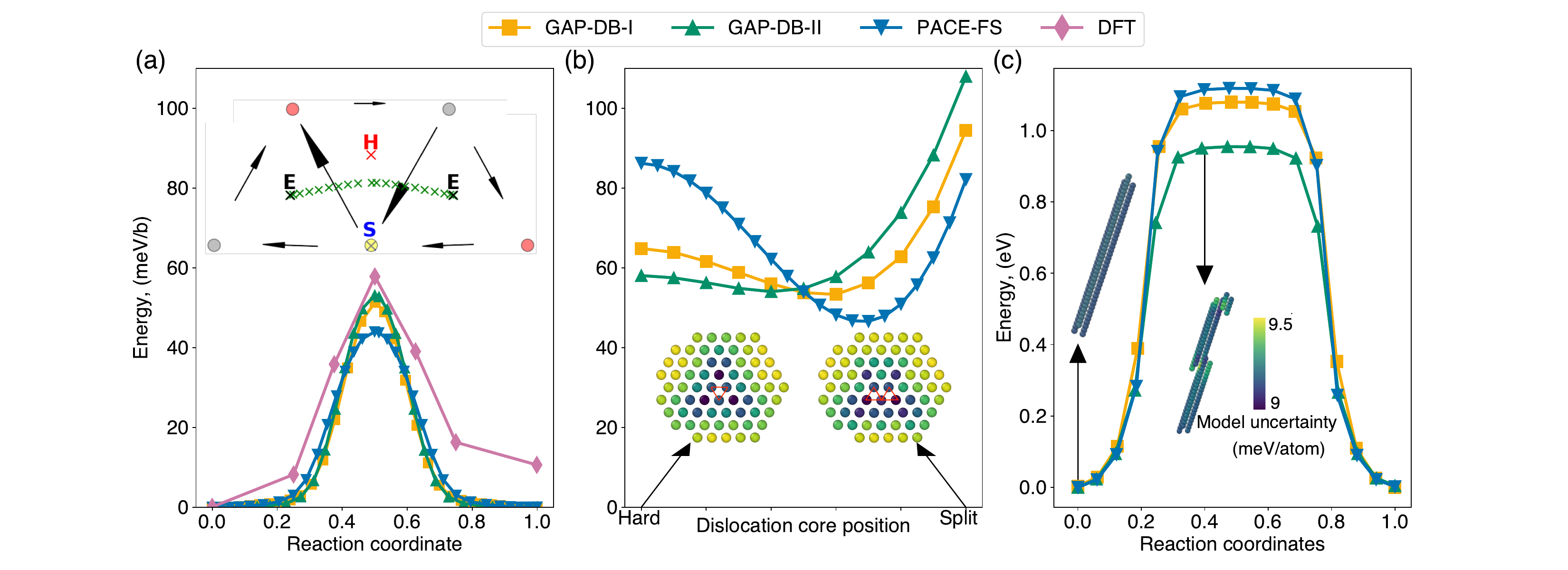}
	\caption{
		\textbf{(a)} Peierls potential computed with the PAD configuration at T=0K and zero applied stress. 
		The DFT results are taken from Ref.\cite{Dragoni2018_PhysRevMaterials.2.013808}, which uses a quadrupolar cell without elastic corrections. 
        The inset shows the Peierls path, and the differential displacement (DD) map for the configuration at the Peierls barrier, predicted by GAP-DB-I. 
        Atoms are colored according to the three possible positions along the dislocation line direction in the pristine bcc crystal, for one periodic unit cell (length b).
        \textbf{E}, \textbf{H}, and \textbf{S} indicate the easy, hard and split core positions, respectively.
        The dislocation trajectory, indicated by the green symbols ``x'' is calculated according to Ref. \cite{itakura2012first,ann_mori2020neural}.
        \textbf{(b)} Energy profile between the hard and the split core positions.
        The insets are the hard and the split dislocation core structures, colored according to the GAP model uncertainty. 
        The red circles indicates the centers of the dislocation cores in the two configurations (two circles in the split configuration indicate the center of the split core partials).
		\textbf{(c)} 
		The energy barrier for the kink-pair nucleation process. 
		The insets show the atoms close to the dislocation core at the initial configuration and in the middle of the transition path.
		Atoms are colored according to the GAP model uncertainty.
	}
	\label{fig:screw_disl}
\end{figure}

Fig. \ref{fig:screw_disl}a shows the Peierls barrier, that is the energy barrier (per unit dislocation length in terms of the Burgers vector magnitude \textit{b}) for a short screw dislocation (here, 2$b$ along $y$) to glide from the easy core position to an adjacent easy core position. 
The reference DFT data is taken from Ref. \cite{Dragoni2018_PhysRevMaterials.2.013808} and is consistent with the DB-I database. 
The energy of the end replica is $\sim$10 meV/b larger than the initial replica because the small DFT quadrupole cell geometry is not optimized with respect to the tilt components along the Burgers vector \cite{ventelon2013ab,ventelon2007core}.
This induces elastic interactions between dislocations in the quadrupole configuration as one of the dislocations moves along the Peierls path \cite{ventelon2013ab,maresca2018screw}. 
Therefore, the actual DFT predicted Peierls barrier \cite{Dragoni2018_PhysRevMaterials.2.013808} lies in between 48 and 58 meV/$b$.
The GAP-DB-I and GAP-DB-II predicted Peierls barriers are within the range from 48 to 54 meV/$b$, therefore within the DFT prediction range. 
The differential displacement map of the screw dislocation is also calculated, showing a compact dislocation core structure (see Fig. S8 of the \textit{Supplementary Material S3.1}).

We further compute the dislocation trajectory along the migration path using GAP-DB-I, as shown in the inset of Fig. \ref{fig:screw_disl}a.
The dislocation core position along the migration path is determined by computing the displacement differences among the three innermost $\langle 111 \rangle$ columns of atoms (see Ref. \cite{ann_mori2020neural}). 
The trajectory shows that the screw dislocation glides along an almost flat path, crossing in between the hard core and the split core configurations, which is in good agreement with other ML-IAPs \cite{ann_mori2020neural} and DFT calculations \cite{ventelon2013ab,dezerald2016plastic}.
This means that the Peierls barrier configuration lies at the saddle point that is located between the higher energy hard and split core configurations \cite{itakura2012first}. 
In order to verify this, we calculate the energy profile between the hard and split core positions. 
The path is obtained by interpolating linearly the in-plane displacements of the three innermost $\langle 111\rangle$ atomic columns between the hard core and the split core configurations. 
The energy profile is evaluated by keeping the in-plane displacements of these three atomic columns fixed while relaxing the rest of the atoms (see \textit{Supplementary Material S3.2} for details).
As shown in Fig. \ref{fig:screw_disl}b, all the tested ML-IAPs (GAP-DB-I, GAP-DB-II and PACE-FS) are able to predict a local minimum along the hard-to-split transition path. 
Moreover, both GAP-DB-I and GAP-DB-II predict that the hard core is lower energy than the split core, consistently with previous DFT calculations \cite{itakura2012first,ventelon2013ab}. 
The computed model uncertainty for GAP-DB-I is less than 10 meV/atom (see insets of Fig. \ref{fig:screw_disl}b). 
Therefore, GAP-DB-I is improved with respect to the previous GAP18 \cite{ann_mori2020neural}, since GAP-DB-I has been trained on a database that has been extended to enable fracture predictions \cite{lei_2022}, and hence it contains a larger amount of distorted DFT supercells compared with the original GAP18 database.
Instead, PACE-FS cannot capture the correct energetic hierarchy (hard core lower energy than split core).
However, we have verified that by fitting PACE-FS to DB-II the correct energy hierarchy is predicted. 
This is consistent with the fact that DB-II includes DFT supercells with various screw dislocation structures. 
In \textit{Supplementary Material S3.3}, the ability of GAP and PACE-FS IAPs to reproduce the DFT energies of the dislocation configurations from DB-II is shown. 
It is also shown that GAP converges to the DB-II hard-to-split path data as the number of DOFs is increased, while PACE-FS predictions do not converge as closely as GAP to the DFT predictions by increasing the number of basis functions. 

The energy to move a dislocation grows proportionally to the length of the dislocation line, which leads to a large barrier to move a long screw dislocation ($\sim$1 eV for 20$b$).
Long-standing theory envisions that the actual glide mechanism of screw dislocations occurs via kink-pair nucleation and migration, which is a thermally activated process \cite{dorn1963nucleation,seeger1956lxv}. 
Here, we compute the kink-formation energy by using CI-NEB method with a long screw dislocation geometry (40 $b$ along $y$), which allows the kink-pair formation and propagation process instead of straight gliding.  
The energy barrier associated with kink-pair nucleation is indicated by the plateau of the energy profile in Fig. \ref{fig:screw_disl}c ($\sim$1 eV), which is in line with Ref. \cite{maresca2018screw}. 
The kink can easily propagate with a small amount of energy once it is formed, since the Peierls barrier for edge dislocation glide is tiny (Fig. \ref{fig:neb_all}). 
The core structures in Fig. \ref{fig:screw_disl}b are colored according to the GAP-DB-I model uncertainty, evidencing that the model uncertainty is converged to less than 10 meV/atom during the kink-formation process. 
Since the kink-pair formation process involves the local tension and compression of $\langle 111\rangle$ atomic columns, kinks possess a vacancy and self-interstitial nature \cite{mrovec2011magnetic}.
The lower kink-pair nucleation barrier predicted by GAP-DB-II compared with GAP-DB-I can be attributed to the fact that the self-interstitial formation energy is 0.551 eV ($\sim 10\%$) lower for DFT calculations associated with DB-II (4.579 eV) than with DB-I (5.13 eV). 

Overall, the analysis shows that the GAP and PACE-FS potentials agree with published DFT and predict glide with limited uncertainty. PACE-FS needs training on more dislocation path-specific data in order to capture the precise hierarchy of dislocation core structures, especially the hard-to-split transition path.

\subsection{Core structures and energy barriers of edge and M111 dislocations} \label{chap4.2}

\begin{figure}
	\centering
	\includegraphics[trim=0 80 0 45, width=16cm]{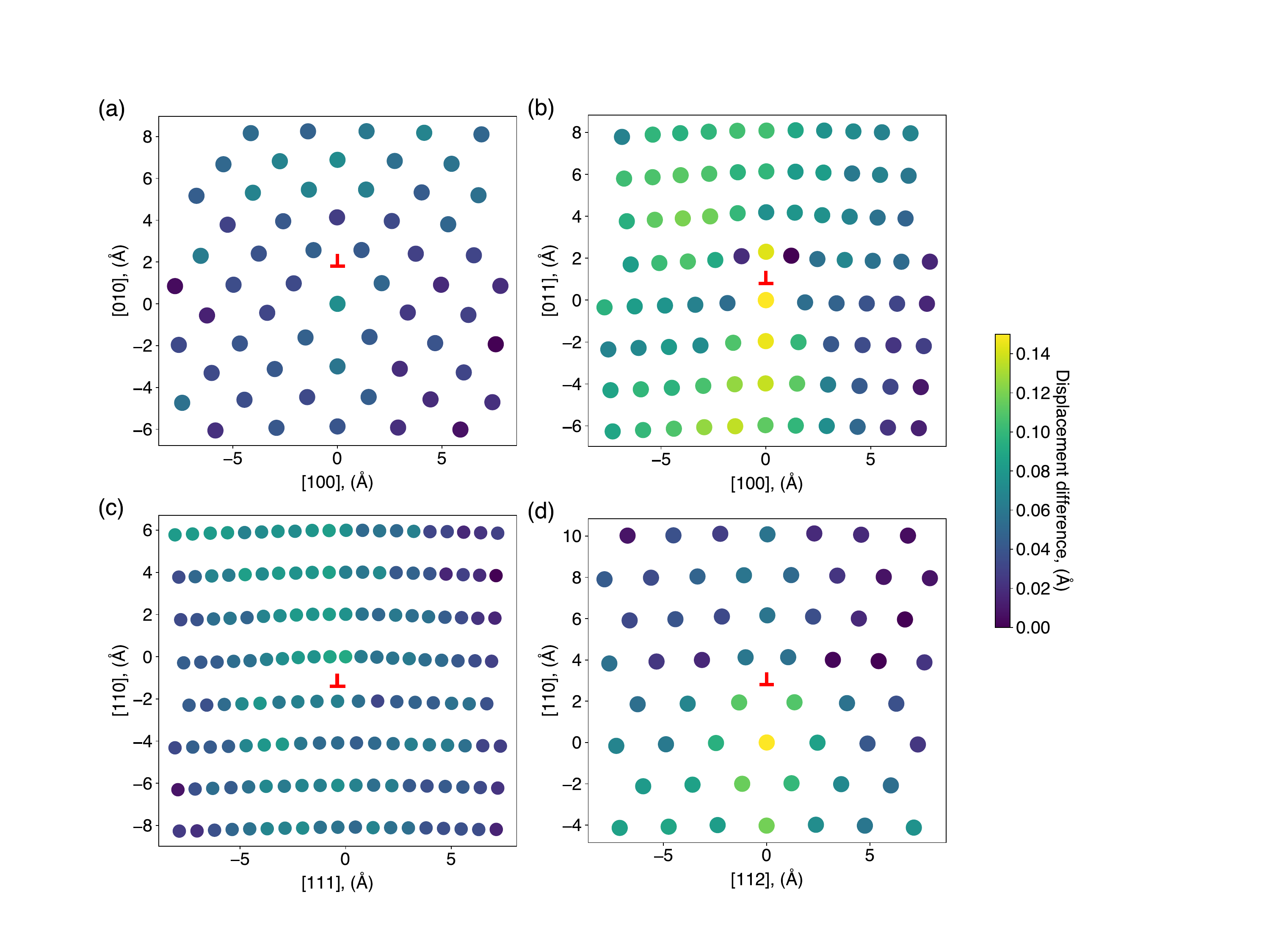}
	\caption{Core structures predicted by GAP-DB-I: \textbf{(a)} $a_0[100](010)$ edge, 
		\textbf{(b)} $a_0[100](011)$ edge, 
		\textbf{(c)} $a_0/2[\bar{1}\bar{1}1](1\bar{1}0)$ edge, and
		\textbf{(d)} $a_0/2[111](1\bar{1}0)$ M111 dislocations.
		Atoms are colored by the displacement difference relative to DFT predictions \cite{dftCore_fellinger2018geometries}.
		The dislocation center is indicated by the red symbol. 
	}
	\label{fig:disl_struc}
\end{figure}

Fig. \ref{fig:disl_struc} shows the core structures of three edge dislocations and the M111 dislocation, predicted by GAP-DB-I using the RB configuration (see \textit{Supplementary Material S4.1.1} for the predictions of GAP-DB-II and PACE-FS). 
The atoms are colored according to the difference between the ML-IAP and the DFT atomic positions, using DFT data from \cite{dftCore_fellinger2018geometries} as a reference, and indicate quantitative agreement with DFT. 
All ML-IAPs are able to predict the DFT core structures except for PACE-FS, which predicts another structure for the $a_0[100](011)$ edge dislocation core \cite{dftCore_fellinger2018geometries} (Fig. S6b). 
$a_0[100](010)$, $a_0/2[\bar{1}\bar{1}1](1\bar{1}0)$ and M111 show errors that are lower than 0.1 Å (Fig. \ref{fig:disl_struc}a and \ref{fig:disl_struc}c). 
The error associated with $a_0[100](011)$ edge is as high as 0.15 Å for the central atoms.
However, the predicted core structure is qualitatively the same as DFT predictions.

\begin{figure}[H]
	\centering
	\includegraphics[width=12cm]{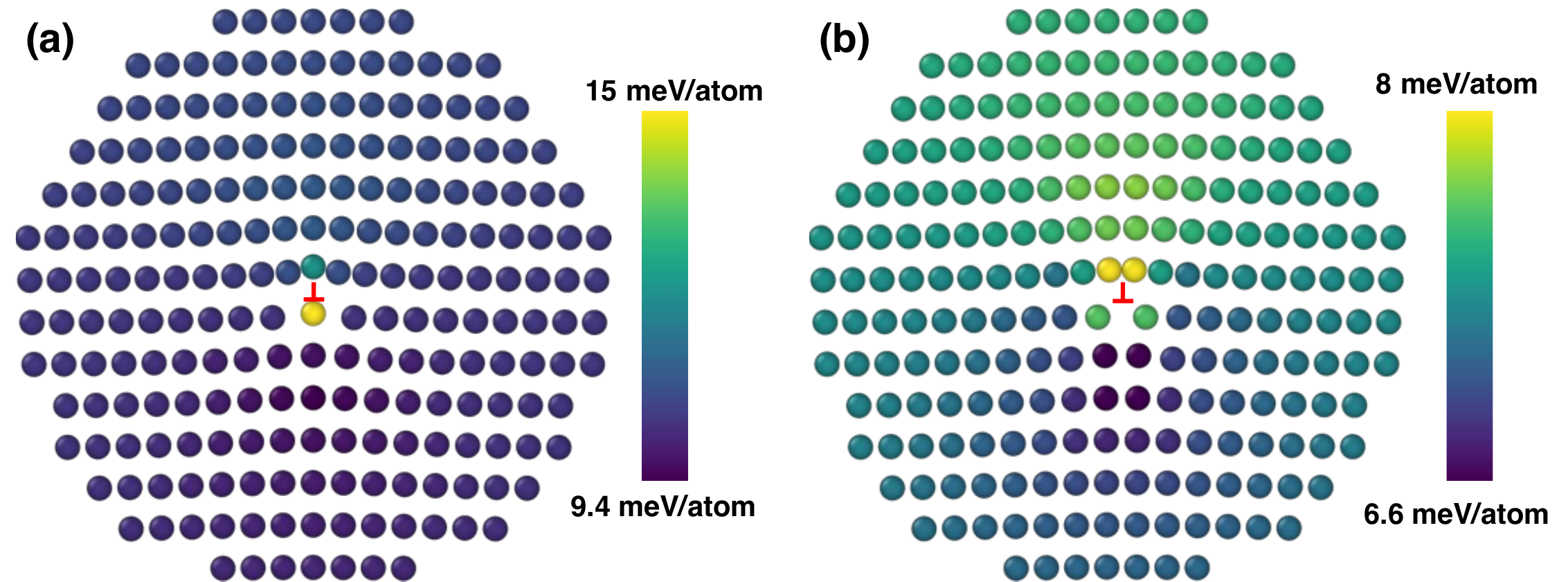}
	\caption{
        Core structure of $a_0[100](011)$ edge dislocation predicted by GAP-DB-I: 
		\textbf{(a)} $R_{\rm cut}=6.5 Å$, and 
		\textbf{(b)} $R_{\rm cut}=5 Å$. 
		The two core structures are referred to as atom-centered (AC) and bond-centered (BC) structures, respectively. 
		The dislocation center is indicated by the red symbol ``$\perp$". 
	}
	\label{fig:edge_110}
\end{figure}

To explore other possible core structures, we further relax the dislocation core by breaking the symmetry of the initial geometry, i.e., by starting from a slightly asymmetric dislocation core configuration.
We find that different core structures are predicted for M111 dislocation by using GAP-DB-I and for $a_0[100](011)$ edge dislocation by using GAP-DB-II (see \textit{Supplementary Material S4.1.2}). 
The discrepancies between the core structures emerging from symmetric and asymmetric initial geometries indicate that multiple local minima core structures exist for these dislocations in GAP-DB-I and GAP-DB-II. 
Moreover, GAP-DB-I and GAP-DB-II with $R_{\rm cut}=5 Å$ are found to yield a bond-centered (BC) rather than an atom-centered (AC) structure for $a_0[100](011)$ edge dislocation, as shown in Fig. \ref{fig:edge_110}b.
Note that the $R_{\rm cut}=6.5Å$ has a larger baseline uncertainty because the local atomic environment is more complex compared to $R_{\rm cut}=5Å$ due to the large cutoff radius. 
Atoms with the largest uncertainty tend to be located at the center of the dislocation, which is expected since the bonds around the central atom are highly distorted. 
The results suggest that $R_{\rm cut}=5 Å$ may not be sufficient to describe the complex energy landscape, and to distinguish between AC and BC core structures.  

The PAD configuration is employed to compute the Peierls barrier, which involves a different process to create the initial dislocation configuration, as described in Ref. \cite{BACON20091}.
The PAD configuration predicts the same equilibrium dislocation core structures as the RB configuration starting from asymmetric initial geometries (see \textit{Supplementary Material S4.1.2}). 
Fig. \ref{fig:neb_all} shows the Peierls barriers of the four dislocations, predicted by two GAPs and PACE-FS.
The per-atom uncertainty of dislocation configurations along the migration path shows only limited extrapolation (see \textit{Supplementary Material S4.2}). 
The NEB calculations reveal that the energy difference between AC and BC core structures are negligibly small (mostly around $\sim$2 meV or less), which indicates that the core structure can be easily transformed under mild external perturbations.
Moreover, three ML-IAPs predict the same trends and similar barriers for the $a_0[100](010)$ and $a_0/2[\bar{1}\bar{1}1](1\bar{1}0)$ edge dislocations. 
For  $a_0[100](011)$ edge, two stable core structures, AC and BC, are predicted by the different potentials, as shown in Fig.\ref{fig:neb_all}b. 
GAP-DB-I predicts the transition pathway AC-BC-AC, where AC is the stable core structure.
However, GAP-DB-II and PACE-FS predict BC-AC-BC, indicating that the BC configuration is more stable. 
The energy differences between the two core structures predicted by the two GAP ML-IAPs are comparable (2-3 meV/b), while PACE-FS predicts a larger barrier (16 meV/b). 
Finally, for the case of M111 dislocations, GAP-DB-I predicts AC core as the stable structure while the other two potentials show that BC is more stable.
GAP-DB-I and PACE-FS predict the same barrier ($\sim$ 2 meV/b). 
The discrepancies among the different ML-IAPs regarding the prediction of the equilibrium core structures suggest the existence of a multi convex hull PES with similar energy minima that ML-IAPs may not be capable of discriminating. 

\begin{figure}[H]
	\centering
	\includegraphics[trim=0 60 0 40, width=12cm]{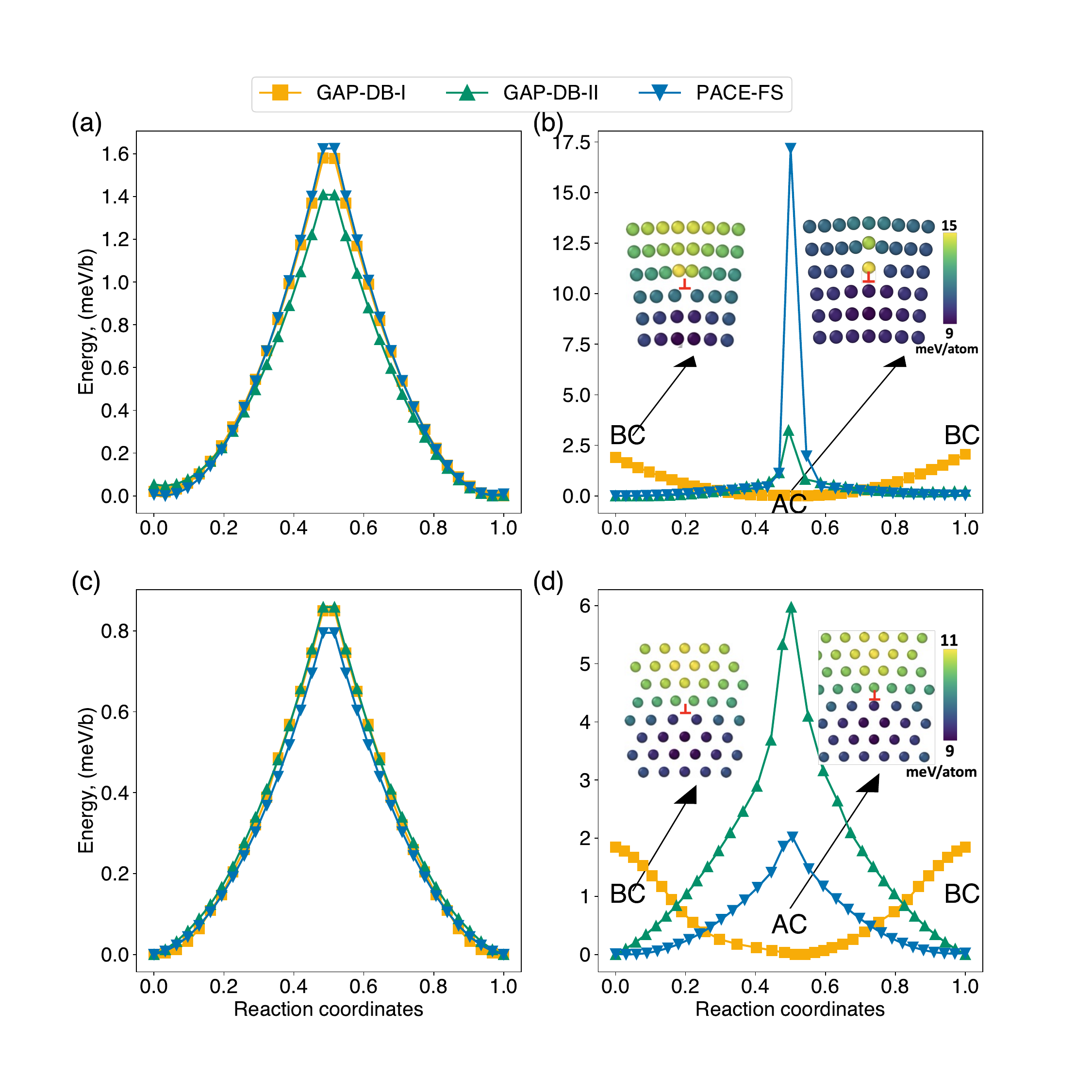}
	\caption{
        Peierls potential of \textbf{(a)} $a_0[100](010)$, 
		\textbf{(b)} $a_0[100](011)$, 
		\textbf{(c)} $a_0/2[\bar{1}\bar{1}1](1\bar{1}0)$ edge dislocations, and 
		\textbf{(d)} M111 dislocation.
		The Peierls barrier is computed using PAD configuration at T=0K under zero applied stress.
		The insets are colored according to the GAP model uncertainty predicted by GAP-DB-I, indicating where the largest extrapolation is located (see \textit{Supplementary Material S4.2.1} for more details). 
		The dislocation center is indicated by the red symbol. 
	}
	\label{fig:neb_all}
\end{figure}

In particular, both the RB and PAD configurations are employed to find the equilibrium dislocation core structures. 
The RB configuration is found to yield different core structures upon relaxing the symmetric and slightly asymmetric initial dislocation geometry. 
GAP-DB-I and GAP-DB-II ($R_{\rm cut}=6.5Å$) with the RB configuration and symmetric initial geometry predict core structures consistent with DFT \cite{dftCore_fellinger2018geometries}. 
However, the core structures predicted by RB configurations with asymmetric initial geometry and PAD configuration are inconsistent with DFT results for $a_0[100](011)$ edge and M111 dislocations, indicating the existence of multiple local minima. 
The results suggest that, if the dislocation has multiple metastable core structures with similar energies, the predicted dislocation core structures can be extremely sensitive to the initial geometry and boundary conditions. 
Therefore, care should be taken when identifying the equilibrium dislocation core structures. 
Breaking the symmetry such as starting from slightly distorted geometry may help to escape from local minima. 

In summary, the predictions of the edge and M111 core structures are consistent across the ML-IAPs considered here, and the per-atom model uncertainty with respect to DFT is low. Where multiple core structures can occur that have small energy differences, close to DFT accuracy ($\sim$1 meV/atom), then the predicted minimum energy core is sensitive to the specific ML-IAP and boundary conditions. Rather than being a ML-IAP limitation, this outcome shows that ML-IAP predictions are limited by the DFT accuracy.

\section{Atomistic fracture}
\label{chap5}
\subsection{Traction-separation process}\label{chap5.1}
The traction-separation (T-S) curve is often used to quantify brittle fracture, since it encodes the surface energy, the cohesive strength and the rigid surface separation process.
We compute the T-S curve based on the ML-IAPs trained in \textit{Section \ref{chap2}.}
The traction-separation profiles are calculated by taking the derivative of the energy-separation curve, which is obtained by rigidly separating a perfect crystal.
During the calculation, a bcc iron single crystal is separated along \{100\} and \{110\} planes using an incremental step of 0.05 Å. 

\begin{figure}[H]
	\centering
	\includegraphics[trim=0 40 0 0, width=14cm]{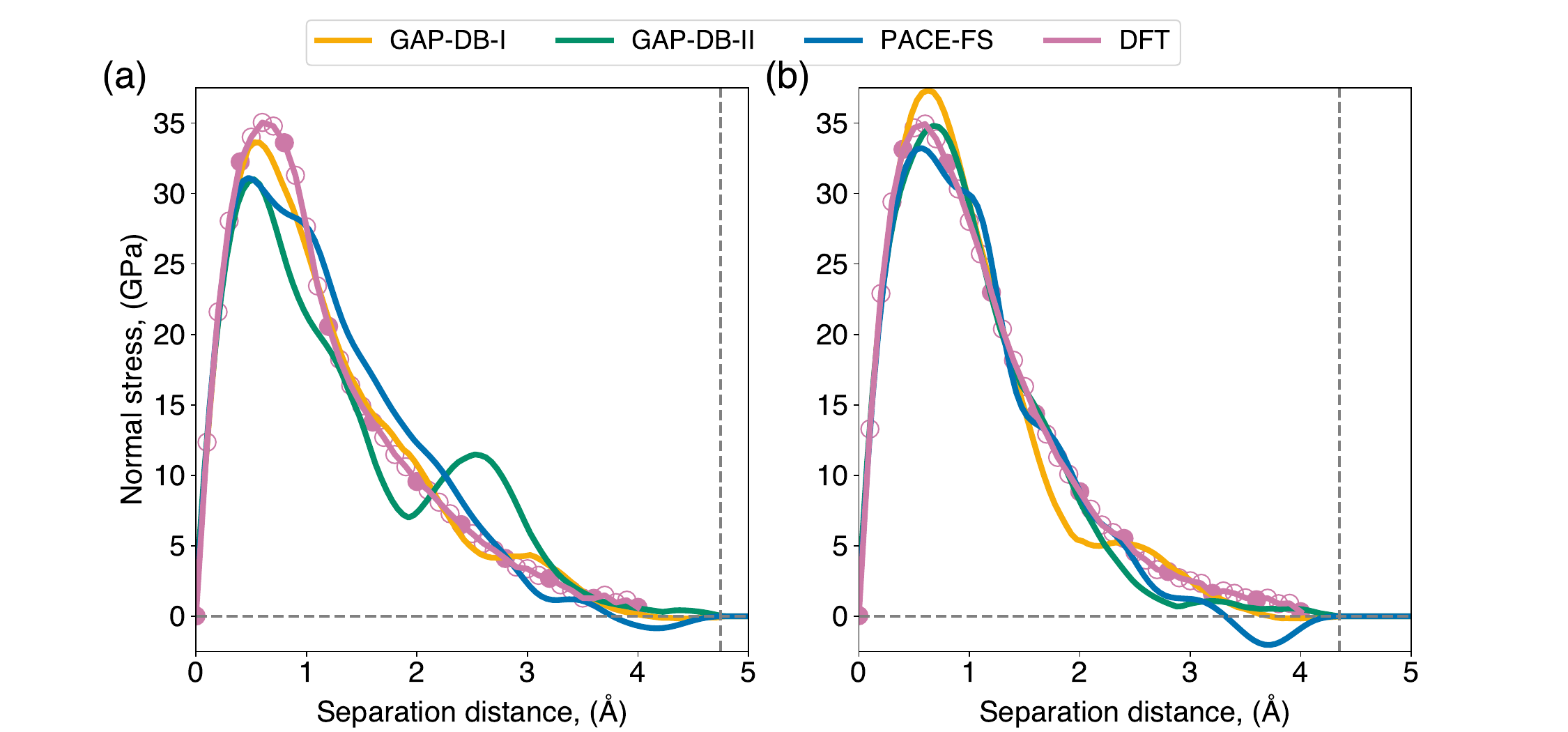}
	\caption{
        Traction-separation curves predicted by ML-IAPs and DFT in bcc iron for
		\textbf{(a)} \{100\} and
		\textbf{(b)} \{110\} plane.
		The solid circles indicate the DFT data used in the training database (DB-I). 
	}
	\label{fig:ts-all}
\end{figure}

Fig. \ref{fig:ts-all} shows the T-S curves predicted by ML-IAPs with $R_{\rm cut}=6.5Å$.
DFT predicts a smooth curve with a single maximum traction (the so-called cohesive strength). 
All ML-IAPs (including MTP, NNP and PACE-L, see \textit{Supplementary Materials S5.1.1}) are able to predict a distinct maximum normal stress, yet, all T-S curves show multiple artificial local minima after reaching the cohesive strength, especially at the end of the separation process. 
Among all potentials, GAP-DB-I and PACE-FS yield the closest predictions compared to DFT. 
The GAP-DB-I predicted T-S curve is smoother than GAP-DB-II since DB-I includes the surface separation process (the solid circle DFT data in Fig. \ref{fig:ts-all}). 
In all cases, the area under the curve is calculated and is equal to twice the surface energy, as expected (see \textit{Supplementary Materials S5.1.2}). 

Since ML potentials are constructed based on the energy localization assumption, the choice of the maximum interaction range ($R_{\rm cut}$) might have consequences on the T-S curve. 
Therefore, we train two sets of PACE-FS potentials (see \textit{Supplementary Material S5.1.3 and S5.1.4}).
The first set is trained with cutoff ranges from 5 to 7 Å to investigate the influence of $R_{\rm cut}$.
The results show that $R_{\rm cut} =6.5\ Å$ has the minimal ``tail" non-smoothness effects while capturing the cohesive strength among all $R_{\rm cut}$ (Fig. S28).
Yet, the potentials are able to capture the essential behaviour (single-peak, skewed shape) of the T-S curve with the proper choice of $R_{\rm cut}$. 
Note that all ML-IAPs are only trained with limited surface separation data (solid circles in Fig. \ref{fig:ts-all}a and Fig. \ref{fig:ts-all}b).
Therefore, the second set is trained by adding four times more separation configurations to the training database (see the computed T-S curves in Fig. S29). 
The results show that the end of the separation is still not smooth, indicating that the oscillation is not primarily caused by the lack of training data.
The many-body interactions nature of ML-IAPs introduces a complex energy landscape, which might induce multiple artificial local minima. 
Such oscillations posit an open question regarding their origin, which might be associated to a poor regularization of the ML-IAPs.  
Attention is thus required when analysing the T-S curves with ML-IAPs. 
Nonetheless, the T-S curve is obtained from rigid separation while the surface is under relaxation during fracture simulation.
Especially after the cohesive strength, the crystal separates physically with a displacement jump. 
Therefore, the end of the T-S is not physically relevant to the fracture behaviour.
We conclude that the oscillations at the tail of the T-S curves have a minor influence on the fracture prediction, as discussed in the next Section. 

\subsection{Atomistic fracture mechanism} \label{chap5.2}

\begin{figure}[H]
	\centering
	\includegraphics[trim=40 0 0 0, width=15cm]{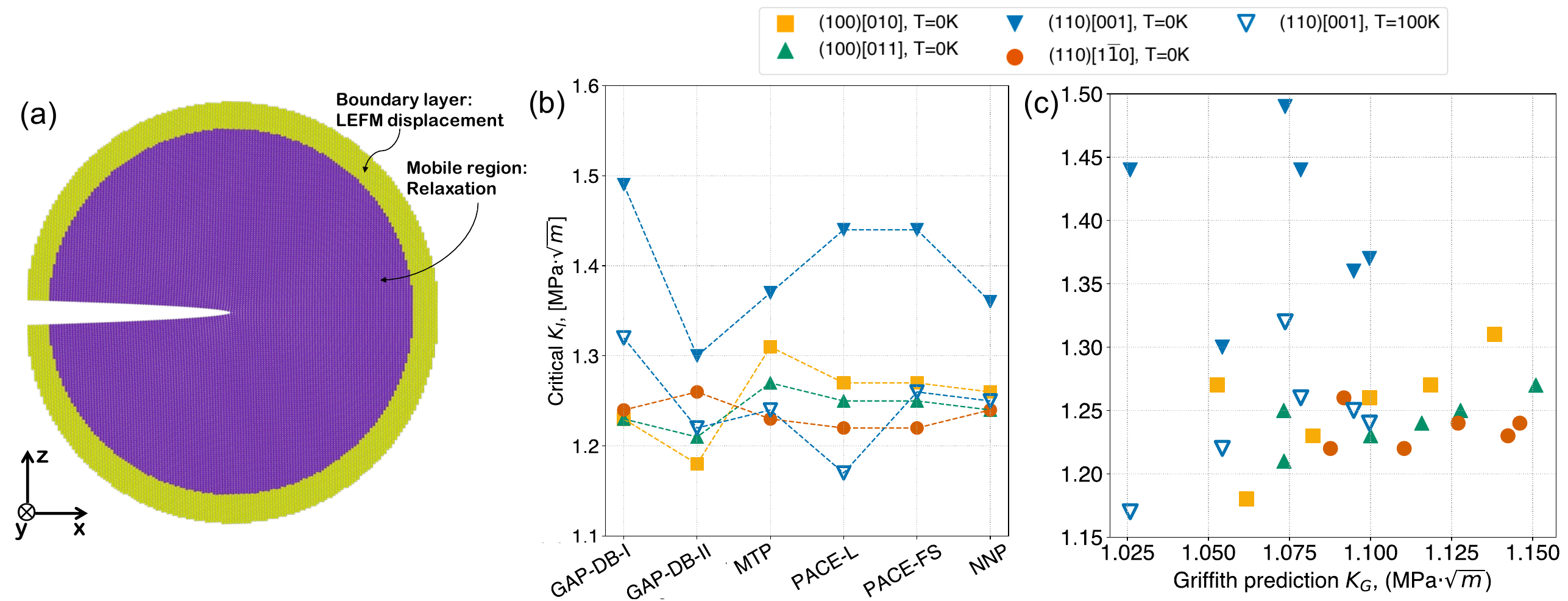}
	\caption{
		\textbf{(a)} Schematic plot of $K$-test fracture simulation setup. $x$, $y$ and $z$ are aligned with the crack propagation direction, crack plane normal and crack front, respectively. 
		\textbf{(b)} Critical $K_{\rm Ic}$ predicted by different ML-IAPs at T=0K for four crack systems and T=100K for (110)[010] crack system. 
		Crack systems are indicated by crack plane/crack front.
        \textbf{(c)} Critical $K_{\rm Ic}$ predicted by ML-IAPs versus Griffith criterion $K_{\rm G}$.
	}
	\label{fig:fracture}
\end{figure}

Based on the ML-IAPs trained in \textit{Section} \ref{chap2}, MS and MD simulations are performed to study the atomistic fracture mechanism and to predict the critical stress intensity factor ($K_{\rm Ic}$) of single crystal bcc iron.
A cylinder-shaped near-crack-tip geometry is used in combination with the $K$-test framework, as illustrated in Fig. \ref{fig:fracture}a \cite{andric2017new}.
The crack is aligned with a Cartesian coordinate system, where $x$, $y$ and $z$ are the crack-propagation direction, the crack-plane normal and the crack-front, respectively. 
The radius of the cylinder is set to 150 Å, which enables converged predictions \cite{lei_2022}. 
A displacement controlled loading process is implemented, whereby the boundary atoms are displaced according to the anisotropic linear elastic solution of an infinite crack subject to a remote $K_{\rm I}$. 
The crack is initially loaded to $K_{\rm init}$, at which the crack tip is maintained at its original position. 
Next, $K_{I}$ is increased with a step of $\Delta K=0.01$ $ \ \rm MPa\sqrt{\rm m}$ until a critical event occurs at the crack tip, i.e., either crack propagation, dislocation emission, phase transition, or the combination of multiple mechanisms.
In the MS simulations, the atoms at the mobile region of the cell ($r<140Å$) are relaxed using a combination of CG and FIRE minimizers \cite{bitzek2006structural} with a force tolerence of $10^{-9}$ eV/Å and $10^{-3}$ eV/Å, respectively, while the rest of the atoms are fixed at each incremental step.
For finite temperature MD simulations, the mobile region is equilibrated for 10 ps at each incremental step with the Nosé-Hoover thermostat.
More details of the $K$-test loading procedure can be found in \cite{lei_2022,Andric_2018}.

We perform $K$-test simulations for four crack systems ((100)[010], (100)[011], (110)[001], and (110)[1$\bar{1}0$]) at T=0K to investigate the atomistic fracture mechanism.
Since the cutoff radius significantly influences the behaviour of the traction-separation curve, all ML-IAPs are trained with two cutoff radii ($R_{\rm{cut}} = 5\ Å$ and $6.5 Å$) to study the effect of the cutoff radius.
Here, we show the results of $R_{\rm{cut}} = 6.5\ Å$, while $R_{\rm{cut}} = 5\ Å$ results are reported in \textit{Supplementary Material S5.3}.
The ML-IAPs trained on DB-I predict cleavage fracture on the pre-cracked plane for all crack systems, for both $R_{\rm cut} = 5\ Å$ and $6.5\ Å$. 
The only exception is that GAP-DB-II with $R_{\rm cut} = 5\ Å$ predicts crack deviation to the (100) plane for the (110)[001] crack system.
The fundamental atomistic fracture mechanism (cleavage) remains unchanged, yet, the predicted $K_{\rm Ic}$ varies with the potential, as shown in Fig. \ref{fig:fracture}b. 
Some fluctuations in the predicted $K_{\rm Ic}$'s are expected because of the different ML-IAP schemes.
For crack system (110)[001], we concluded in our previous study \cite{lei_2022} that the large $K_{\rm Ic}$ at T=0K compared to $K_{\rm G}$ is due to the existence of a small lattice trapping barrier, which is introduced by the rough PES. 
Therefore, we perform MD simulations for (110)[001] at T=100K, which allows the system to overcome the small activation energy barrier and reduce the predicted $K_{\rm Ic}$. 
As indicated by the open triangles in Fig. \ref{fig:fracture}b, $K_{\rm Ic}$'s for (110)[001] drop significantly at T=100K, which is consistent with the existence of a small lattice trapping barrier.
The predicted $K_{\rm Ic}$'s for crack system (110)[1$\bar{1}0$] are nearly the same for all potentials. 
All ML-IAPs with $R_{\rm cut}=6.5 \AA$ predict cleavage on the pre-cracked plane for all crack systems, confirming that the atomistic fracture mechanism is cleavage. 
Moreover, the per-atom model uncertainty predicted by the GAP variance and the extrapolation grade $\gamma$ are consistent with each other.
No large extrapolation ($\gamma>3$, see \cite{podryabinkin2023mlip}) is detected during the fracture process (see \textit{Supplementary Material S5.2}). 
Fig. \ref{fig:fracture}c reveals the relation between MS predicted $K_{\rm Ic}$'s and the Griffith prediction \cite{Griffith} 
	\begin{equation}
		K_{\rm G}=\sqrt{\frac{2\gamma_{\rm s}}{B}},
	\end{equation}
where $\gamma_{\rm s}$ is the surface energy.
$B$ is a constant that is expressed as
	\begin{equation}
		B=\sqrt{\frac{b_{11}b_{22}}{2}\left(\frac{2b_{12}+b_{66}}{2b_{11}}+\sqrt{\frac{b_{22}}{b_{11}}}\right)},
	\end{equation}
where $b_{ij}$ are determined by the elastic constants.
Since $K_{\rm G}$ only depends on surface energy and elastic constants, all ML-IAPs predict similar results, ranging from 1.025-1.15 $\rm MPa\sqrt{m}$.  
$K_{\rm Ic}$ is always larger than $K_{\rm G}$, indicating that all ML-IAPs predict lattice trapping effects.
For \{100\} crack plane, (110)[1$\bar{1}0$] system at T=0K and (110)[001] at T=100K,  $K_{\rm Ic}$ and $K_{\rm G}$ show an approximately linear correlation.
This result reveals that the critical $K$ is correlated with the surface energy in the presence of lattice trapping effects.
Crack system (110)[001] at T=0K exhibits an artificial trapping barrier for part of the ML-IAPs, leading to the scattered distribution. 

Our analysis shows that the trained ML-IAP potentials are transferable to crack propagation simulations, and predictions are qualitatively consistent. Limited quantitative differences in the predictions ($K_{\rm Ic}$, lattice trapping) can be encountered based on the choice of the cutoff radius (which has been optimized here) and the ML-IAP package.

\section{Discussion}
\label{chap6}

\subsection{Complementing accuracy analysis with transferability assessment via model uncertainty quantification} \label{chap6.1}

We have proposed a three-step procedure to achieve DFT accuracy, optimize the efficiency, and verify the transferability of ML-IAPs.
This procedure includes extensive use of training/testing RMSE, quality factor evaluation, and model uncertainty quantification.
As a matter of fact, the RMSE and quality factor are commonly used during the validation of ML-IAPs \cite{zuo2020performance,ann_mori2020neural,ogata_nnp_meng2021general}. 
However, instead of using an average quality factor/score, the existing literature compares the individual properties directly to DFT, e.g. by looking at the equation of state, the elastic constants, the surface energies, the vacancy formation and migration energies, and the GSFE curves \cite{zuo2020performance,lysogorskiy2021performant}. 
Here, we use an aggregate index $Q$ that is defined in Eq. (\ref{eq:q_factor}) as the average relative error of the predicted physical properties. 
The relation between RMSE and $Q$ reveals the existence of an RMSE limit ($\sim 5$ meV/atom) under which lower RMSE does not correspond to smaller $Q$ (Fig. \ref{fig:qfactor}), which underlines the limitation of using RMSE as the only accuracy measure. 
Therefore, an optimum between accuracy and efficiency can be found by selecting potentials with RMSE around 5 meV/atom, and we propose to use model uncertainty quantification in the context of benchmark simulations of extended defects to further assess the ML-IAP transferability.

In this work, we have calculated and demonstrated the usage of two different model uncertainty quantification measures to validate the transferability of the ML-IAPs to dislocation and cracks. 
These two measures are correlated to each other, in the sense that they identify consistently the highest uncertainty in the same atomistic regions (see \textit{Supplementary Material S4.2.2 and S5.2}).
The properties of screw dislocations in bcc metals, i.e., compact core structure, Peierls barrier, and kink-pair nucleation/migration, are considered to be essential aspects that ML-IAPs should be able to capture \cite{szlachta2014accuracy,Dragoni2018_PhysRevMaterials.2.013808,alam2021artificial_Molybdenum,wang2022classical_vanadium,freitas2022machine,ann_mori2020neural,goryaeva2021efficient}, and they cannot be computed directly with DFT supercells.  
Here, we have shown that the GAP variance and extrapolation grade $\gamma$ from PACE-FS indicate interpolation for screw dislocation glide, including Peierls barrier and kink pair nucleation. 
Furthermore, $K$-tests have been performed to investigate the atomistic fracture mechanisms of single crystal bcc iron, whereby the model uncertainty is converged within the threshold.
We highlight that, when applying the ML-IAP for large-scale MD simulations, the extrapolation degree (i.e., predicted variance for GAP and extrapolation grade $\gamma$ for ACE) should be monitored to further verify the transferability of the potential and the reliability of the simulations.

\subsection{The choice of the DFT database}
\label{chap6.2}
Generating a database is the most time-consuming part of training ML-IAPs, since quantum mechanical calculations (typically DFT) are computationally expensive. 
The conventional way is to create a general database that contains equilibrium and perturbed configurations of possible phases and various defects (surface, point defects, lattice expansion/compression, isolated clusters, grain boundaries \textit{etc.}) \cite{Dragoni2018_PhysRevMaterials.2.013808,botu2017machine,deringer2019machine}.
The perturbed configurations can be generated either by random rattling of atomic displacements or by high-temperature MD using some preliminary version of the potential. 
For specific applications, user experience is still needed to design the database that consists of relevant configurations. 
However, any hand-built database is user-biased and sometimes redundant. 
To reduce both human intervention and usage of computational resources, (hyper)active learning (combined with model uncertainty quantification) is being developed and applied to different ML frameworks, which enables fast creation of the database \cite{bochkarev2022efficient,novikov2020mlip,cvd2022hyperactive,zhang2019active}.

As detailed in \textit{section \ref{chap2.1}}, DB-I was initially constructed in a conventional manner and subsequently extended with an active-learned fracture-relevant database, where small crack-tip configurations were added based on GAP variance predictions \cite{Dragoni2018_PhysRevMaterials.2.013808,lei_2022}.
DB-II has been designed with an attempt to cover \textit{a priori} all possible defects, leading to a 5 times larger number of LAEs compared with DB-I.
DB-I contains no dislocation structures while DB-II includes $a_0/2[111]$ screw dislocation and edge dislocations. 
Yet, GAP IAPs trained on either DBs can predict DFT-accurate Peierls barrier and kink pair nucleation barrier of screw dislocations, as well as core structures of edge and M111 dislocations. 
This finding is consistent with previous work, where an ANN potential was shown to accurately predict the dislocation core structure and Peierls barrier, despite being trained on a DFT database without any dislocation structures \cite{ann_mori2020neural}.
This suggests that the $\gamma$-surface is enough for the prediction of the compact core structure and single hump Peierls barrier for screw dislocation \cite{ogata_nnp_meng2021general,maresca2018screw}, a result that is consistent with earlier analysis by Duesbery and Vitek \cite{duesbery1998plastic} , where authors show that correct reproduction of the $\gamma$-surface gives correct core structure.
It has been shown in Ref. \cite{freitas2022machine,szlachta2014accuracy} that direct inclusion of dislocation core configurations in the database can lead to a better accuracy in predicting the core structure. 
A detailed study, i.e., training and testing ML-IAPs with different sub-databases, would be required to quantitatively examine the influence of  directly including the relevant configurations. 

GAP-DB-II predicts cleavage on the pre-cracked plane to be the atomistic fracture mechanism on \{100\} and \{110\} planes, which is consistent with the prediction of GAP-DB-I. 
However, DB-I encompasses several crack-tip configurations that are obtained through active learning while DB-II does not include any crack-tip configurations \cite{lei_2022}. This result suggests that an extensive but also less redundant DFT database can be obtained via specific design and active learning, which is a key implication for researchers interested in developing, using and/or adapting ML potentials. 
The comparison of two databases suggests that inherent liquid state structures or grain boundaries included in DB-II may play an important role when predicting fracture. 
A separate study would be needed to identify the extra data that enables the fracture prediction without direct inclusion of the crack tips, which is beyond the scope of the present work. 

\subsection{Variability in the predictions of dislocation core structures and Peierls barriers} 
\label{chap6.3}
Thermally activated $a_0/2 \langle 111 \rangle$ screw dislocation controls the low-temperature plasticity (brittle-ductile transition) of bcc metals. 
Most of the classical potentials fail to predict a compact core structure and a single hump Peierls potential \cite{proville2012quantum}. 
Therefore, the core structure and Peierls barrier of screw dislocations are first benchmarked for bcc ML-IAPs \cite{szlachta2014accuracy,wang2022classical_vanadium,goryaeva2021efficient,Dragoni2018_PhysRevMaterials.2.013808}. 
In this work, we have shown that all ML-IAPs are able to predict a compact screw dislocation core structure for bcc iron. 
Peierls barrier calculated via NEB approach ranges from 48 to 54 meV/b, depending on the choice of the ML-IAP.
DFT calculations which are consistent with DB-I predict a Peierls barrier between 48 and 58 meV/b \cite{Dragoni2018_PhysRevMaterials.2.013808}. 
We also show that GAP-DB-II predicts a Peierls barrier in this same range, and it converges to DFT data of DB-II which include configurations that are relevant to the Peierls path (see \textit{Supplementary Materials S3.3}). 
Using the same geometry, other independent DFT calculations predict a Peierls energy barrier of $40\pm 5$ meV/b based on a plane wave approach \cite{ventelon2013ab}. 
Such discrepancies are induced by the model geometry and DFT calculation details, e.g., code, convergence setup, exchange-correlation function, and pseudopotential.
Therefore, the Peierls barrier of screw dislocation ranges from 35 to 58 meV/b according to different DFT calculations \cite{ventelon2007core,ventelon2013ab,Dragoni2018_PhysRevMaterials.2.013808}. 
Table \ref{tab:peierls_screw} lists the Peierls barrier and double kink formation energy predicted by DFT and ML-IAPs, including two NNPs \cite{ann_mori2020neural,ogata_nnp_meng2021general}, an MTP \cite{mtp_wang2022machine}, a LML model, and a QNML Potential \cite{goryaeva2021efficient}. 
From Table \ref{tab:peierls_screw}, all available ML-IAPs are able to quantitatively predict the Peierls barrier and kink-pair mechanism irrespective of the ML approach and DFT database, which makes ML-IAPs suitable candidates for studying screw dislocation behaviour. 
Furthermore, the energy profiles along the cross section of hard to split core path predicted by GAP-DB-I and GAP-DB-II show the correct energy hierarchy, i.e., the energy of split core is larger than hard core, while PACE-FS can be improved to predict the correct energy hierarchy by training on DB-II.

\begin{table}[H]
	\begin{center}
		\caption{\label{tab:peierls_screw} Peierls barrier and double kink-pair formation energy predicted by DFT, line tension (LT) model, and different ML-IAPs.}
		\begin{tabular}{c|c|c} 
			\hline
			\hline
			Potential & Peierls barrier (meV/b) & Kink-pair formation (eV) \\
			\hline 
			\hline
			DFT \cite{ventelon2013ab,Dragoni2018_PhysRevMaterials.2.013808,ventelon2007core,dezerald2014ab} & 35-58 & - \\
			line-tension (LT) models \cite{proville2013prediction,itakura2012first} & - & 0.73-0.91 \\
			ANN (\textit{aenet}) \cite{ann_mori2020neural} & 35.3 & 0.94 \\
			NNP (\textit{n2p2}) \cite{ogata_nnp_meng2021general} & 38.2 & 0.70 \\
			MTP \cite{mtp_wang2022machine} & 30.5  & - \\
			LML \cite{goryaeva2021efficient} & 41.9 & 0.77 \\
			QNML \cite{goryaeva2021efficient} & 38.2 & 0.84 \\
			GAP-DB-I & 54.1 & 1.07 \\
			GAP-DB-II & 53.4 & 0.95 \\
			PACE-FS & 48.7 & 1.12 \\
			\hline
			\hline
		\end{tabular}
	\end{center}
\end{table}

The core structure of edge ($a_0[100](010)$, $a_0[100](011)$, and $a_0/2[\bar{1}\bar{1}1](1\bar{1}0)$) and M111 dislocations have also been used to benchmark the predictability of the ML-IAPs \cite{ann_mori2020neural,mtp_wang2022machine,dftCore_fellinger2018geometries}. 
For example, it is shown that NNP and an MTP are capable of reproducing these dislocation core structures with DFT accuracy \cite{ann_mori2020neural,mtp_wang2022machine}. 
In the current study, both RB and PAD configurations have been employed to reveal the dislocation core structures and Peierls barriers. 
We have shown that GAP-DB-I and GAP-DB-II along with the RB configuration are able to reproduce DFT-predicted core structures for all dislocations.  
Discrepancies are found between different initial geometries, which is the results of multiple metastable core structures with similar energies. 
Indeed, DFT calculations with flexible boundary conditions predict BC core for M111 \cite{dftCore_fellinger2018geometries}. 
However, a different DFT calculation \cite{romaner2021theoretical} with a rectangular arrangement of cells shows that the energy difference between AC and BC core is essentially zero, indicating that M111 dislocation in bcc iron has no clear core preference. 
Since the energy difference between AC and BC core structures is so small that it cannot be captured reliably by small-cell DFT calculations, it is not surprising that ML-IAPs yield different results. 

As for edge dislocations, our NEB results showed that the Peierls barriers are negligibly small ($\sim$2 meV/b) except for PACE-FS which predicts an activation barrier of 16 meV/b for $a_0[100](011)$. 
This is expected to be an incorrect extrapolation of PACE-FS. 
These negligible activation barriers can be easily overcome by mild thermal fluctuations, thus explaining why edge dislocations in bcc iron move without thermal activation. 
Previous calculations of $a_0/2[\bar{1}\bar{1}1](1\bar{1}0)$ edge dislocation based on three EAM potentials also predict Peierls barrier of 0.1-2 meV/b \cite{haghighat2014influence}. 
However, in the same study, another EAM potential (Chiesa09 \cite{chiesa2009free}) predicts a Peierls barrier of 9 meV/b, which is expected to be an artifact of the potential \cite{haghighat2014influence}. 
Such inconsistent outcomes of EAM potentials again highlight the need of accurate IAPs that capable of providing consistent predictions. 

It has been shown that M111 dislocations can play a role in dislocation mobility at low temperature for bcc Ta \cite{kang2012singular}.
Here, our NEB calculations predict a negligible Peierls barrier under zero applied stress at T=0K, which is consistent with recent DFT calculations \cite{romaner2021theoretical}. 
The DFT calculations show that M111 dislocation in bcc iron has a negligible energy difference between AC and BC core structures \cite{romaner2021theoretical}. 
Therefore, M111 dislocation in bcc iron is expected to move without thermal activation/kink-pair nucleation and propagation, which indicates that its contribution to low-temperature plasticity is negligible.

\subsection{Challenges associated with the prediction of the traction-separation law}
\label{chap6.4}
We have shown that the influence of $R_{\rm cut}$ on the traction-separation profile is significant, i.e., the inappropriate $R_{\rm cut}$  introduces multiple peak stress and fluctuations at the end of the separation process. 
Ideally, a larger cutoff radius is able to contain more information about the LAE, which in principle yields more accurate predictions, at increased computational cost. 
In practice, the cutoff radius should be chosen to reproduce the elementary properties shown in \textit{Section} \ref{chap3} while keeping $R_{\rm cut}$ as small as possible. 
As also pointed out in Ref. \cite{ko2014origin}, unrealistic stresses are predicted at the end of the separation process based on MEAM potential, which can be addressed by increasing the $R_{\rm cut}$. 
Hiremath \textit{et. al.} also fitted MEAM to predict a smooth T-S curve by optimizing the cutoff and smoothing ranges \cite{hiremath2022effects}. 
Another study on bcc vanadium shows that an extended MEAM potential is able to predict a smooth T-S curve without small fluctuations \cite{wang2022classical_vanadium}. 
On the contrary, predictions of GAP exhibit an artificial peak stress at the end of decohesion, which is not observed here. 
In the same study, a deep NN potential is also employed, which predicts the T-S curve with fluctuations \cite{wang2022classical_vanadium}. 
It is argued here that the T-S curve cannot be entirely captured by ML-IAPs because of the complex energy landscape introduced by many-body interactions and the regularization of the potential, which underlines an open challenge for ML-framework developers. 

\label{chap6.5}

\section{Conclusions} \label{chap7}

In this work, we have trained and benchmarked several ML-IAPs based on two independent large DFT databases. High accuracy with respect to the training database has been achieved by extensive hyperparameter optimization. The computational cost has been assessed and the optimal potentials in terms of computational speed (PACE-FS) and accuracy (GAP) have been identified. These potentials have been successfully tested on a broad number of benchmark simulations including screw, mixed M111 and several edge dislocations; as well as the challenging crack propagation.
By making extensive use of model uncertainty quantification and direct comparison with consistent DFT calculations, we have shown that the optimized ML-IAPs are capable of reproducing a broad range of properties with DFT accuracy, including the structure and Peierls barriers of five dislocation characters, the traction-separation law and crack propagation behaviour of bcc iron.
The following salient conclusions can be drawn:

\begin{enumerate}

	\item The three-step validation procedure adopted in this work enables the assessment of the accuracy and transferability of several state-of-the-art ML IAPs to model dislocations and cracks. 
    The procedure has been applied to the case of bcc iron. 
    By optimizing the model parameters and considering different ML packages, the ML-IAPs efficiency can be increased by two orders of magnitude in terms of computational time. 
    We find that GAP occupies the Pareto front in terms of accuracy, while PACE-FS is the most efficient ML-IAP.
	\item Depending on the specific application, MS/MD simulations beyond DFT supercell size need to be employed to validate the transferability of ML-IAPs (e.g., dislocations and cracks). 
    This assessment has been performed by using model uncertainty quantification, which is implemented in the state-of-the-art ML-IAP packages and requires limited computing time.
    \item Both GAP and PACE-FS ML-IAPs are capable of reproducing the key features of screw dislocations in bcc iron (Peierls barrier and kink-pair nucleation), which is a necessary condition for large-scale plasticity simulations of bcc crystals. 
    With both databases, GAP can reproduce the details of the screw dislocation migration path including the correct energetic hierarchy of the dislocation cores (easy, hard and split). 
    The PACE-FS ML-IAP optimized in this manuscript can also reproduce this hierarchy, if DFT data relevant to the hard-to-split migration path are included in the training database.
	\item $a_0[100](011)$ edge and M111 dislocation cores show multiple structures that have a similar energy, which is further verified by NEB calculations. 
    The predicted minimum energy structures can differ depending on the ML-IAP (GAP or PACE-FS) and the training database. This finding shows that, when the energy difference of two core configurations is close to DFT accuracy, ML-IAPs cannot discriminate between the two configurations. We also find that M111 dislocation, which is shown to be the second immobile dislocation character at low temperature in bcc Ta,  has negligible Peierls barrier in bcc iron.
	\item All the ML-IAPstrained in this work and two independent databases confirm that the atomistic fracture mechanisms in bcc iron at T=0K under mode-I loading is cleavage on the pre-cracked plane, irrespective of the crack front. 
	\item GAP-DB-I predicts dislocation and fracture properties that are in good agreement with GAP-DB-II yet using only $1/5$ of the data, showing that the same accuracy and transferability can be achieved by training the ML-IAP to a much smaller database. 
    This finding is important since the DFT database construction involves the most part of the ML-IAP training time.
	Thus, DFT computational resources can be reduced significantly by careful design of the training database and by using active learning techniques.
\end{enumerate}

\section*{Data availability} 

The training scripts and potentials are available on the Github page \url{https://github.com/leiapple/ML-IAPs_iron}. 
The physical properties benchmarking workflow and the LAMMPS script for dislocations are available at \url{https://github.com/leiapple/Potential_benchmark_iron}.

\section*{Acknowledgement} 

This work made use of the Dutch national e-infrastructure with the support of the SURF Cooperative using grant no. EINF-3104. 
We thank the Center for Information Technology of the University of Groningen (UG) for their support and for providing access to the Peregrine and Hábrók high performance computing cluster. 
LZ would like to thank Cas van der Oord and Christoph Ortner for useful discussions. 
FM acknowledges the support through the start-up grant from the Faculty of Science and Engineering at the University of Groningen. \\

\bibliographystyle{elsarticle-num} 
\bibliography{ref}

\end{document}

\setstretch{1.5}
\maketitle

\tableofcontents

\section{Training of the Machine Learning Interatomic Potentials (ML-IAPs)}
\subsection{Summary of the packages considered in this work}

\indent The development of ML-IAPs has become a highly dynamic field, and new packages keep being released. 
Here, we have employed four widely used open-source packages to train ML-IAPs based on two comprehensive bcc iron DFT databases. 
The features of each package are briefly summarized in this section. 
We emphasize that this summary only provides a few fundamental features of the packages employed in the current study and the details can be found in the original publications associated with the packages.

\textit{QUIP} provides a set of tools to run classical MD simulations or tight binding quantum mechanics, and to train GAP potentials. 
It is also interfaced with other packages (LAMMPS, CP2K) and is able to call external packages via \textit{Python} interface. 
Based on the Gaussian process, \textit{QUIP} provides the predicted variance on each atom, which serves as model uncertainty for active learning. 
It is worth mentioning that a massively parallel fitting version of GAP has been developed recently, which enables significant speed-up of the training \cite{klawohn2023massively}. 
\textit{QUIP} also allows to adjust the weights of energy/force/virial for different configurations, which offers the flexibility to emphasize on particular properties during the training. 

\textit{PACEMAER} offers the tool for fitting ML-IAPs based on a generic nonlinear ACE model \cite{drautz2019atomic,lysogorskiy2021performant,bochkarev2022efficient}. 
Both linear and Finnis-Sinclair (FS) nonlinear models can be specified by the user. 
Tensorflow is applied to accelerate the training, especially on GPU architectures \cite{bochkarev2022efficient}. 
The body-order, radial and angular expansion degree, and weights can be customized in the training script. 
Recently, D-optimality (MaxVol algorithm) has been applied to assess the model uncertainty, referred to as the extrapolation grade \cite{lysogorskiy2023active}. 
The calculation of the extrapolation grade is integrated with LAMMPS, which allows the implementation of active learning without heavy coding.  

\textit{MLIP-2} is mainly developed to train MTP, which is accompanied by other functions such as geometry optimization, calculating the extrapolation grade \cite{shapeev2016moment,novikov2020mlip}. 
Since \textit{MLIP-2} implements an empirical optimal relation of the moment tensor descriptor, the users only need to specify the \textit{level} of expansion and the radial basis function form/size  \cite{gubaev2019accelerating,novikov2020mlip}. 
The extrapolation grade can be calculated with the LAMMPS interface, which also provides active learning functionality. 
Recently, a new version (\textit{MLIP-3}) is available, which provides several working modes for different application scenarios.
\textit{MLIP-3} is well-integrated with both VASP and LAMMPS, which may help to reduce the learning curve for the new users.  

\textit{n2p2} is designed for training high-dimensional NNPs, and it offers a set of tools for training ML-IAPs, such as file format conversion and dataset normalization \cite{morawietz2016van,singraber2019parallel}. 
\textit{n2p2} also provides multiple symmetry functions, cutoff functions, and activation functions.  
The model uncertainty can be obtained with the query by committee approach, and can be used in active learning.  

\subsection{Optimal ML-IAP hyperparameters}
\subsubsection{GAP}
\begin{figure}[H]
	\centering
	\includegraphics[trim=0 20 0 10, width=15cm]{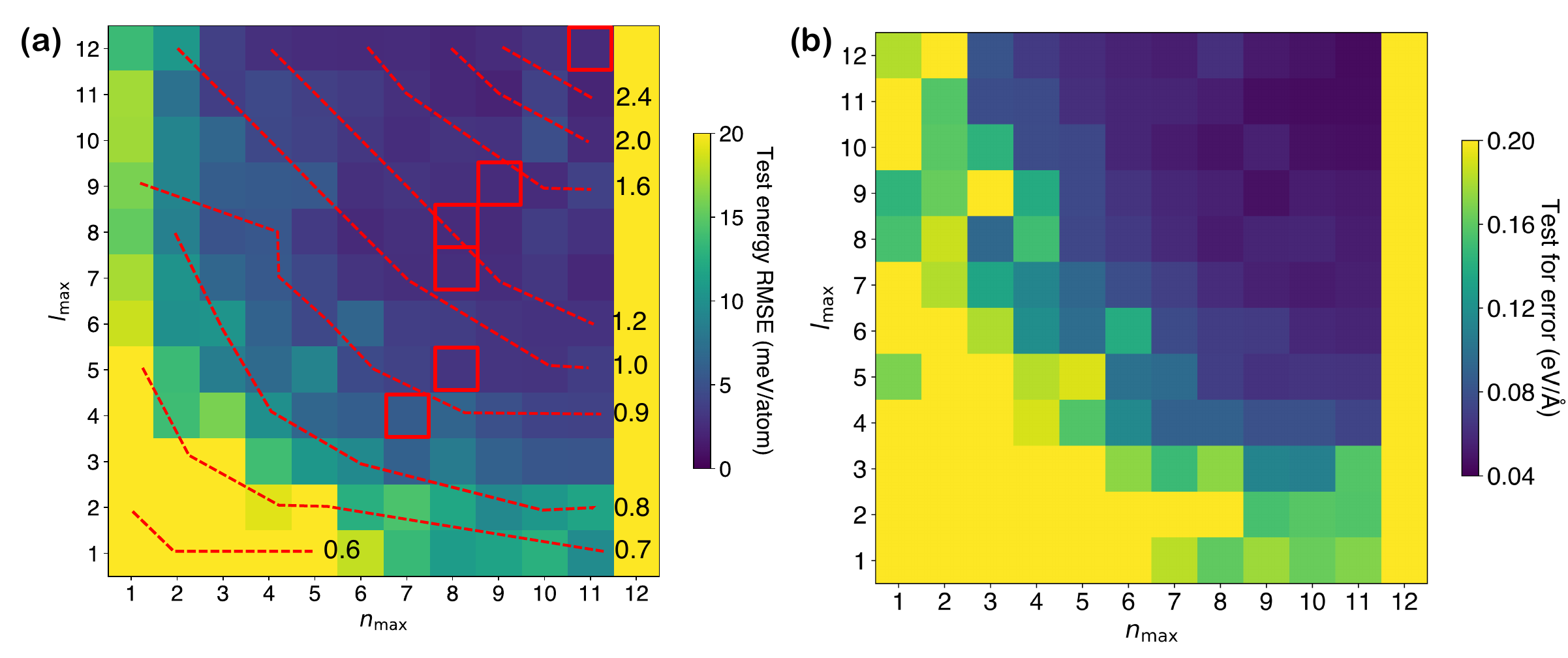}
	\caption{
		Test RMSE on \textbf{(a)} energy and \textbf{(b)} force of GAP (Turbo-SOAP descriptors) with different radial ($n_{\rm max}$) and angular ($l_{\rm max}$) expansion levels. 
		Dashed red lines indicate the same level of computational speed. 
		The numbers next to the lines are the computational cost (unit: seconds) for 16,000 atoms to run 100 timesteps using 32 cores on AMD Rome CPU. 
		The red boxes highlight the optimal combination of radial $n_{\rm max}$ and angular $l_{\rm max}$ expansion levels.
	}
	\label{fig:gap_energy_force}
\end{figure}

The expansion of the LAE in GAP is determined by the radial ($n_{\rm max}$) and angular ($l_{\rm max}$) expansion degree. 
To find the optimal combination of $n_{\rm max}$ and $l_{\rm max}$, a grid search is conducted by varying  $n_{\rm max}$ and $l_{\rm max}$ from 1 to 12.
We choose the maximum value to be 12 since the expansion becomes numerically unstable for $n_{\rm max}>12$ of the TurboSOAP according to Ref. \cite{turboSOAP}. 
As shown in Fig. \ref{fig:gap_energy_force}b, the testing RMSE reduces with the increase of $n_{\rm max}$ and $l_{\rm max}$.
However, the RMSE increases significantly when $n_{\rm max}=12$ for all $l_{\rm max}$, which is due to numerical instability.
With the exception of $n_{\rm max}=12$, the testing RMSE and the computational cost are quasi-symmetric with respect to the diagonal of the contour (Fig. \ref{fig:gap_energy_force}a).
The force RMSE also converges by increasing $n_{\rm max}$ and $l_{\rm max}$ (Fig. \ref{fig:gap_energy_force}b).
Compared with previous results of the original SOAP, where the optimal choice of $n_{\rm max}$ is greater than $l_{\rm max}$ \cite{patrick_2020_gap}, the behavior of TurboSOAP is fundamentally different.
This is because TurboSOAP implements Chebyshev polynomials with recursive relation as radial expansion functions, which is more accurate and efficient than the Gaussian functions from the original SOAP.
The optimal combination of hyperparameters that yields the lowest energy and force RMSE are highlighted by the red boxes in \ref{fig:gap_energy_force}a and listed in Table \ref{tab:gap_error}. 

Table \ref{tab:gap_error}-\ref{tab:n2p2_error} list the optimized ML-IAPs at a given energy and force RMSE. 
The feature that defines the degree of freedom (DOF) of the ML-IAP is also presented.

\begin{table}[H]
	\begin{center}
		\caption{\label{tab:gap_error}
        Optimal combination of hyperparameters at different levels of accuracy for GAP.}
		\begin{tabular}{ c|c|c|c|c } 
			\hline
			Energy RMSE (meV/atom) & Force RMSE (eV/Å) &  $n_{\rm max}$ & $l_{\rm max}$ & DOF \\
			\hline 
			2.49 & 0.045 & 11 & 12 & 8000 \\ 
			2.51 & 0.047 & 9 & 9 & 8000  \\ 
			2.53 & 0.052 & 8 & 8 & 8000  \\
			2.80 & 0.059 & 8 & 7  & 8000 \\
			3.11 & 0.070 & 8 & 5  & 8000 \\
			5.8 & 0.09 & 7 & 4  & 8000 \\
			\hline
		\end{tabular}
	\end{center}
\end{table}

\subsubsection{PACE}
\begin{table}[H]
	\begin{center}
		\caption{\label{tab:pace_error}
  Optimal combination of hyperparameters at different levels of accuracy for PACE-FS and PACE-L.}
		\begin{tabular}{ c|c|c|c|c|c}
			\hline
			Energy RMSE (meV/atom) & Force RMSE (eV/Å) &  Function form & $\kappa$ & $B$  & DOF \\
			\hline 
			4.68 & 0.043 & FS & 0.008 & 200 & 712 \\ 
			4.81 & 0.043 & L & 0.008 & 400 & 760 \\ 
			6.07 & 0.049 & FS & 0.008 & 100 &  464 \\
			8.34 & 0.058 & FS & 0.008 & 70 & 380 \\
			12.87 & 0.074 & FS & 0.008 & 50 & 316 \\
			\hline
		\end{tabular}
	\end{center}
\end{table}

\subsubsection{MTP}
\begin{table}[H]
	\begin{center}
		\caption{\label{tab:mtp_error}
        Optimal combination of hyperparameters at fifferent levels of accuracy for MTP.}
		\begin{tabular}{ c|c|c|c|c } 
			\hline
			Energy RMSE (meV/atom) & Force RMSE (eV/Å) &  \textit{level} & E/F weights  & DOF \\
			\hline 
			4.536 & 0.053 & 22 & 0.001 & 548 \\ 
			5.631 & 0.057 & 22 & 0.01 & 548  \\ 
			5.683 & 0.043 & 22 & 0.05 & 548 \\
			10.6125 & 0.061& 16 & 0.05 & 124  \\
			18.3344 & 0.076 & 14 & 0.05& 84 \\
			\hline
		\end{tabular}
	\end{center}
\end{table}

\subsubsection{NNP}
\begin{table}[H]
	\begin{center}
		\caption{\label{tab:n2p2_error}
  Optimal combination of hyperparameters at different levels of accuracy for NNP.}
		\begin{tabular}{ c|c|c|c|c } 
			\hline
			Energy error (meV/atom) & Force error (eV/\AA) &  angular expansion & No. of neurons & DOF \\
			\hline 
			4.903 & 0.1139 & 16 & 40 & 2361 \\
			6.625 & 0.0769 & 24 & 20 & 941 \\
			7.877 & 0.0761 & 24 & 10 & 371 \\ 
			9.791 & 0.0911 & 8 & 10 & 211 \\
			\hline
		\end{tabular}
	\end{center}
\end{table}

\subsection{Details of energy and force convergence}

\begin{figure}[H]
	\centering
	\includegraphics[trim=0 20 0 20, width=15cm]{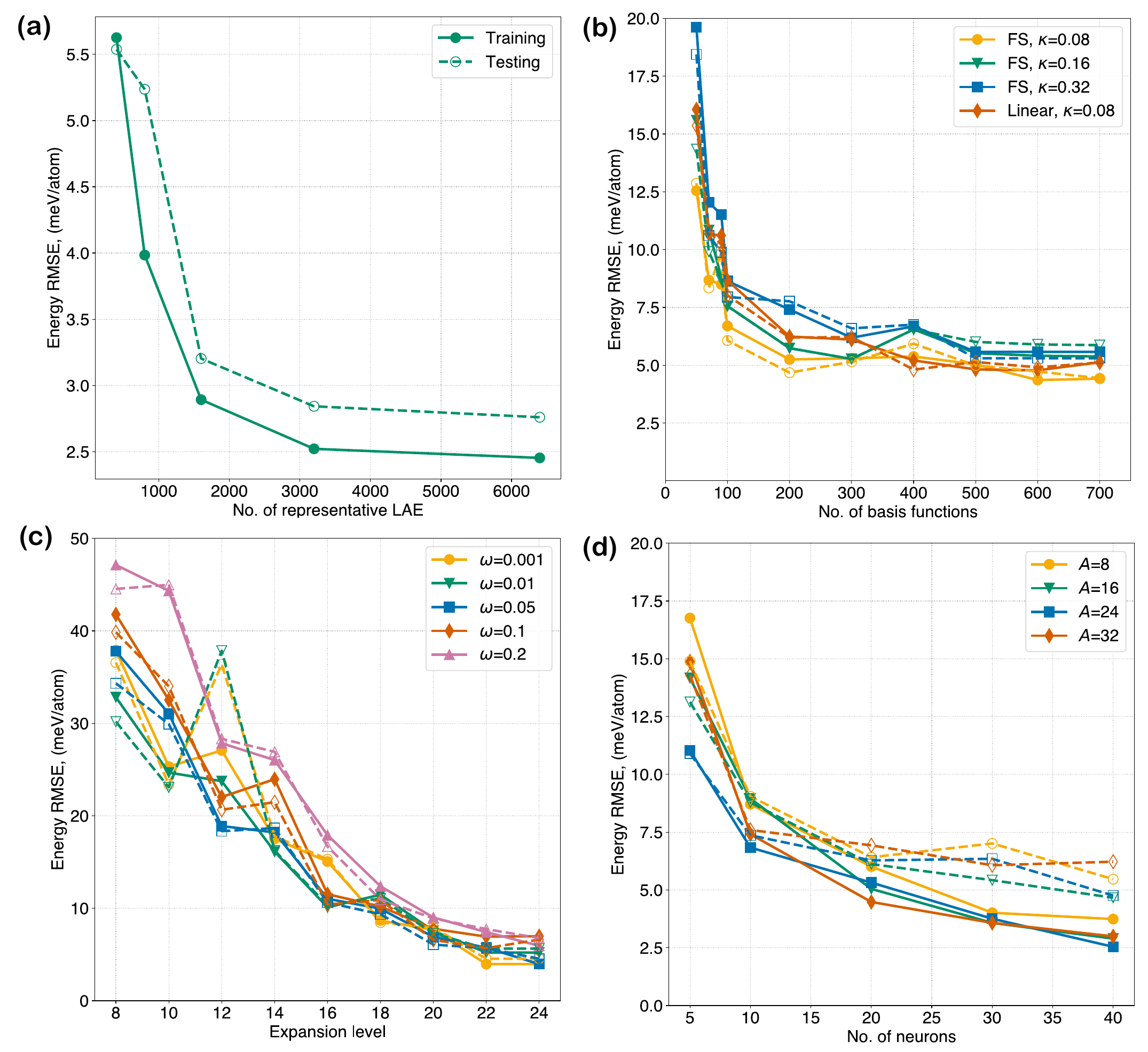}
	\caption{Training and testing RMSE on energy for
		\textbf{(a)} GAP
		\textbf{(b)} PACE-FS and PACE-L,
		\textbf{(c)} MTP, 
		and \textbf{(d)} NNP.
		Solid and dashed lines indicate the training and testing error, respectively.
		$\kappa$ is the relative energy/force weights ($1/\kappa-1$) of PACE. 
		$\omega$ is the energy/force weights of MTP.
		$A$ is the number of angular basis function of NNP. 
	}
	\label{fig:convergence_panel}
\end{figure}

Fig. \ref{fig:convergence_panel} shows the full convergence of training/testing energy RMSE for all ML-IAPs, supplementing \textit{Fig. 2 of the main manuscript}. 
A fluctuation within 1 meV/atom is observed by changing the weights for PACE and MTP, showing that the RMSE can be converged for all weights considered here (Fig.\ref{fig:convergence_panel}b and c).
However, the convergence of the potential with respect to the prediction of benchmark properties and extended defect configurations remains to be validated.
As the number of angular basis functions increases, the training and testing energy RMSE does not decrease significantly. 
Yet, the flexibility of the potential enhances upon increasing the number of DOFs.
This highlights the necessity of the potential validation via approaches beyond RMSE.

\begin{figure}[H]
	\centering
	\includegraphics[trim=0 20 0 20, width=15cm]{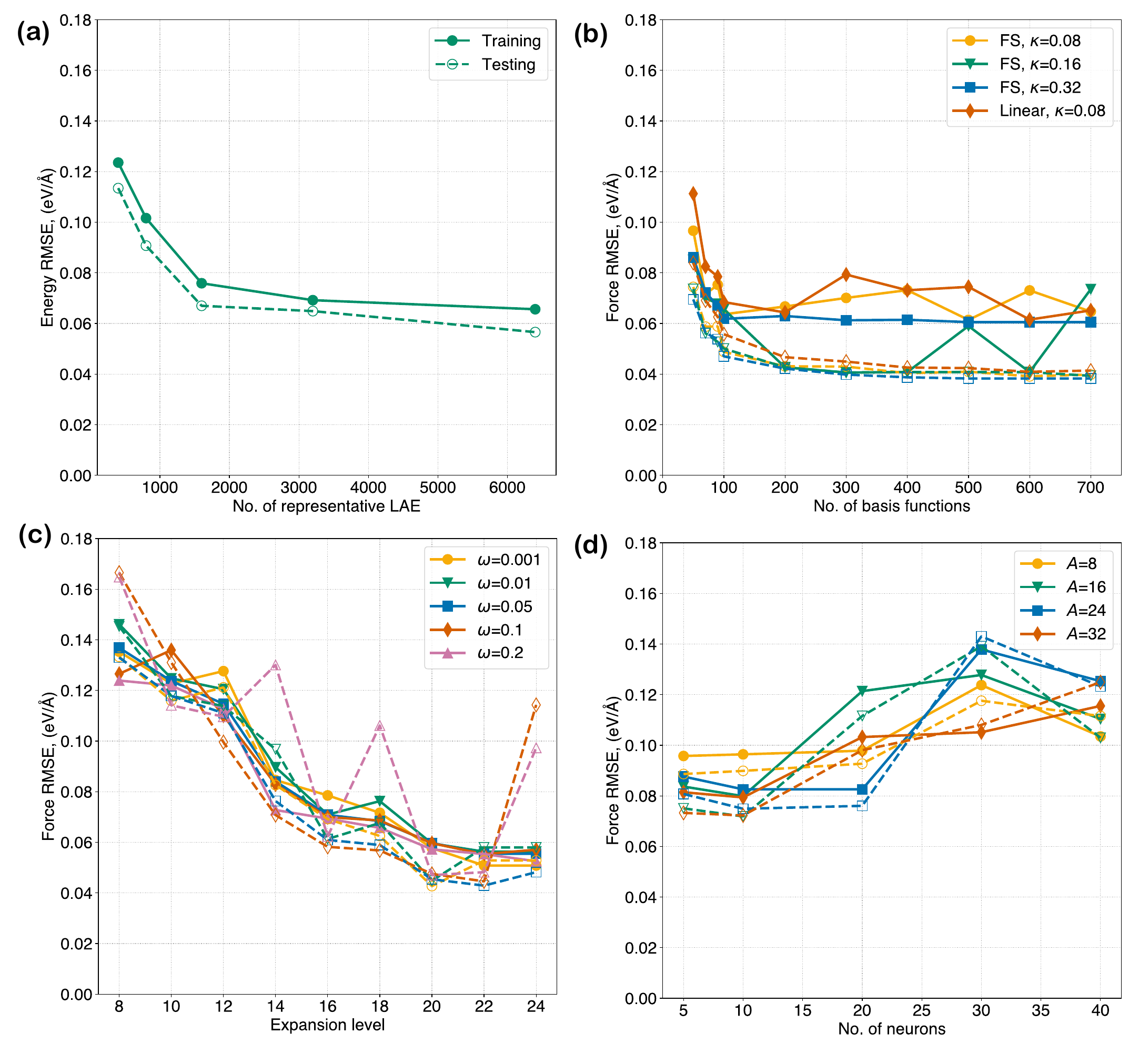}
	\caption{
		Training and testing RMSE on force for
		\textbf{(a)} GAP
		\textbf{(b)} PACE-FS and PACE-L,
		\textbf{(c)} MTP, 
		and \textbf{(d)} NNP.
		Solid and dashed lines indicate the training and testing error, respectively.
		$\kappa$ is the relative energy/force weights ($1/\kappa-1$) of PACE. 
		$\omega$ is the energy/force weights of MTP.
		$A$ is the number of angular basis function of NNP. 
	}
	\label{fig:convergence_panel_force}
\end{figure}

Fig. \ref{fig:convergence_panel_force} is the convergence of force RMSE that corresponds to Fig. \ref{fig:convergence_panel}.
PACE-L and PACE-FS converge as the number of basis function increases (Fig. \ref{fig:convergence_panel}a).
Some of the testing force RMSE of MTP $\omega=0.2$ is 0.04 eV/Å larger than the training, indicating possible overfitting.
The force RMSE is converged to 0.08 eV/Å with only 5 neurons for NNP.
Further increase of number of neurons leads to an increase of force RMSE, which is caused by the relative small weight of force in the loss function. 
Compared with other ML-IAPs, the force RMSE is not converged to the same error (0.04 eV/Å) in NNP. 

\begin{figure}[H]
	\centering
	\includegraphics[trim=0 20 0 40, width=14cm]{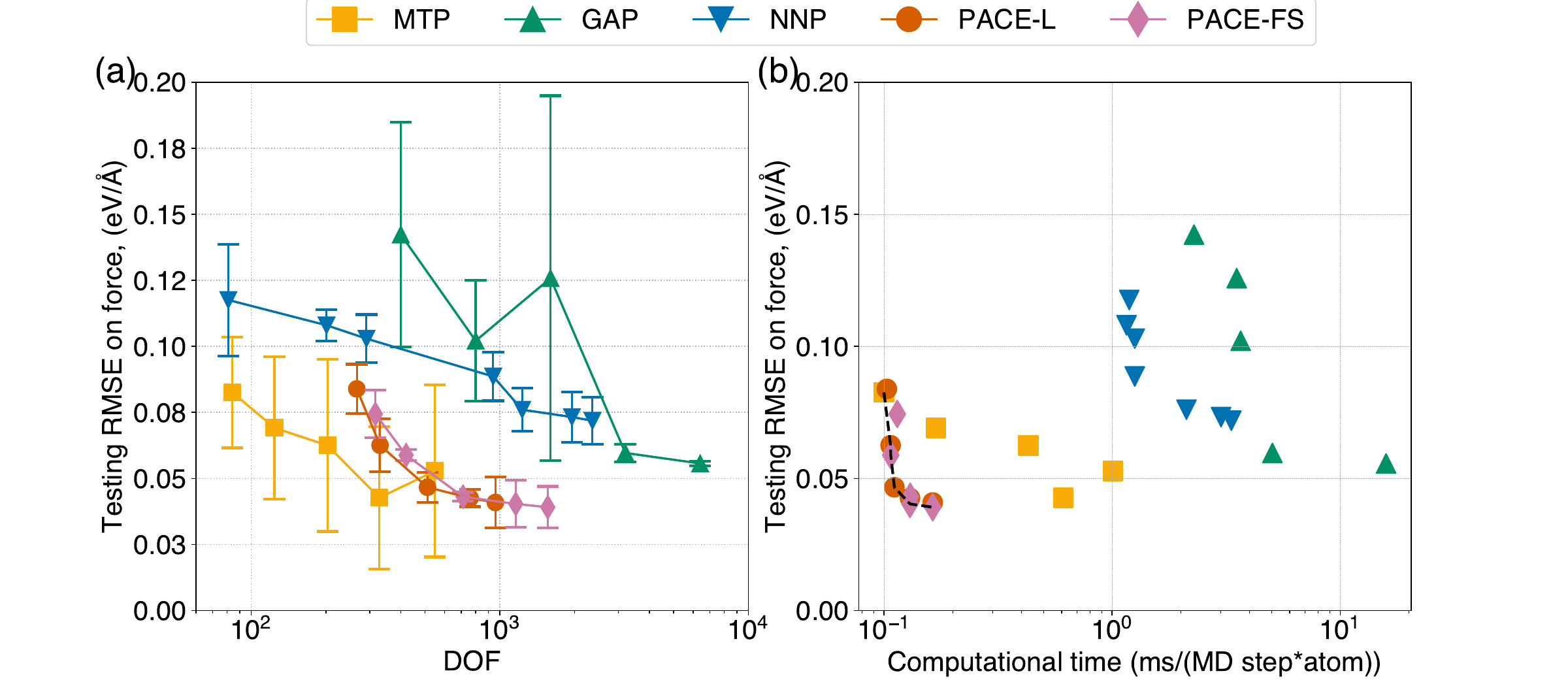}
	\caption{
		\textbf{(a)} Testing force RMSE as a function of DOFs for five ML-IAPs considered in the current study. 
		\textbf{(b)} Testing force RMSE versus computational cost different MLIAPs. 
		The grey dashed line indicates an approximate Pareto frontier formed by the convex hull of points lying on the bottom left of the chart, which approximates an optimal trade-off between accuracy and computational cost.
		Timings are performed by LAMMPS calculations on a paralleled using 32 cores on a AMD Rome CPU, 7H12 (2x), 64 Cores/Socket, 2.6GHz, 280W.
	}
	\label{fig:pareto_front}
\end{figure}

Fig. \ref{fig:pareto_front}a plots the force RMSE as functions of DOF for the optimized ML-IAPs (\textit{Fig. 3 of the main manuscript}).
GAP is able to reach 0.06 eV/Å at the largest DOF. 
MTP, PACE-L, and PACE-FS can reach 0.04 eV/Å for training and testing RMSE respectively.
For NNP, the force RMSE is larger than other ML-IAPs, i.e., 0.08 eV/Å for both training the testing. 
Fig. \ref{fig:pareto_front}b shows that PACE-L and PACE-FS occupy the Pareto front, which is also indicated by\textit{ Fig. 3 of the main manuscript}.

\subsection{Convergence of GAP trained on DB-II}

\begin{figure}[H]
	\centering
	\includegraphics[trim=0 20 0 40, width=8cm]{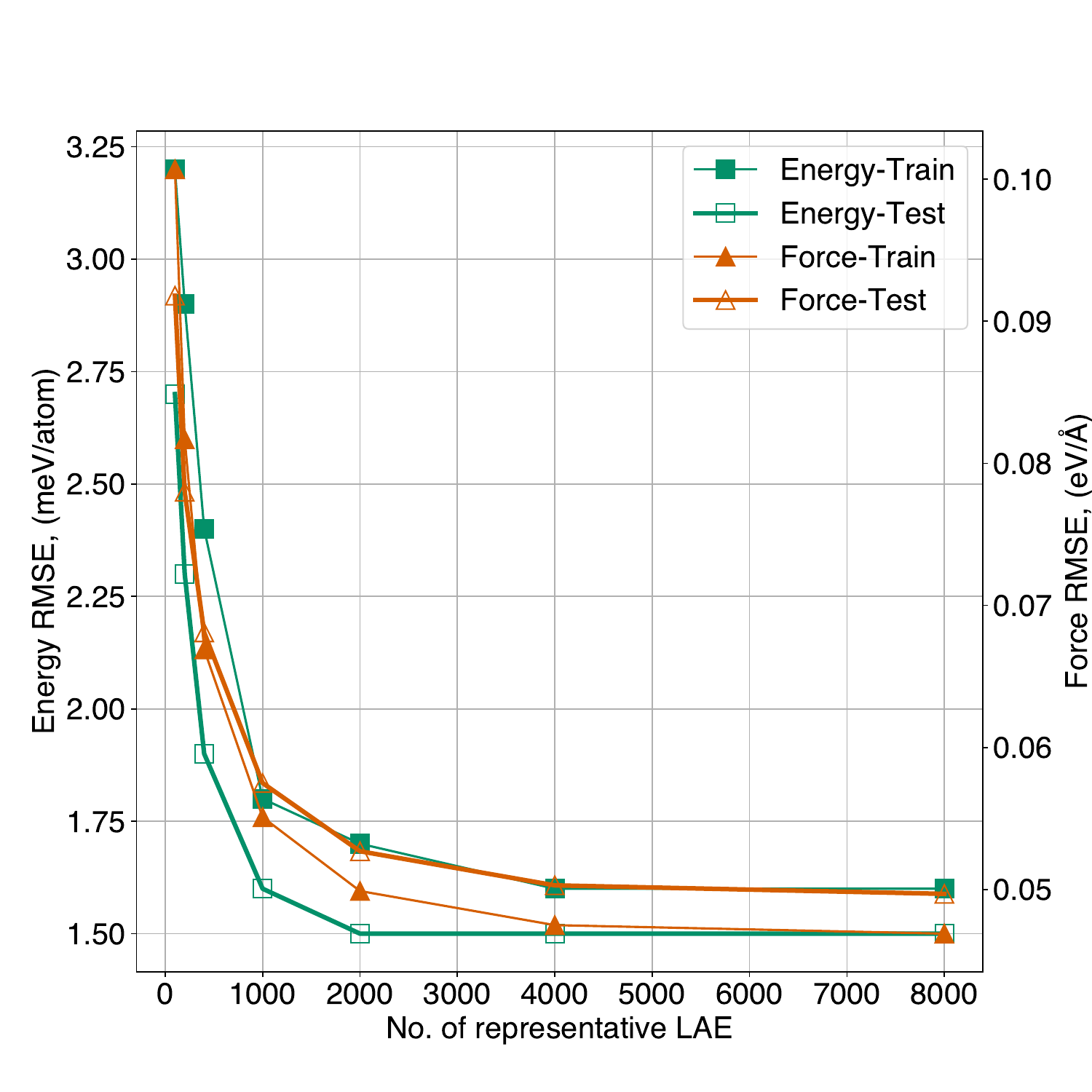}
	\caption{
		Training/Testing energy/force RMSE as a function of the number of representative points for GAP trained on DB-II.
	}
	\label{fig:gap_meng}
\end{figure}

To investigate the influence of the database, we train GAP on another independent database, i.e. DB-II from the Fe-H database developed in \cite{ogata_nnp_meng2021general}.
 \textit{QUantum mechanics and Interatomic Potentials (QUIP) } is employed to train the potential. 
A distance-based 2-body descriptor and two TurboSOAP descriptors are employed \cite{bartok2013representing,turboSOAP}, i.e., one “inner” TurboSOAP with $r_{\rm cut}=3Å$ and another “outer” TurboSOAP with $r_{\rm cut}=6.5Å$.
To ensure  convergence of the sparse GP, we investigate the influence of $M$ on RMSE by fixing the LAE expansion ($n_{\rm max}=8,\ l_{\rm max}=8$). 
We train an array of GAP IAPs on DB-II using different numbers of representative points ($M$), as shown in Fig. \ref{fig:gap_meng}. 
The energy and force RMSE converge to 1.6 meV/atom and 0.05 eV/Å, respectively. 
The results shown in \textit{Section 4 and 5} of the main manuscript are derived using GAP-DB-II with $M=8000$. 

\section{Benchmark properties}
\subsection{Generalized stacking fault energy profile}

\begin{figure}[H]
	\centering
	\includegraphics[trim=0 0 0 20, width=14cm]{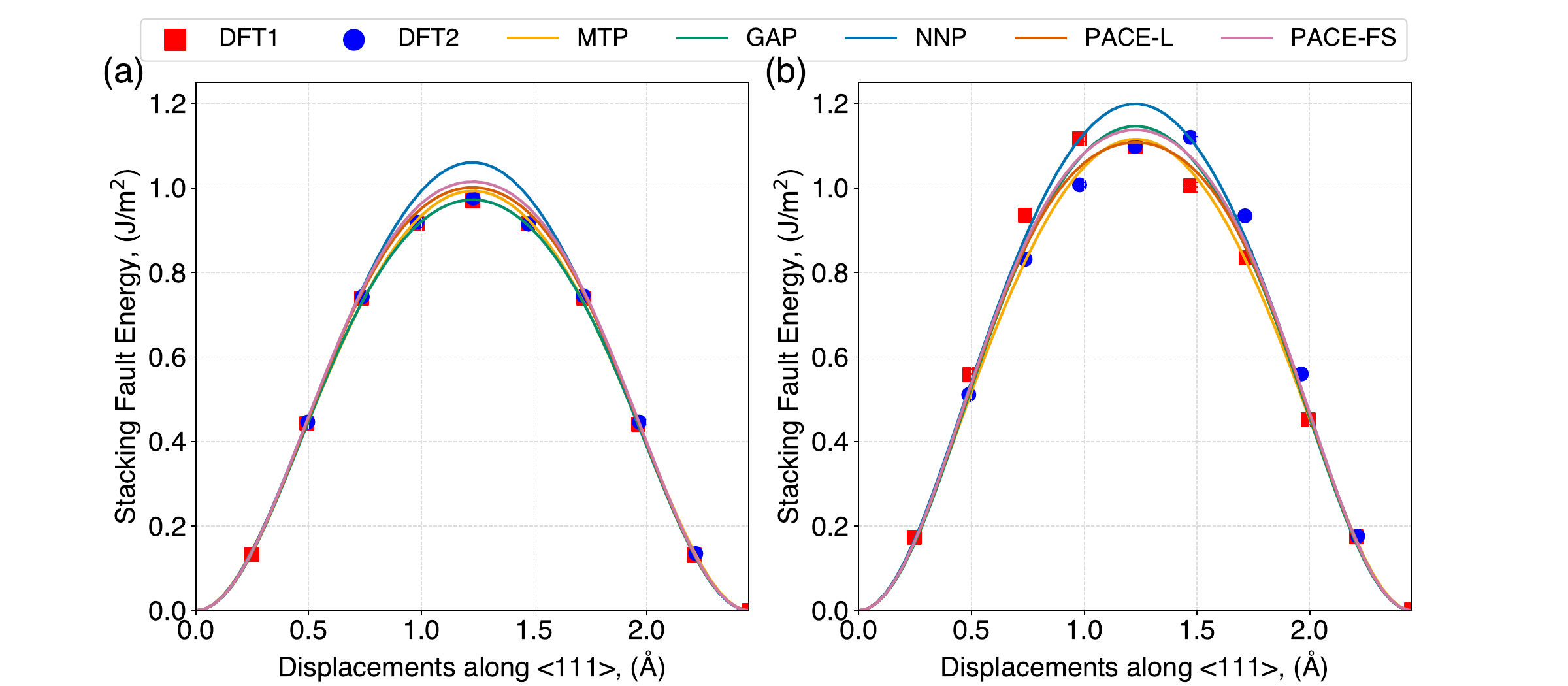}
	\caption{Generalized stacking fault energy predicted by different ML-IAPs: 
		\textbf{(a)} \{100\} and 
		\textbf{(b)} \{110\} planes. 
		DFT1 and DFT2 are taken from Ref. \cite{mtp_wang2022machine} and \cite{ventelon2010generalized}.
    }
	\label{fig:gsfe}
\end{figure}

Fig. \ref{fig:gsfe} shows the generalized stacking fault energy curves predicted by all ML-IAPs shown in \textit{Fig. 4a of the main manuscript}. 
All ML-IAPs shows good agreement with DFT predictions. 

\subsection{Quality factor and force RMSE}

\begin{figure}[H]
	\centering
	\includegraphics[trim=0 20 0 20, width=8cm]{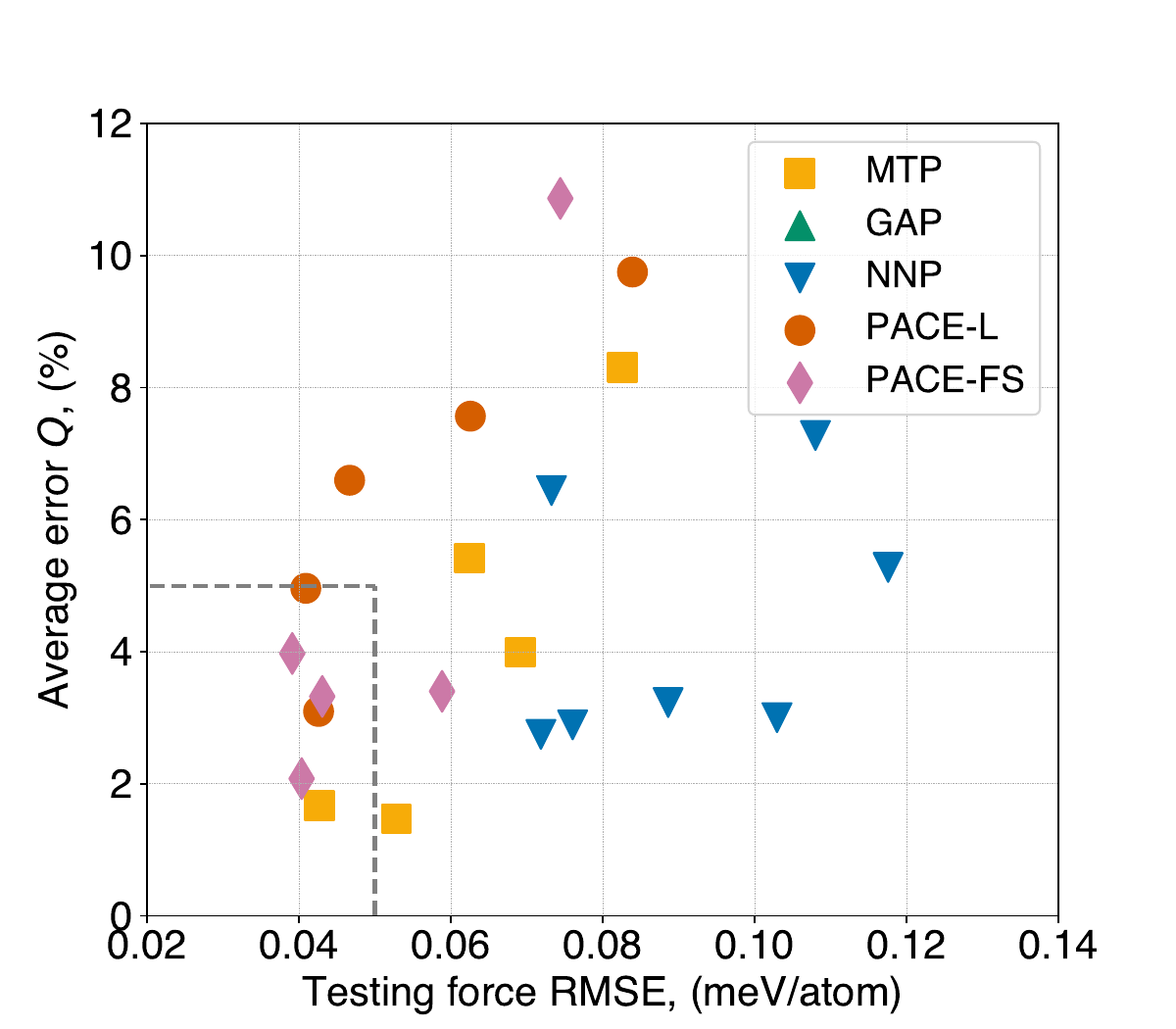}
	\caption{
		Error index $Q$ versus testing force RMSE. 
	}
	\label{fig:quality_factor}
\end{figure}

Fig. \ref{fig:quality_factor} plots the predicted quality factor $Q$ as a function of the testing force RMSE.
The data show the general trend that larger RMSE corresponds to a larger average error.
However, no linear correlation can be derived from the data. 

\section{Screw dislocation}

\subsection{Details about differential-displacement maps}

\begin{figure}[H]
    \centering
    \includegraphics[trim=0 0 0 0,width=16cm]{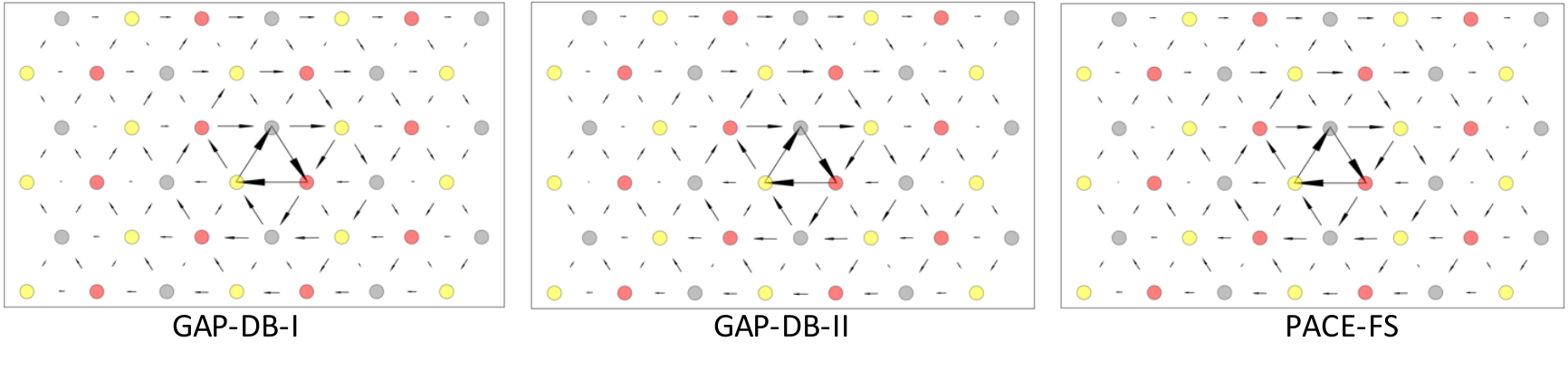}
    \caption{DD map of screw dislocation core predicted by the ML-IAPs.}
    \label{fig:dd_map}
\end{figure}

Fig. \ref{fig:dd_map} shows the  differential displacement (DD) maps of screw dislocation predicted by three ML-IAPs, where  the arrows reveal the relative out-of-plane displacement of neighboring atomic columns. 
The arrows are normalized such that the magnitude equals $b/3$ (where $b$ is the Burgers vector magnitude) is the atomic centers are connected. 
Atoms are colored according to the three possible positions along the dislocation line direction in the pristine bcc crystal, for one periodic unit cell (length $b$). 
The DD maps indicate a compact  screw dislocation core structure, which is consistent with the DFT literature \cite{ventelon2010generalized,itakura2012first,ventelon2013ab}.

\subsection{Method to compute hard-to-split core path}

\begin{figure}[H]
	\centering
	\includegraphics[trim=100 160 100 190,width=12cm]{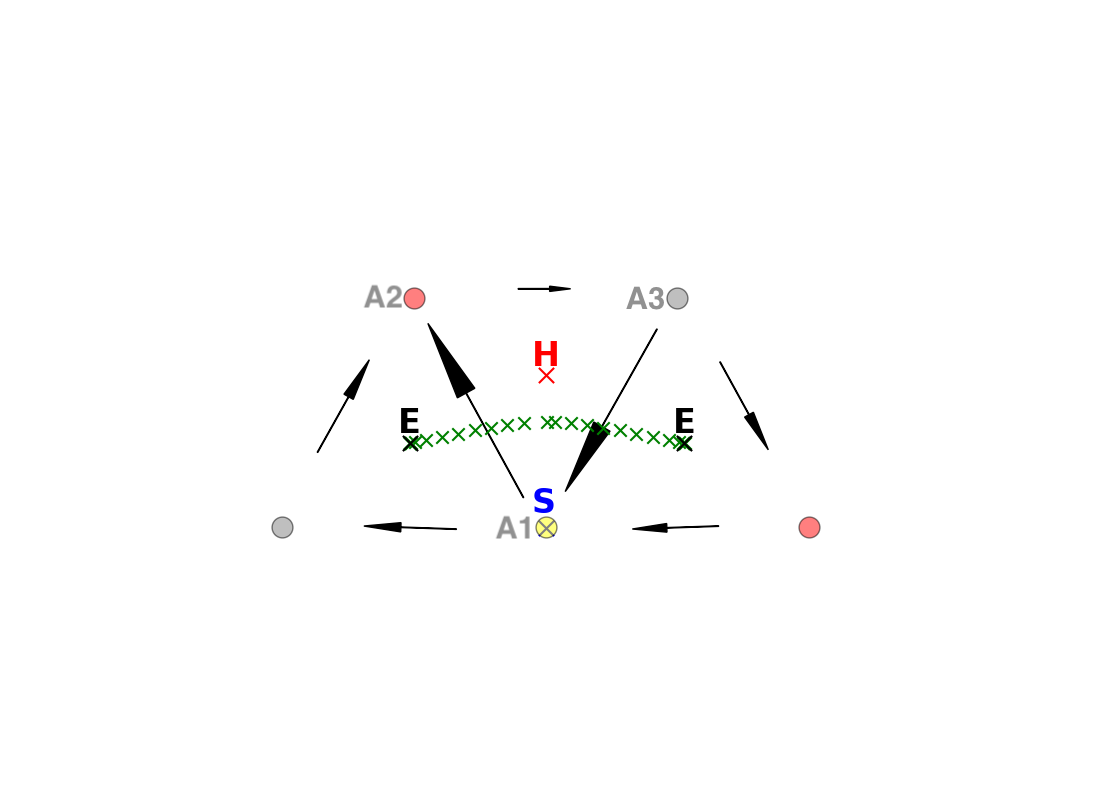}
	\caption{
	Atomic configurations of the screw dislocation core with the differential displacement (DD) map.
	Black \textbf{E}, red \textbf{H}, and blue \textbf{S} crosses indicate the positions of the easy core, hard core, and split core, respectively.
	Green crosses indicate the dislocation core trajectory calculated according to Ref. \cite{ann_mori2020neural}.	
	\textbf{A1}, \textbf{A2}, and \textbf{A3} refer to the three atomic columns that are the closest to the dislocation core along its migration path from easy to easy core configurations}. 
	\label{fig:h2s_trajectory}
\end{figure}

We use the rigid boundary (RB) configuration (\textit{Fig. 5 of the main manuscript}) to calculate the energetics of the screw dislocation in bcc iron. 
The hard core configuration is obtained by centering the anisotropic elastic field at the center of the downward triangle, indicated by the red ``\textbf{H}" in  Fig. \ref{fig:h2s_trajectory}. 
The split core configuration is generated based on hard core configuration, by adding a displacement along the dislocation line direction of $b/6$ to \textbf{A2} and $-b/6$ to \textbf{A3} (Fig. \ref{fig:h2s_trajectory}) \cite{ventelon2013ab}.
The configurations along the hard-to-split core path are generated by interpolating linearly the in-plane displacements of the three innermost atomic columns (\textbf{A1-A3 }in Fig. \ref{fig:h2s_trajectory}) between the hard core and the split core. 
The energy profile is obtained by relaxing the dislocation structures. 
Conjugate gradient algorithm with a force tolerance of 10$^{-12}$ eV/Å is applied during the relaxation.
Note that the displacements along the dislocation line of atomic columns \textbf{A1-A3} are fixed during the relaxation. 

\subsection{Analysis of the hard-to-split core energy profile}

\begin{figure}[H]
	\centering
	\includegraphics[trim=0 0 0 0,width=14cm]{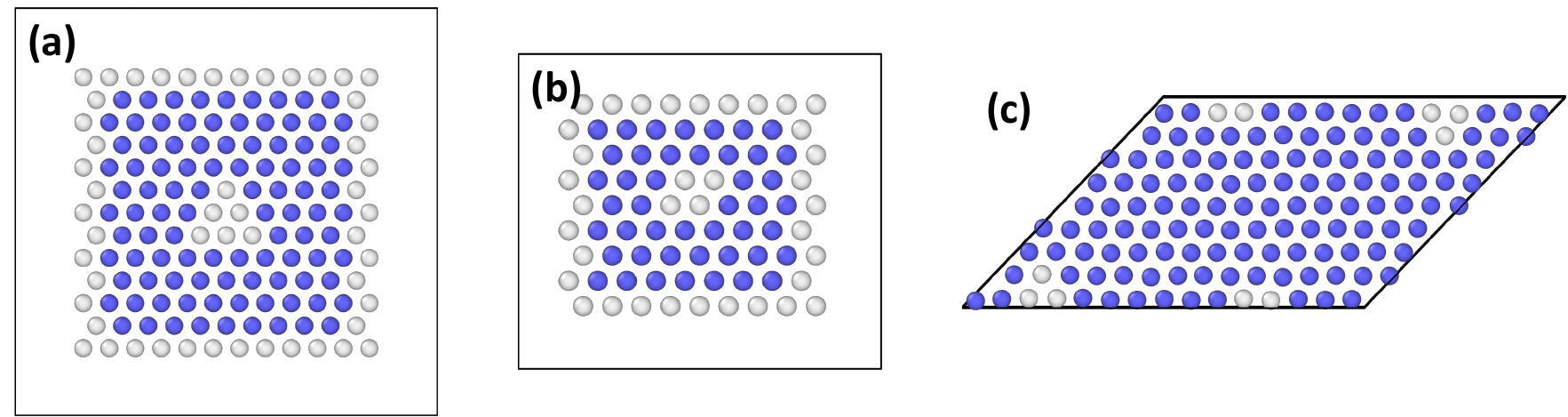}
	\caption{
		Screw dislocation configurations extracted from DB-II \cite{ogata_nnp_meng2021general}. 
		\textbf{(a)} 150 atoms cluster at T=0K,
		\textbf{(b)} 81 atoms cluster at T=0K, 
		and \textbf{(c)} 135 atoms dislocation dipole configurations at T=100K and 300K.
	}
	\label{fig:screw_configs}
\end{figure}

Since DB-II includes various screw dislocation configurations, we trained a spectrum of GAP and PACE-FS based on DB-II to understand their interpolation ability. 
As detailed in Ref. \cite{ogata_nnp_meng2021general}, DB-II includes three different screw dislocation configurations (Fig. \ref{fig:screw_configs}).
The clustered configurations are obtained by self-consistent calculations at T=0K while the dislocation dipole configuration is obtained using AIMD at T=100K and 300K. 
We extract the energies from one set of easy-to-easy configurations (Fig. \ref{fig:screw_configs}a) and the two sets of hard-to-split core configurations (Fig. \ref{fig:screw_configs}a and b). 
We predict the energies of those dislocation configurations from DB-II and compare the ML-IAP predictions with these DFT results to reveal the accuracy of GAP, PACE and NNP. 

\begin{figure}[H]
	\centering
	\includegraphics[trim=0 0 0 0,width=18cm]{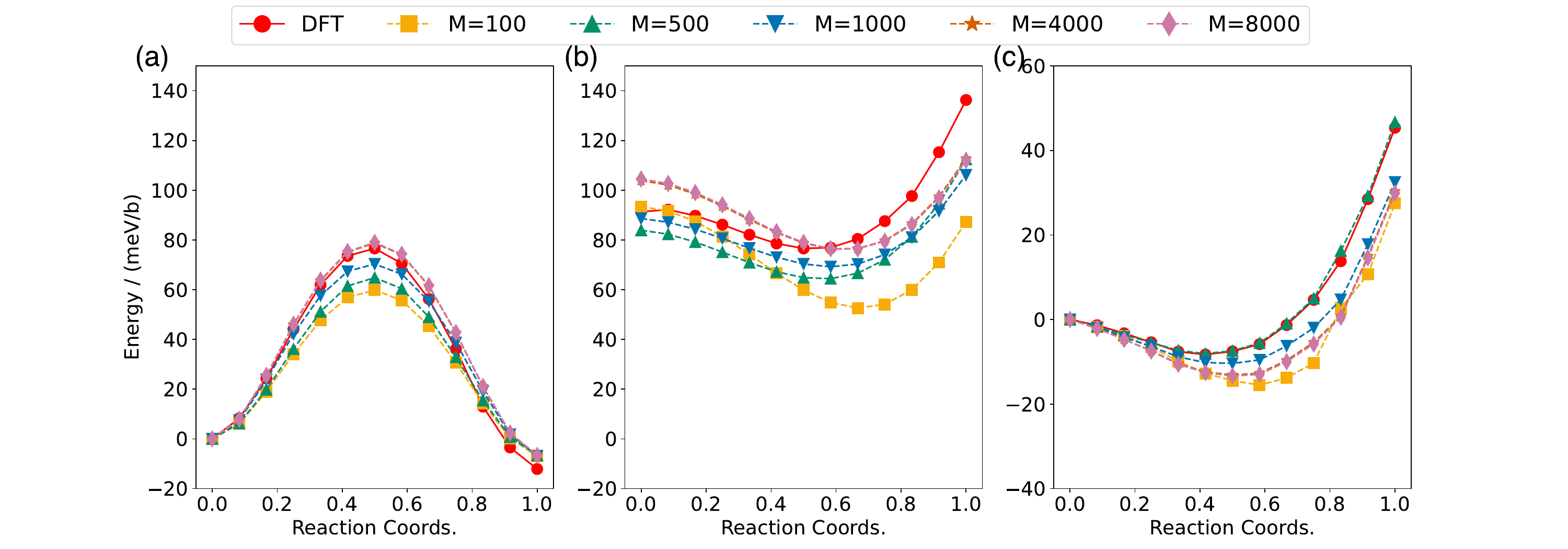}
	\caption{
		Energy of clustered screw dislocation configurations in DB-II predicted by DFT and GAP potentials with different number of DOF.
        Note that the energy differences in \textbf{(a)} and \textbf{(b)} are computed with respect to the easy core energy since they are both predictions based on the 81 atoms clustered configuration.
        The energy differences in \textbf{(c)} are computed with respect to the hard core configuration. 
	}
	\label{fig:h2s_gap}
\end{figure}

\begin{figure}[H]
	\centering
	\includegraphics[trim=0 0 0 0,width=18cm]{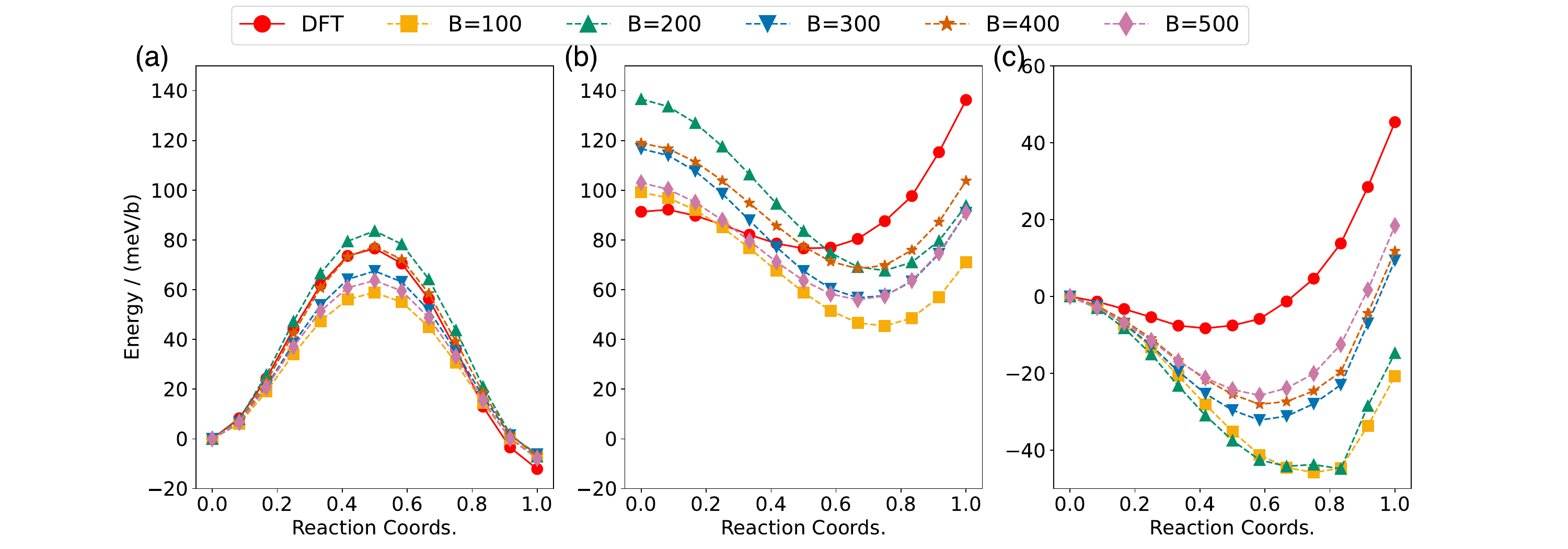}
	\caption{
		Energy of clustered screw dislocation configurations in DB-II predicted by DFT and PACE-FS potentials with different number of basis functions.
         Note that the energy differences in \textbf{(a)} and \textbf{(b)} are computed with respect to the easy core energy since they are both predictions based on the 81 atoms clustered configuration.
        The energy differences in \textbf{(c)} are computed with respect to the hard core configuration.
	}
	\label{fig:h2s_pace}
\end{figure}

Figs. \ref{fig:h2s_gap}-\ref{fig:h2s_pace} plot the predictions of GAP and PACE-FS for the screw configurations in DB-II. 
The prediction of both ML-IAPs is improved by increasing the DOFs, i.e. $M$ in GAP and $B$ in PACE.
The prediction of GAP $M=4000$ overlaps with $M=8000$, indicating a converged prediction (\ref{fig:h2s_gap}).
However, the PACE prediction does not converge with the number of basis functions. 
The current result confirms (like the RMSE data) that GAP is capable of reproducing DFT energies that are more accurate than PACE.


\begin{figure}[H]
	\centering
	\includegraphics[trim=0 0 0 0,width=18cm]{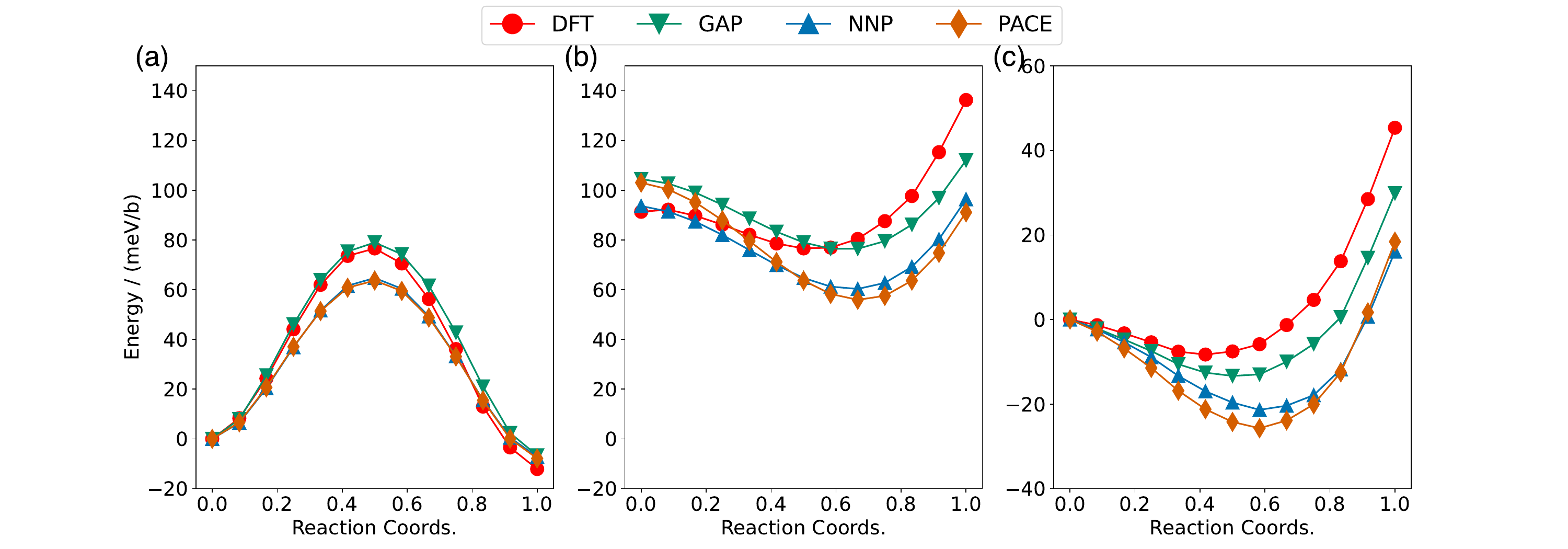}
	\caption{
		Energy of clustered screw dislocation configurations (Fig. \ref{fig:screw_configs}a and b) in DB-II predicted by DFT, GAP (M=8000), original NNP, and PACE-FS (B=500).
        Note that the energy differences in \textbf{(a)} and \textbf{(b)} are computed with respect to the easy core energy since they are both predictions based on the 81 atoms clustered configuration.
        The energy differences in \textbf{(c)} are computed with respect to the hard core configuration.
	}
	\label{fig:h2s_all}
\end{figure}

\begin{figure}[H]
	\centering
	\includegraphics[trim=0 0 0 0,width=10cm]{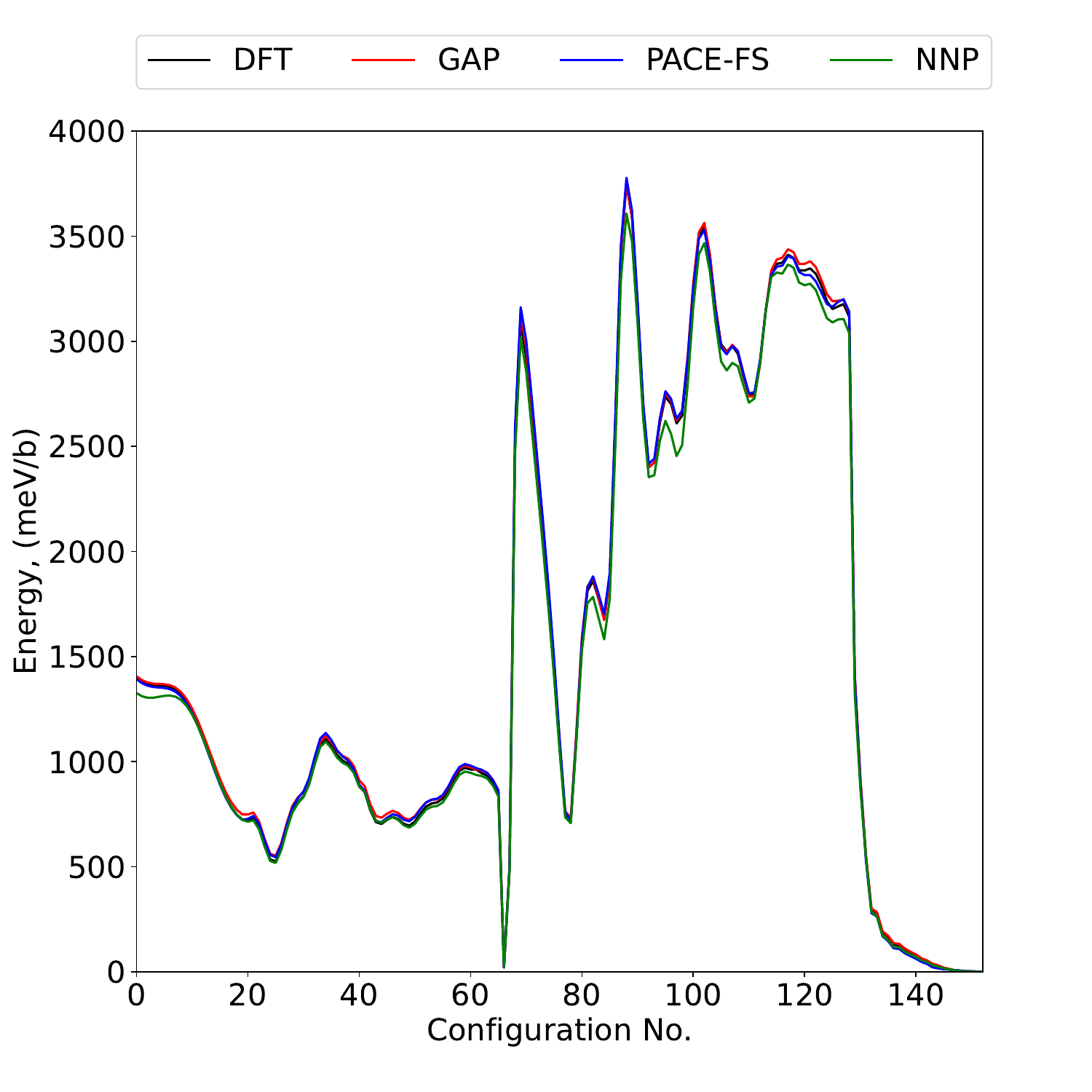}
	\caption{
		Energy of screw dislocation dipole configurations (Fig. \ref{fig:screw_configs}c) in DB-II predicted by DFT, GAP (M=8000), original NNP, and PACE-FS (B=500).
	}
	\label{fig:disl_dipole}
\end{figure}

Fig. \ref{fig:h2s_all} plots the predictions of NNP (original, from Ref. \cite{ogata_nnp_meng2021general}), GAP (M=8000) and PACE-FS (B=500) for the clustered screw configurations in DB-II. 
Surprisingly, NNP underestimates the DFT energies, especially at the maximum and minimum of the transition path. 
The prediction of PACE-FS is very close to NNP while GAP predictions are the most accurate with respect to DFT energies.
This trend is confirmed by direct comparison between ML-IAP and DFT energies related to the dipole configurations in Fig. \ref{fig:screw_configs}c. 
Fig. \ref{fig:disl_dipole} shows that GAP and PACE-FS are consistently close to DFT, while NNP deviates the most from the DFT predictions.
Therefore, the accuracy of PACE-FS is comparable or better than NN, yet at a fraction of the computational cost (see Pareto front).

\section{Non-screw dislocations}

\subsection{Dislocation core structures}
\subsubsection{Predictions of GAP-DB-II and PACE-FS, using RB configuration}

\begin{figure}[H]
	\centering
	\includegraphics[trim=0 20 0 20,width=14cm]{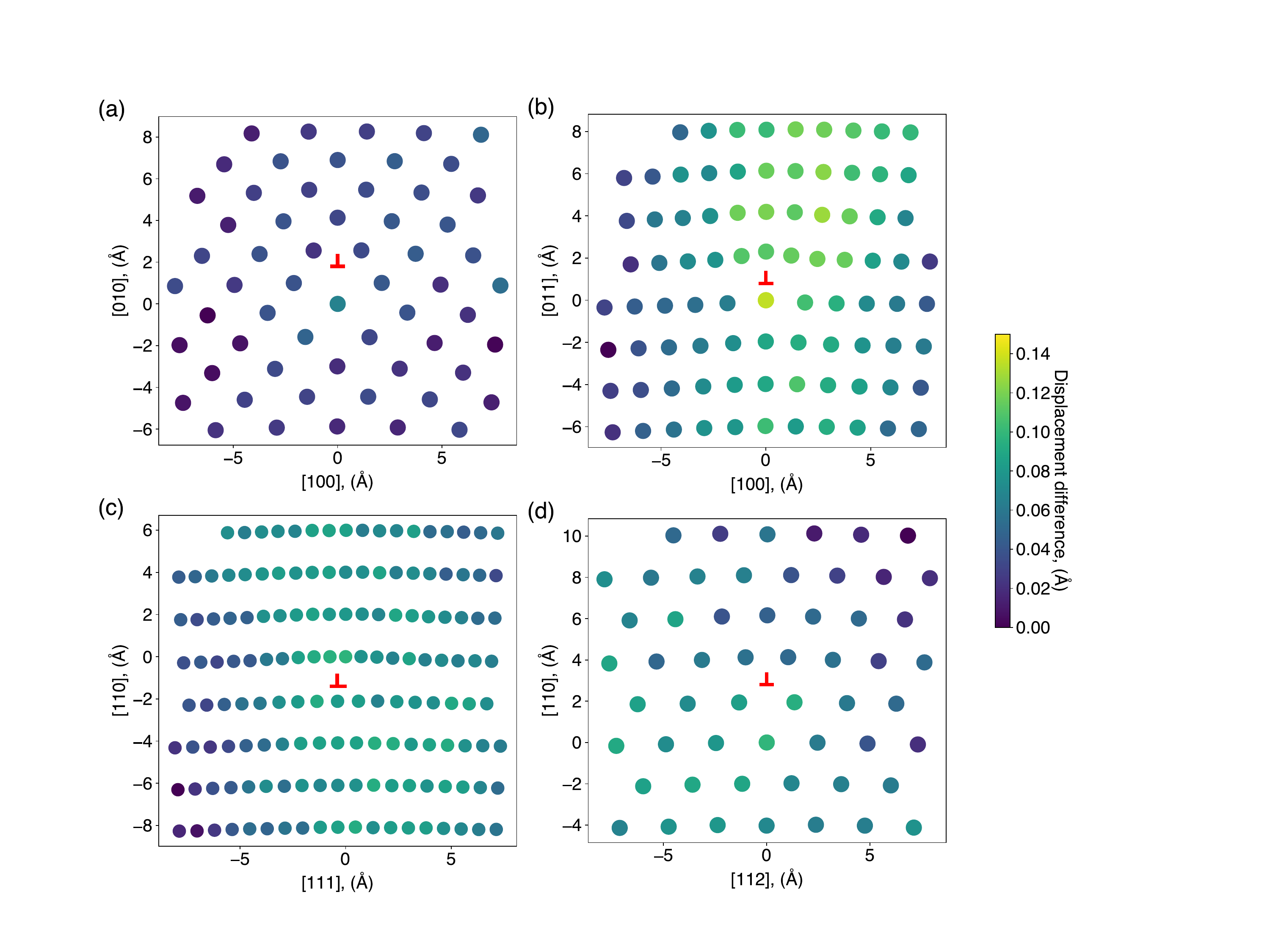}
	\caption{Core structure predicted by GAP-DB-II: \textbf{(a)} $a_0[100](010)$ edge, 
		\textbf{(b)} $a_0[100](011)$ edge, 
		\textbf{(c)} $a_0/2[\bar{1}\bar{1}1](1\bar{1}0)$ edge, and
		\textbf{(d)} $a_0/2[111](1\bar{1}0)$ 70.5$^\circ$  mixed dislocations.
		Atoms are colored by the displacement difference between GAP-DB-II and DFT predictions \cite{dftCore_fellinger2018geometries}.
		Dislocation center is indicated by a dislocation symbol ``$\mathbf{\perp}$''. }
	\label{fig:disl_gap_db2}
\end{figure}

\begin{figure}[H]
	\centering
	\includegraphics[trim=0 80 0 50,width=14cm]{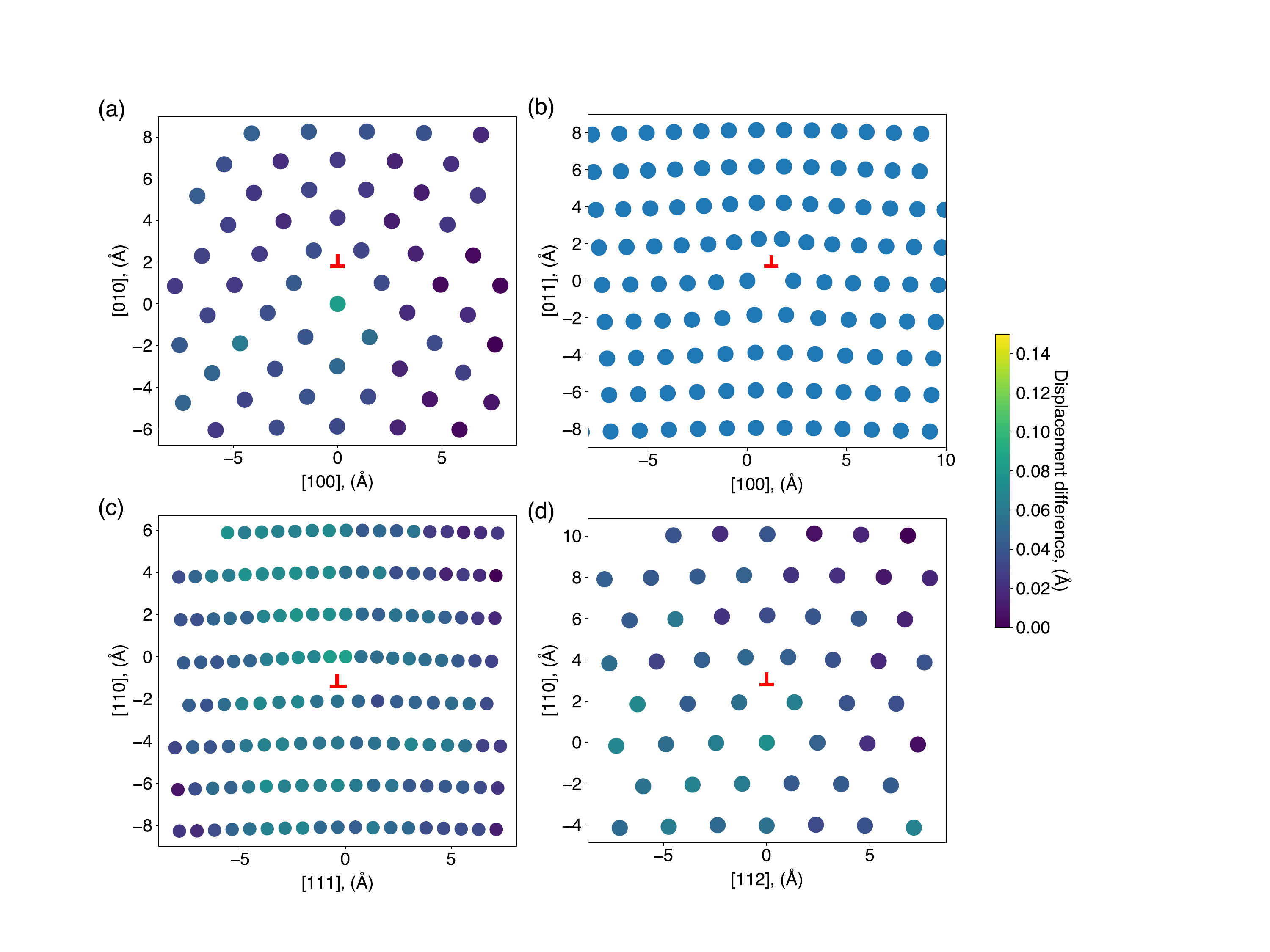}
	\caption{Core structure predicted by PACE-FS: \textbf{(a)} $a_0[100](010)$ edge, 
		\textbf{(b)} $a_0[100](011)$ edge, 
		\textbf{(c)} $a_0/2[\bar{1}\bar{1}1](1\bar{1}0)$ edge, and
		\textbf{(d)} $a_0/2[111](1\bar{1}0)$ 70.5$^\circ$  mixed dislocations.
		Atoms are colored by the displacement difference between GAP-DB-I and DFT predictions \cite{dftCore_fellinger2018geometries}.
		Dislocation center is indicated by a dislocation symbol ``$\mathbf{\perp}$''.
		Note that \textbf{(b)} is not colored according to the scalebar because the core structure is different from DFT prediction. 
	}
	\label{fig:disl_pacefs}
\end{figure}

Fig. \ref{fig:disl_gap_db2} and \ref{fig:disl_pacefs} plot the core structure predicted by GAP-DB-II and PACE-FS respectively. 
The atoms are colored according to the displacement difference between DFT and ML-IAP predictions. 
All core structures predicted by GAP-DB-II have errors less than 0.15 Å. 
In particular, the error of M111 dislocation is smaller than the prediction of GAP-DB-I, which may be the results of including dislocation configurations in the database. 
PACE-FS predicts a bond-centered structure (BC) for $a_0[100](011)$ edge dislocation (Fig. \ref{fig:disl_pacefs}), in contrast with the DFT prediction. 

\subsubsection{Impact of initial configuration asymmetry on the predicted core structures}

\begin{figure}[H]
	\centering
	\includegraphics[trim=0 -10 0 0,width=10cm]{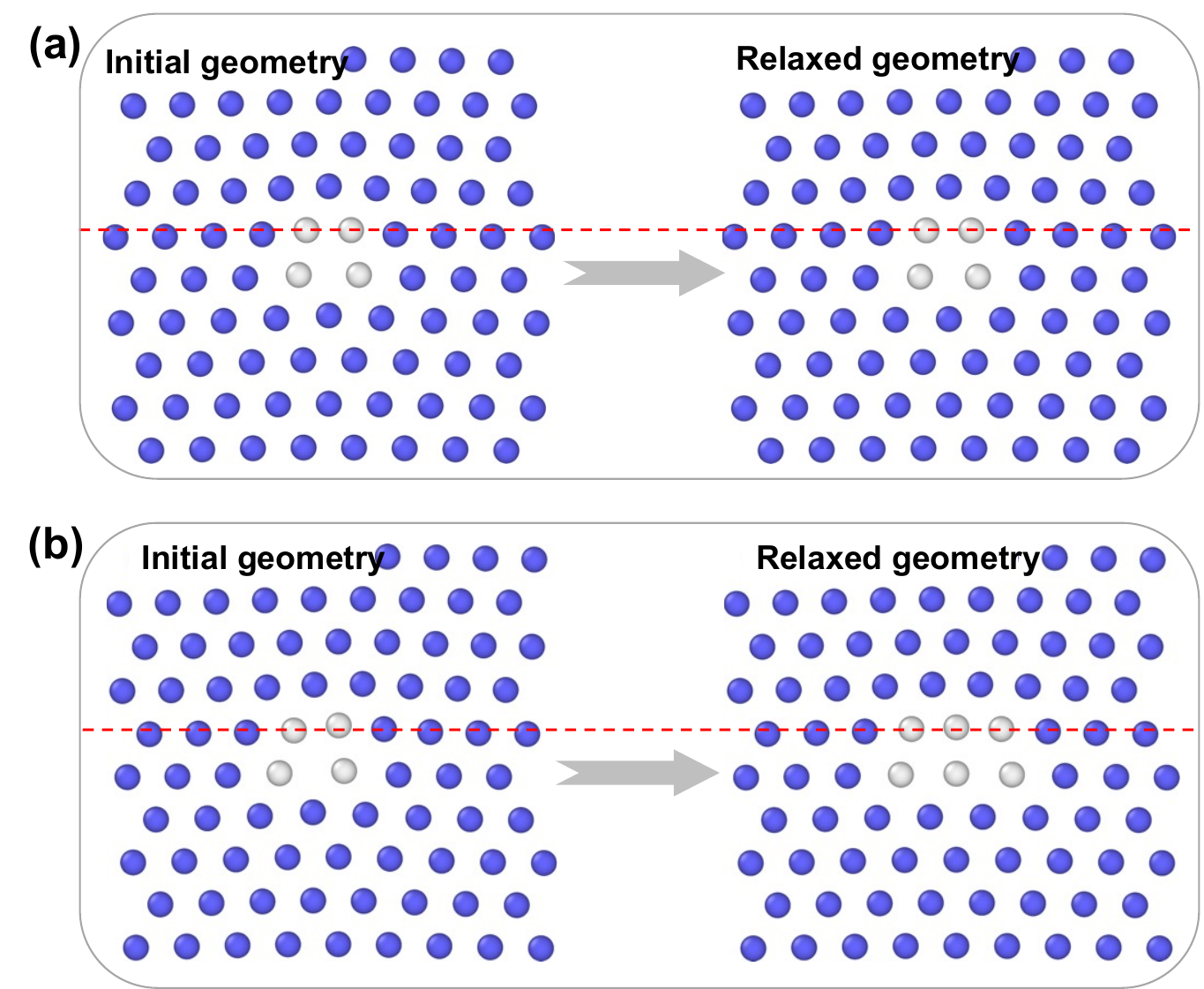}
	\caption{M111 dislocation core structure predicted by GAP-DB-I by starting from \textbf{(a)} symmetric, and 
		\textbf{(b)} asymmetric initial geometries.  
		Atoms are analyzed based on common neighbor analysis (CNA).
		Blue and grey indicate bcc and disordered atoms respectively. 
		The horizontal red dashed lines are added to visualise the initial asymmetric core structure in \textbf{(b)}. 
	}
	\label{fig:m111_diff}
\end{figure}

\begin{figure}[H]
	\centering
	\includegraphics[trim=0 -10 0 0,width=10cm]{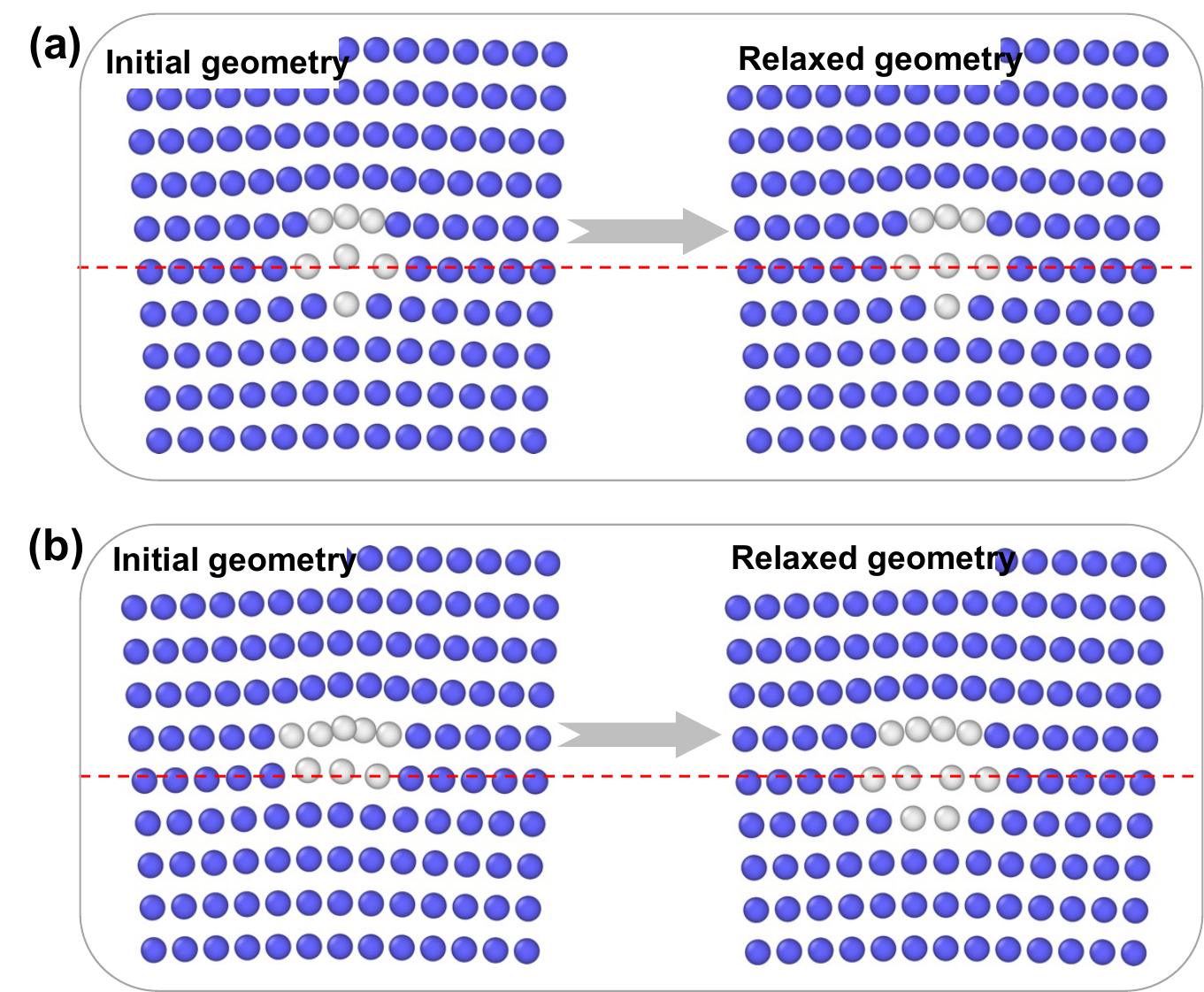}
	\caption{$a_0[100](011)$ edge dislocation core structure predicted by GAP-DB-II by starting from \textbf{(a)} symmetric, and 
		\textbf{(b)} asymmetric initial geometries.  
		Atoms are analyzed based on CNA.
		Blue and grey indicate bcc and disordered atoms respectively. 
		The horizontal red dashed lines are added to visualise the initial asymmetric core structure in \textbf{(b)}. 
	}
	\label{fig:e100_110_diff}
\end{figure}

Fig. \ref{fig:m111_diff} shows the equilibrium core structure of M111 dislocation relaxed from a different starting geometry. 
BC core and AC core structures are predicted based on symmetric and asymmetric geometries, respectively. 
Similarly, Fig. \ref{fig:e100_110_diff} shows the equilibrium core structure of $a_0/2[\bar{1}\bar{1}1](1\bar{1}0)$  edge dislocation relaxed from different starting geometry. 
AC core and BC core structures are predicted based on symmetric and asymmetric geometries, respectively.

\subsubsection{Predictions of GAP-DB-I, GAP-DB-II, and PACE-FS, using PAD configuration}

\begin{figure}[H]
	\centering
	\includegraphics[trim=0 -10 0 0,width=10cm]{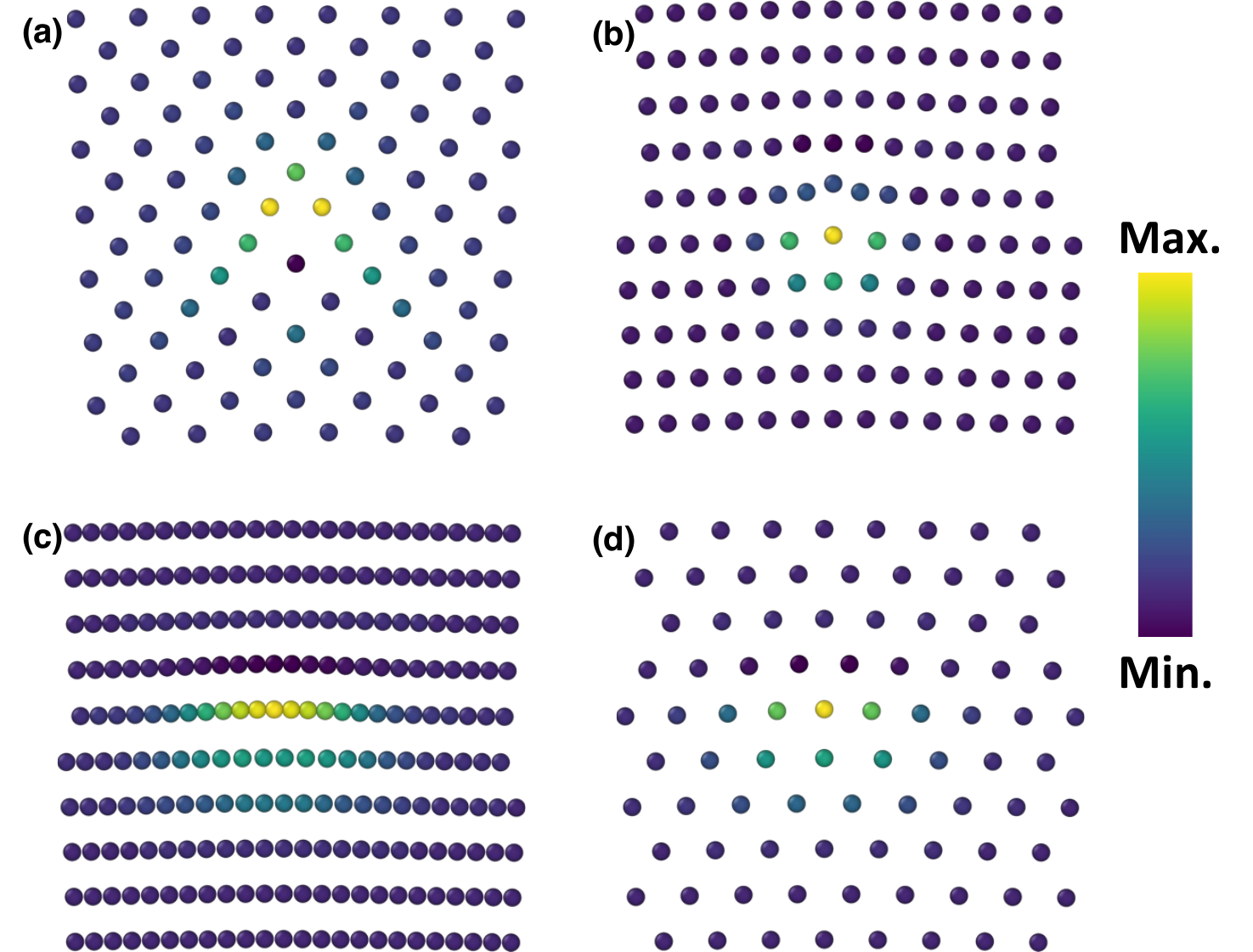}
	\caption{Core structure predicted by GAP-DB-I: \textbf{(a)} $a_0[100](010)$ edge, 
		\textbf{(b)} $a_0[100](011)$ edge, 
		\textbf{(c)} $a_0/2[\bar{1}\bar{1}1](1\bar{1}0)$ edge, and
		\textbf{(d)} $a_0/2[111](1\bar{1}0)$ 70.5$^\circ$  mixed dislocations.
		Atoms are colored according to their potential energy, which is normalized by the largest value (at the core).
	}
	\label{fig:disl_gap_pad}
\end{figure}

\begin{figure}[H]
	\centering
	\includegraphics[trim=0 -10 0 0,width=10cm]{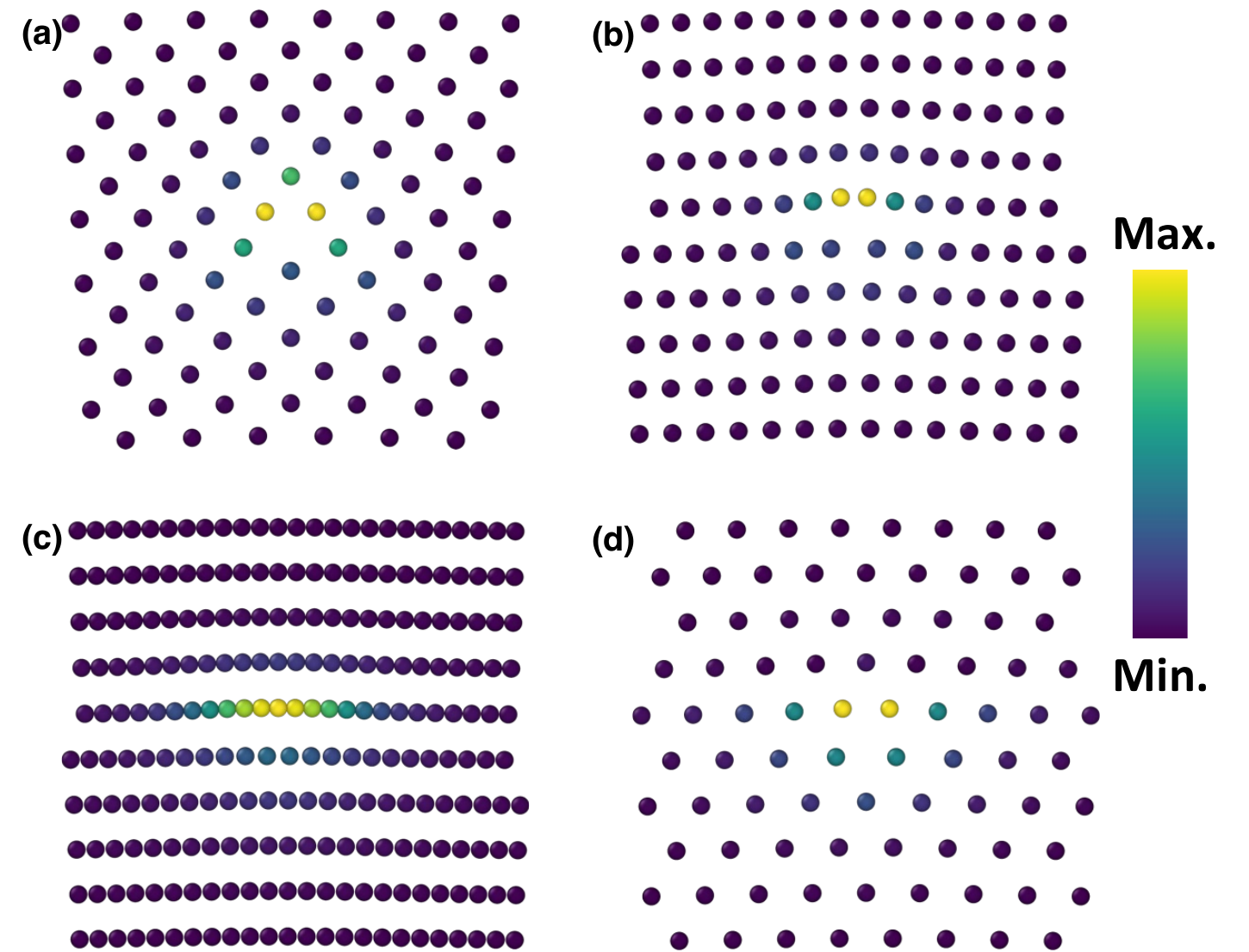}
	\caption{Dislocation core structures predicted by GAP-DB-II: \textbf{(a)} $a_0[100](010)$ edge, 
		\textbf{(b)} $a_0[100](011)$ edge, 
		\textbf{(c)} $a_0/2[\bar{1}\bar{1}1](1\bar{1}0)$ edge, and
		\textbf{(d)} $a_0/2[111](1\bar{1}0)$ 70.5$^\circ$  mixed dislocations.
        Atoms are colored according to their potential energies, which is normalized by the largest value (at the core).
	}
	\label{fig:disl_gap_db2_pad}
\end{figure}

\begin{figure}[H]
	\centering
	\includegraphics[trim=0 -10 0 0,width=10cm]{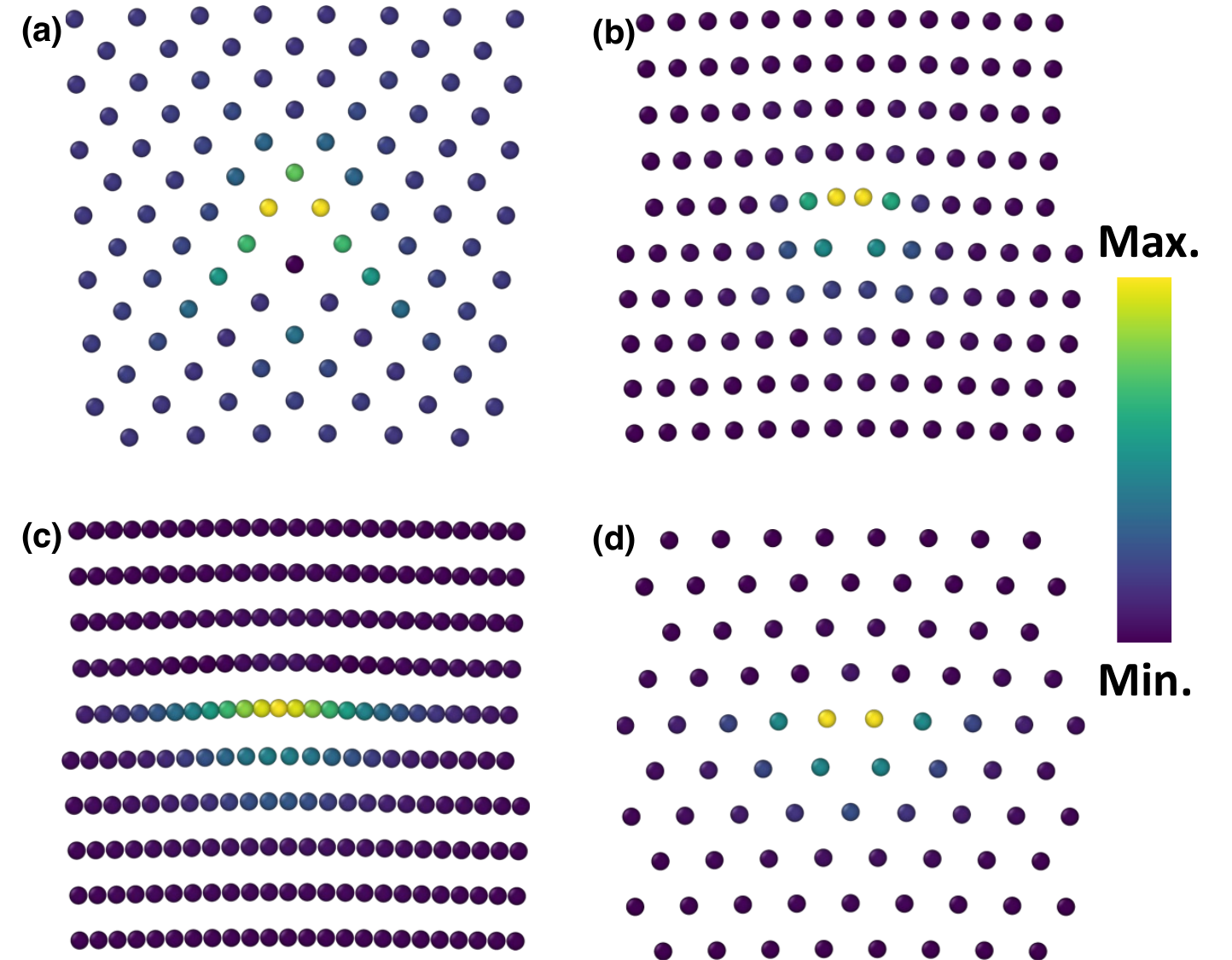}
	\caption{Dislocation core structures predicted by PACE-FS: \textbf{(a)} $a_0[100](010)$ edge, 
		\textbf{(b)} $a_0[100](011)$ edge, 
		\textbf{(c)} $a_0/2[\bar{1}\bar{1}1](1\bar{1}0)$ edge, and
		\textbf{(d)} $a_0/2[111](1\bar{1}0)$ 70.5$^\circ$  mixed dislocations.
        Atoms are colored according to their potential energy, which is normalized by the largest value (at the core).
	}
	\label{fig:disl_pace_pad}
\end{figure}

Fig. \ref{fig:disl_gap_pad}, \ref{fig:disl_gap_db2_pad}, and \ref{fig:disl_pace_pad} are the dislocation core structures predicted by GAP-DB-I, GAP-DB-II, and PACE-FS based on PAD configurations. 
Atoms are colored according to their potential energies to reveal the core position. 

\subsection{Model uncertainty predicted by GAP and PACE}
\subsubsection{GAP model uncertainty of $a_0[100](011)$ edge and M111 dislocations along the NEB path}

\begin{figure}[H]
	\centering
	\includegraphics[trim=0 -10 0 0,width=14cm]{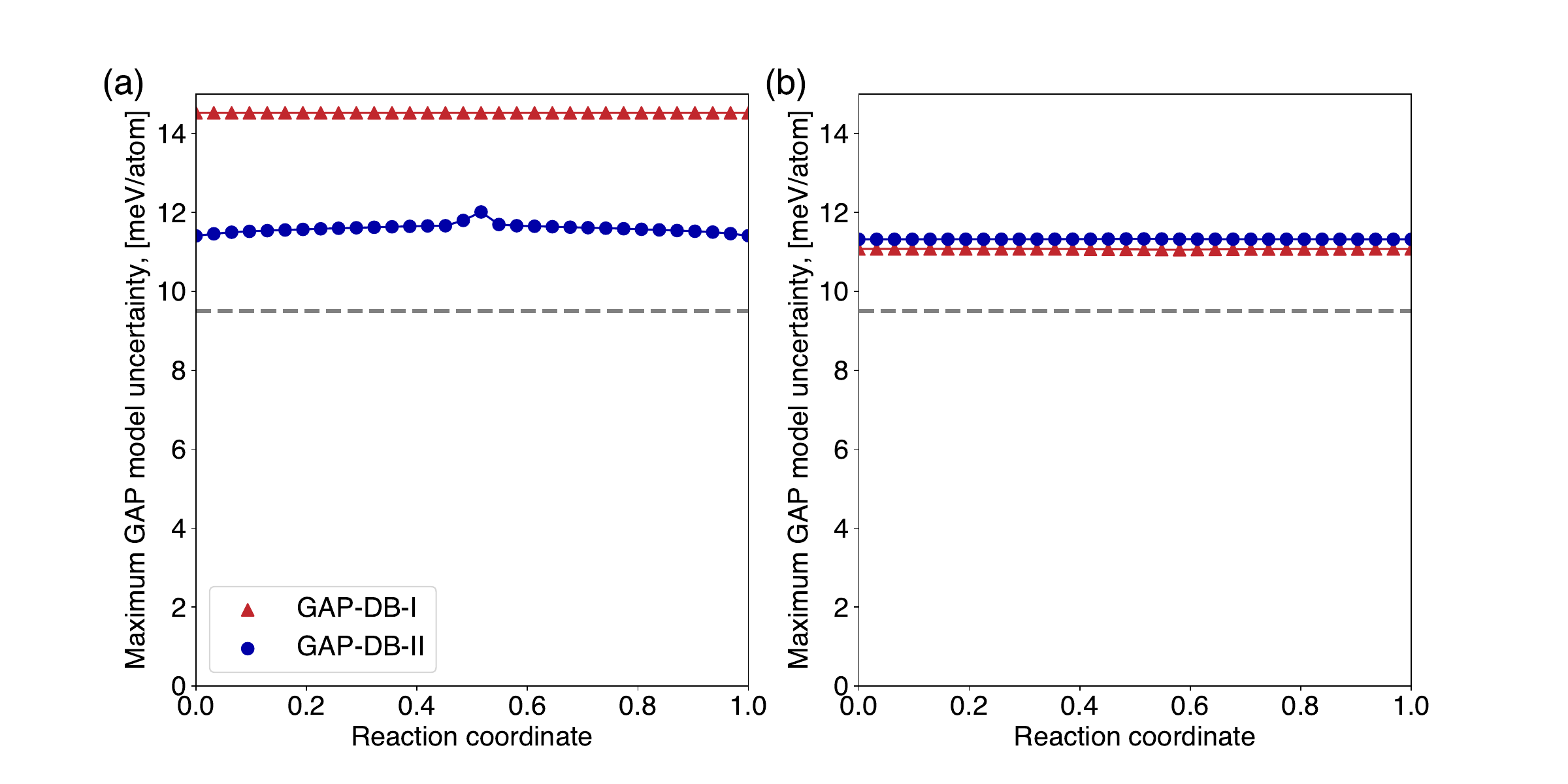}
	\caption{Maximum GAP model uncertainty during NEB calculation predicted by GAP-DB-I and GAP-DB-II: \textbf{(a)} $a_0[100](011)$ edge, and
		\textbf{(b)} M111 dislocations.
	}
	\label{fig:gap_uncertainty}
\end{figure}

Fig. \ref{fig:gap_uncertainty} plots the maximum GAP model uncertainty of the dislocation configurations along the NEB path. 
Note that the dashed line represents the model uncertainty of the perfect bulk atom, which is the accuracy limit of the model. 
Both GAP IAPs predict a converged model uncertainty within 5 meV/atom difference compared with the model uncertainty lower-bound. 
For $a_0[100](011)$ edge dislocation, GAP-DB-II predicts a lower uncertainty, which is likely to be the result of including the dislocation structures in the database. 

\begin{figure}[H]
	\centering
	\includegraphics[trim=0 -10 0 0,width=12cm]{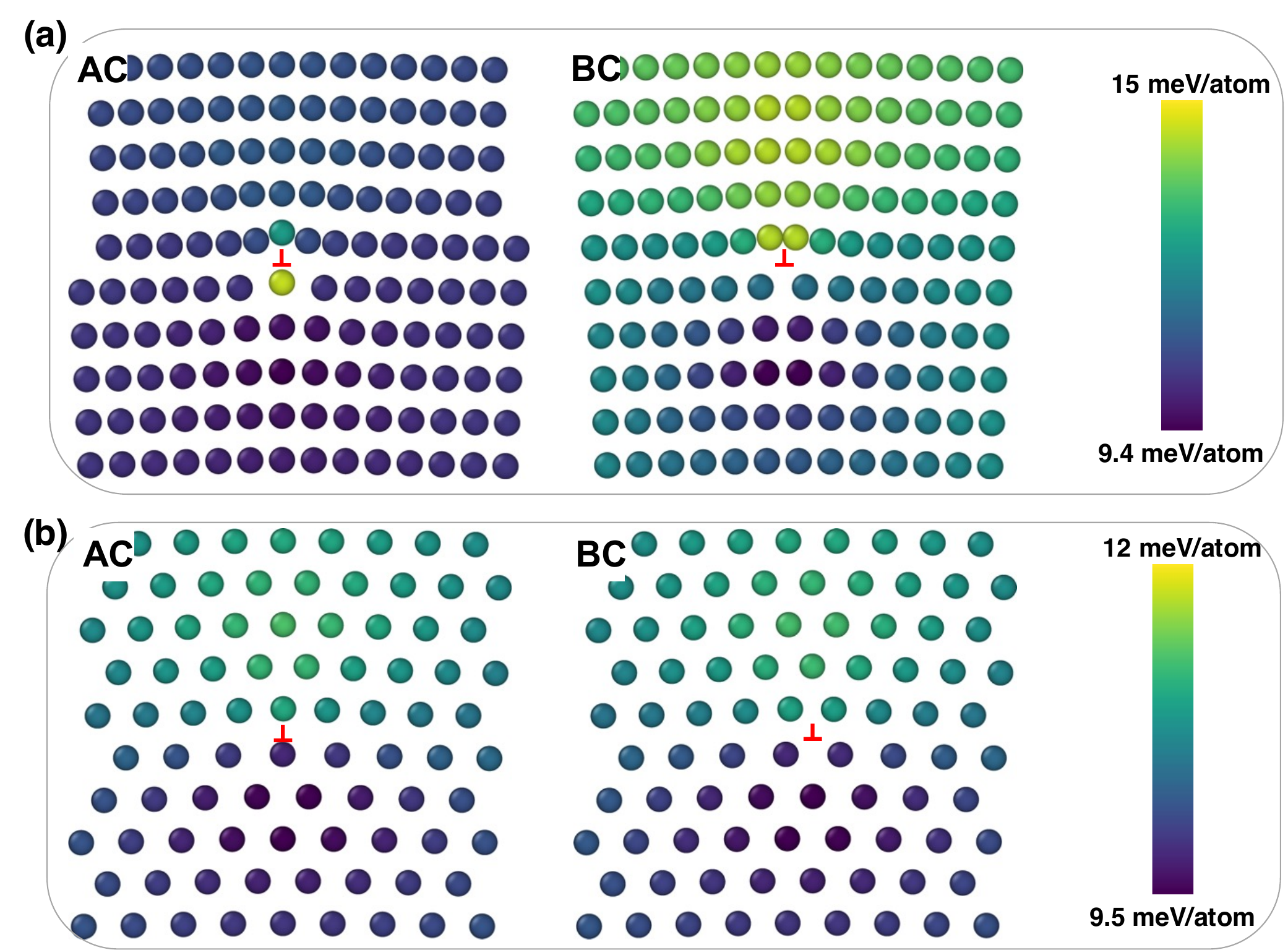}
	\caption{Dislocation core structure change during NEB calculation predicted by GAP-DB-I, corresponding to \textit{Fig. 9 of the main manuscript}.
		 \textbf{(a)} $a_0[100](011)$ edge (Fig. 9b), and
		\textbf{(b)} M111 (Fig. 9d) dislocations.
		Atoms are colored according to the GAP model uncertainty.
	}
	\label{fig:gap_scale}
\end{figure}

Fig. \ref{fig:gap_scale} shows the atomic dislocation core structure colored according to the model uncertainty predicted by GAP-DB-I.
The configurations are taken from NEB calculation, corresponding to \textit{Fig. 9 of the main manuscript}.

\subsubsection{GAP and PACE model uncertainty of non-screw dislocations along the NEB path}

\begin{figure}[H]
	\centering
	\includegraphics[trim=0 0 0 0,width=14cm]{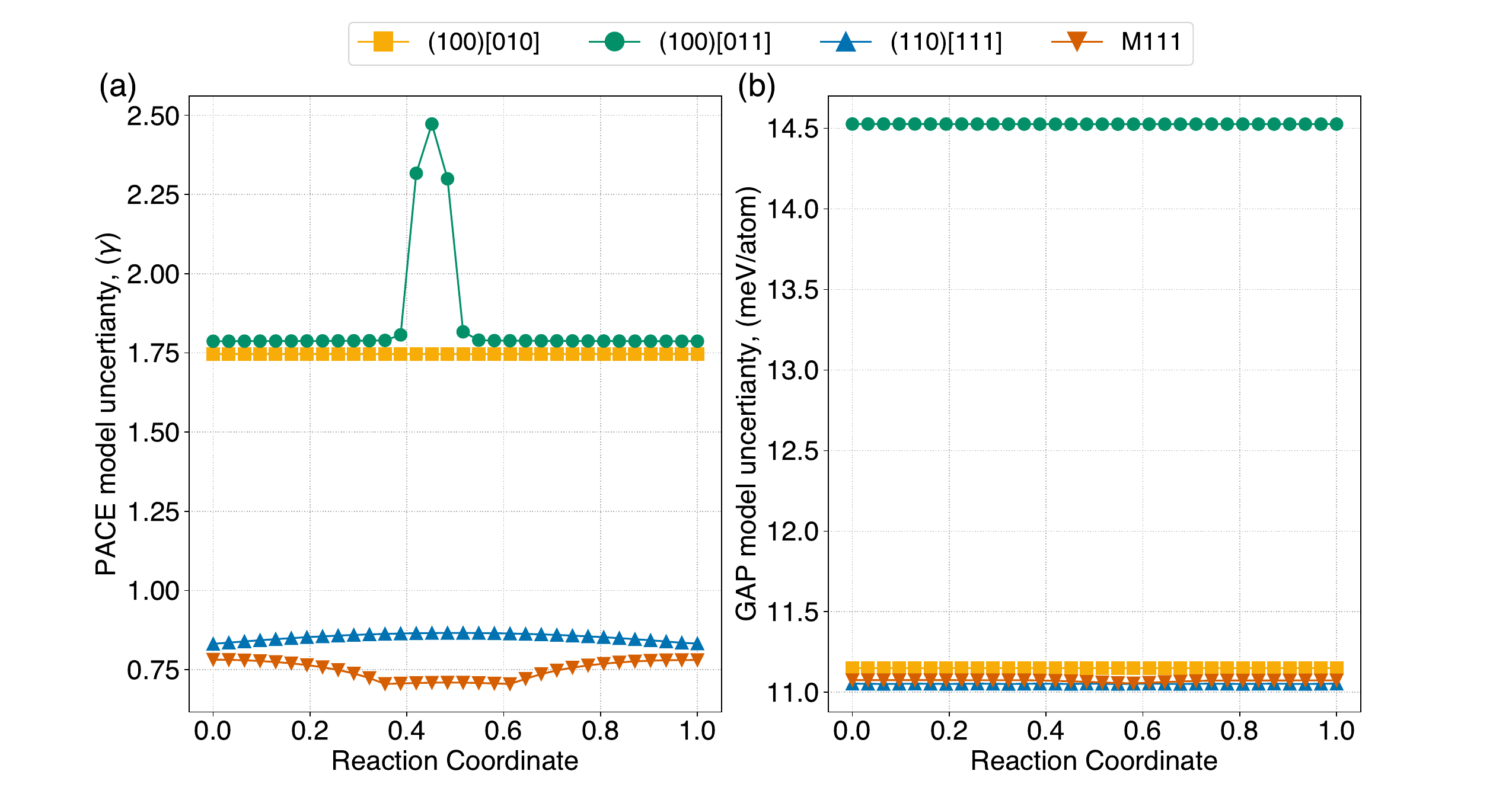}
	\caption{ Maximum model uncertainty around the dislocation core of the replicas along NEB path predicted by
		\textbf{(a)} GAP-DB-I and
		\textbf{(b)} PACE-FS. 
		Four dislocations are considered, i.e., $a_0[100](010)$, $a_0[100](011)$, $a_0/2[\bar{1}\bar{1}1](1\bar{1}0)$ edge dislocations, and M111 (\textit{Fig. 9 of the main manuscript}).
	}
	\label{fig:disloc_uncertainty}
\end{figure}

We have calculated the per-atom model uncertainty of non-screw dislocations along the transition path using both GAP and PACE trained on DB-I, i.e., GAP-DB-I and PACE-DB-I in \textit{Fig. 9 of the main manuscript}.
Fig. \ref{fig:disloc_uncertainty}a shows the maximum $\gamma$ of each replica along the transition path for four dislocations ($a_0[100](010)$, $a_0[100](011)$, $a_0/2[\bar{1}\bar{1}1](1\bar{1}0)$ edge dislocations, and M111). 
The maximum $\gamma$ of $a_0/2[\bar{1}\bar{1}1](1\bar{1}0)$ edge dislocation and M111 are less than 1 along the NEB path, showing interpolation of the PACE potential. 
$a_0[100](010)$ and $a_0[100](011)$ dislocation show mild extrapolation.
In particular, $a_0[100](011)$ shows a peak signal, which indicates the prediction is less reliable. 
The GAP predicted uncertainty converges along the transition path for all dislocations.
However,  $a_0[100](011)$ has a higher value than other three dislocations, which qualitatively agrees with the larger $\gamma$ predicted by PACE.

\begin{figure}[H]
	\centering
	\includegraphics[trim=0 0 0 0,width=12cm]{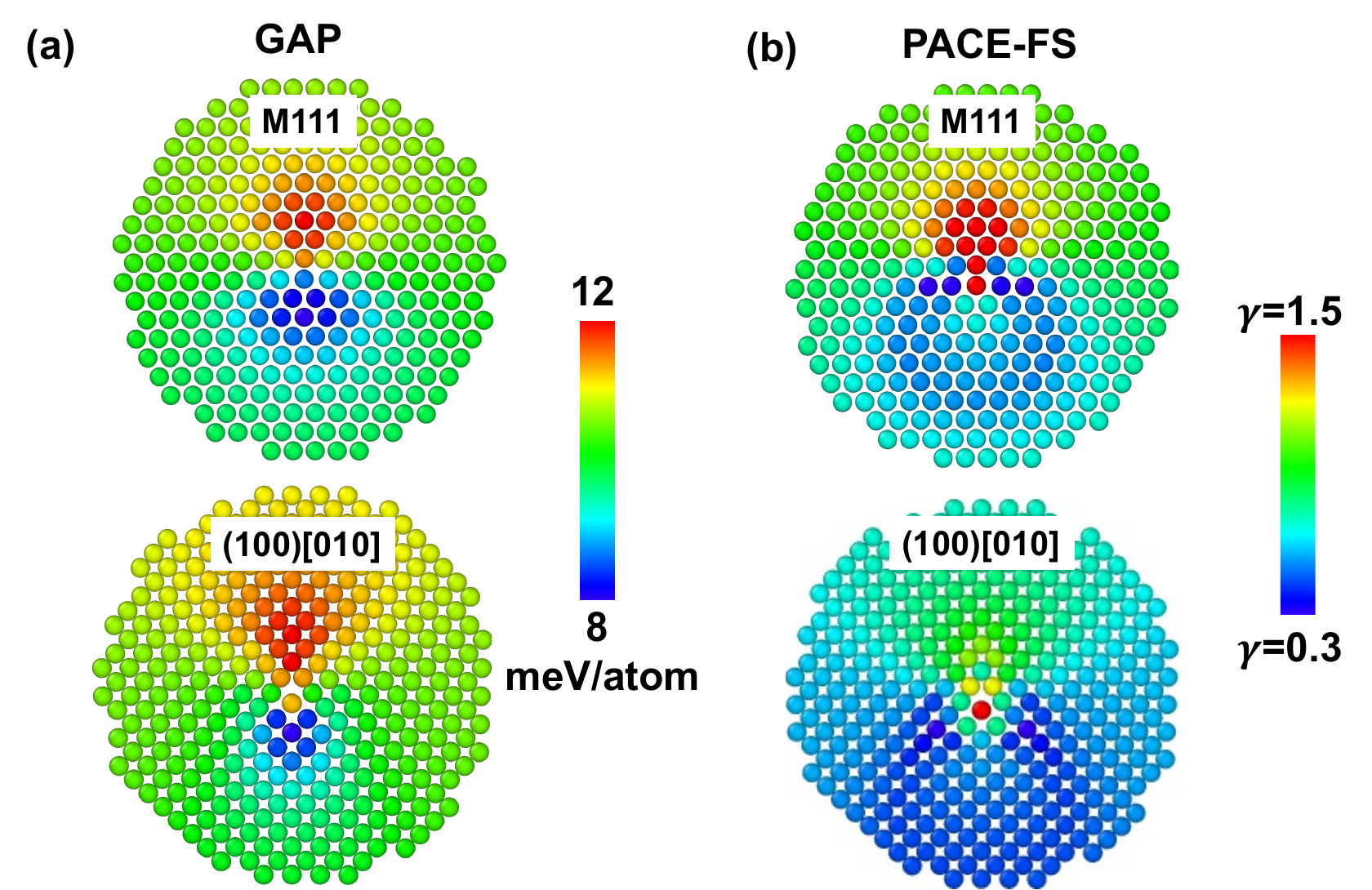}
	\caption{ Per-atom uncertainty for M111 and (100)[010]edge dislocations predicted by two model uncertainty quantification approaches. 
        \textbf{(a)} GAP-DB-I and
		\textbf{(b)} PACE-FS.
        Atoms are colored according to the square root of the GP variance and extrapolation grade $\gamma$ in GAP and PACE-FS, respectively. 
	}
	\label{fig:disloc_gamma_variance}
\end{figure}

Fig. \ref{fig:disloc_gamma_variance} shows a qualitative correlation of the position of the most uncertain atoms using GAP and PACE-FS model uncertainties.
Both uncertainty measures predict the upper core region to more uncertain than the lower region for M111 dislocation and (100)[010] edge dislocation.
Quantitative differences are expected since GAP uncertainty is related to the variance of the predicted energy while the D-optimality in PACE is related to the difference in atomic positions between the predicted configurations and the configurations in the database. 

\section{Atomistic fracture}
\subsection{Traction-separation curves}

\subsubsection{T-S curve predicted by MTP, NNP and PACE-L}

\begin{figure}[H]
	\centering
	\includegraphics[trim=0 0 0 0, width=16cm]{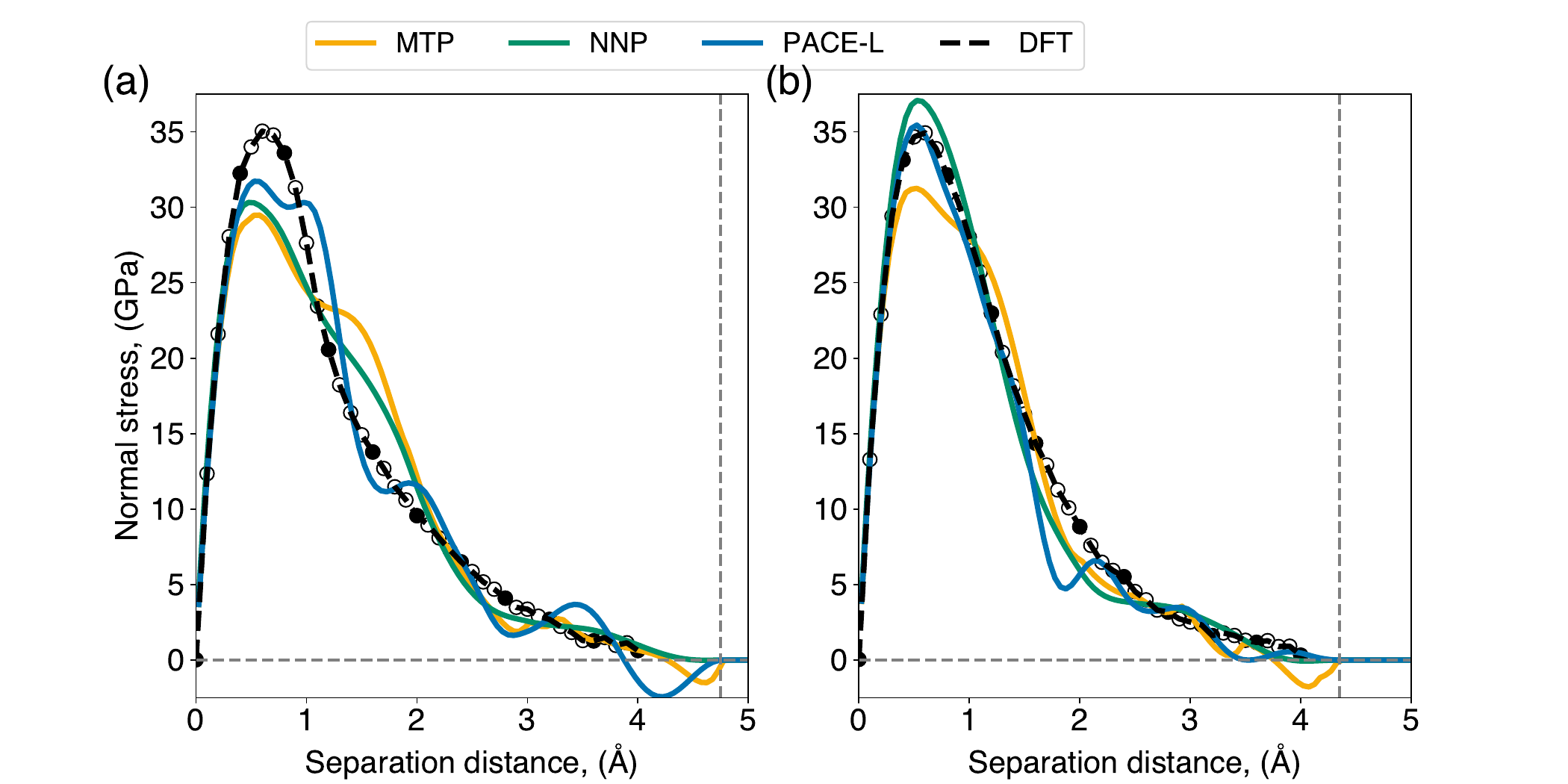}
	\caption{T-S profile predicted by  MTP, NNP and PACE-L with $R_{\rm cut}$=6.5 Å.
		\textbf{(a)} \{100\} and
		\textbf{(b)} \{110\} plane.
	}
	\label{fig:ts-mliaps}
\end{figure}

Fig. \ref{fig:ts-mliaps}shows the T-S curves of (100) and (110) planes predicted by MTP, NNP and PACE-L. 
All ML-IAPs are able to predict a peak normal stress at similar distance compared with DFT prediction. 
However, NNP shows two peak values around the maximum normal stress.
All ML-IAPs show the fluctuation at the end of the separation.

\subsubsection{Area under T-S curve versus surface energy (unrelaxed)}

\begin{figure}[H]
	\centering
	\includegraphics[trim=0 20 0 45, width=7cm]{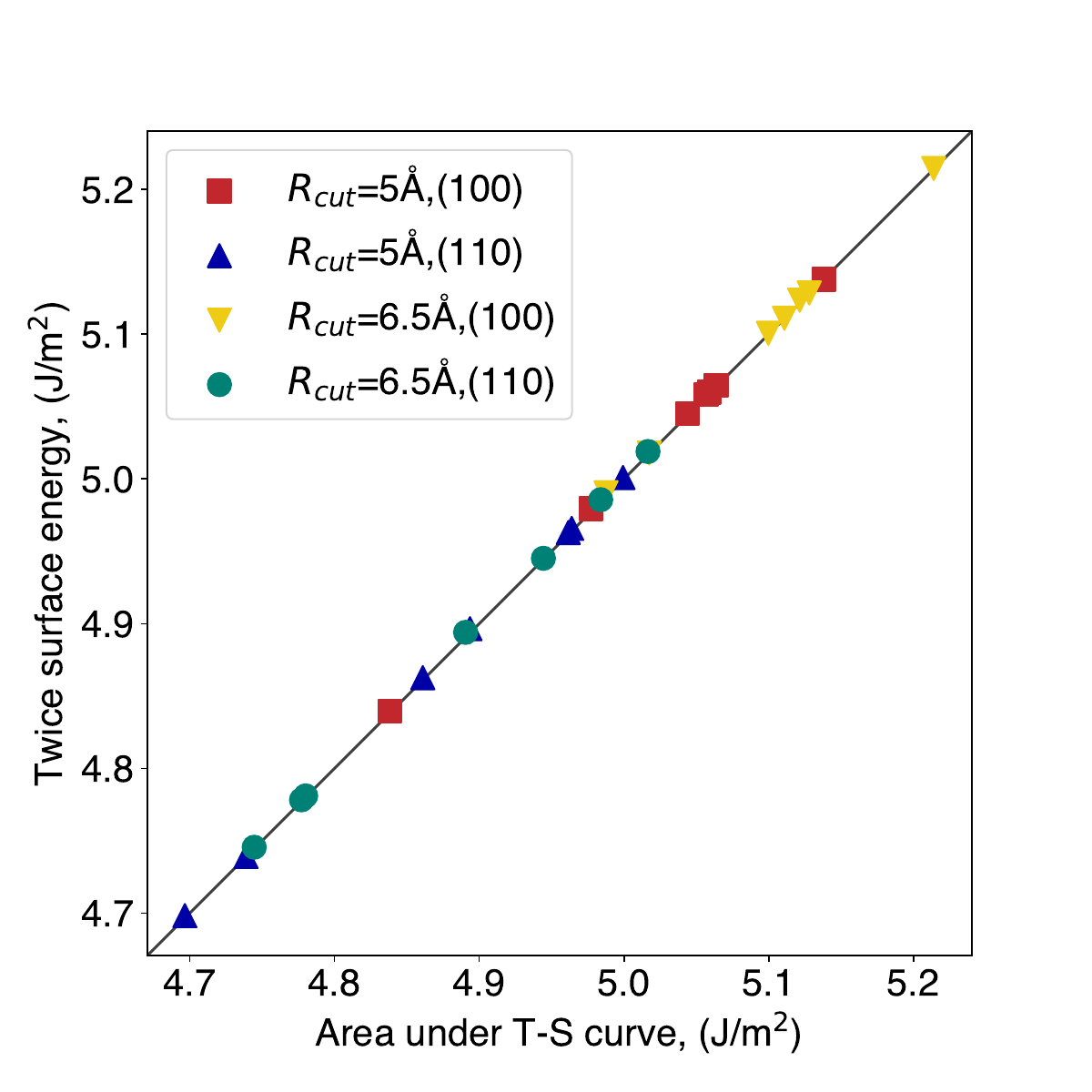}
	\caption{Area under T-S curve versus unrelaxed surface energy for ML-IAPs with $R_{\rm cut}=5$ and 6.5 Å .
	}
	\label{fig:area_gama}
\end{figure}

As shown in Fig. \ref{fig:area_gama}, the area under the T-S curve is exactly twice the unrelaxed surface energy.

\subsubsection{T-S curve predicted by PACE-FS with different cutoff radii}

\begin{figure}[H]
	\centering
	\includegraphics[trim=0 20 0 45, width=16cm]{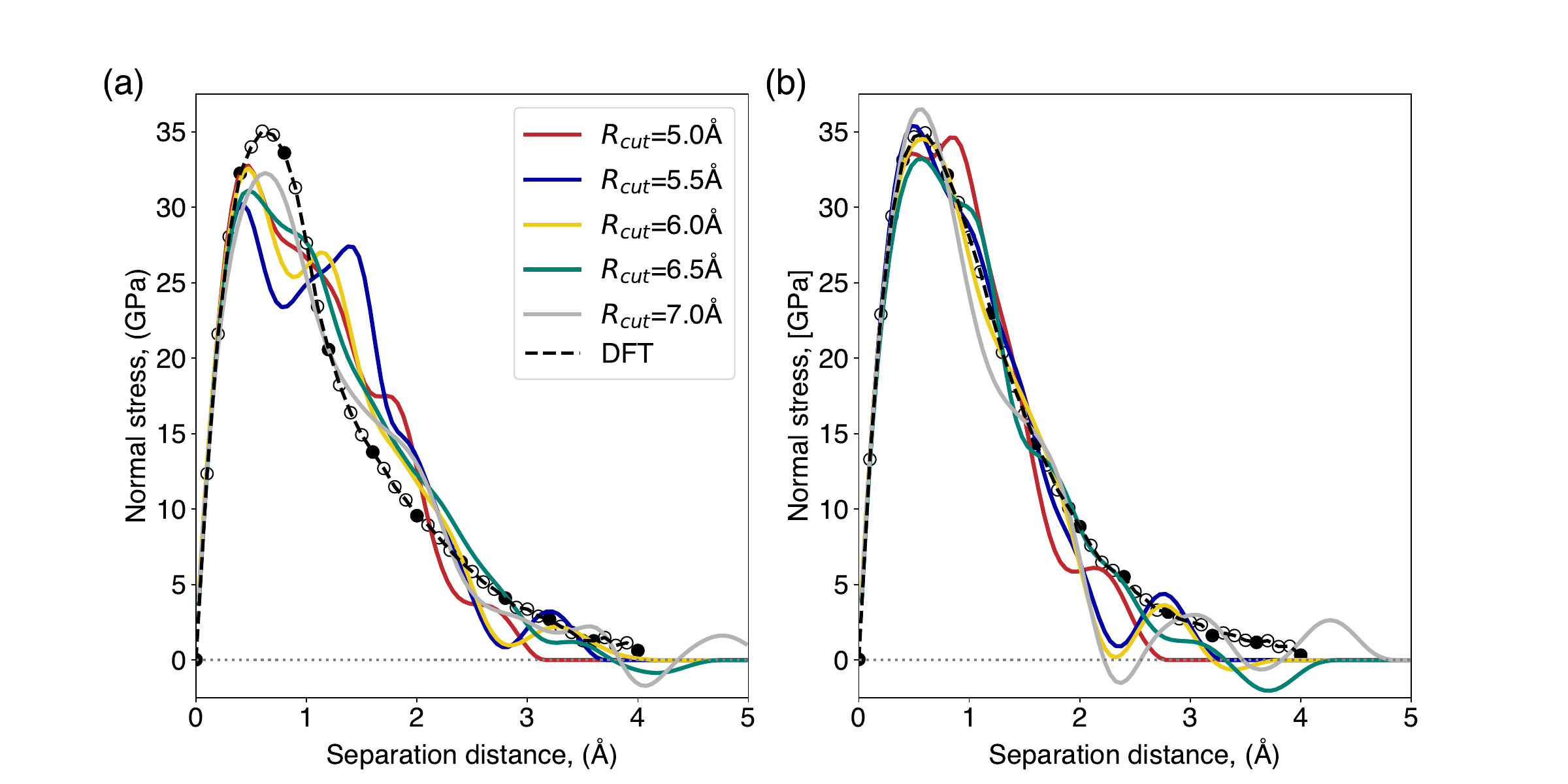}
	\caption{T-S profile predicted by PACE-FS with different cutoff radii. 
		Cutoff ranges from $R_{\rm cut}$=5-7 Å with a interval of 0.5 Å.
		\textbf{(a)} \{100\} and
		\textbf{(b)} \{110\} plane.
		Note that the potential is trained on the solid circle DFT data.
	}
	\label{fig:ts-pace}
\end{figure}

Fig. \ref{fig:ts-pace}shows the T-S curves of (100) and (110) planes with different $R_{\rm cut}$. 
Note that Fig. \ref{fig:ts-pace}a and b are predicted by PACE-FS trained on part of the DFT data indicated by the solid circles.
For (100) plane, $R_{\rm cut} = 6.5$ Å and 7 Å display a single peak and $R_{\rm cut} = 6.5$ Å is smoother at the end of separation.
The predicted maximum stresses are all lower than DFT results. 
The cohesive strength of (110) plane can be well described by all $R_{\rm cut}$.

\subsubsection{T-S curve predicted by PACE-FS with increased separation data}

\begin{figure}[H]
	\centering
	\includegraphics[trim=0 20 0 25, width=16cm]{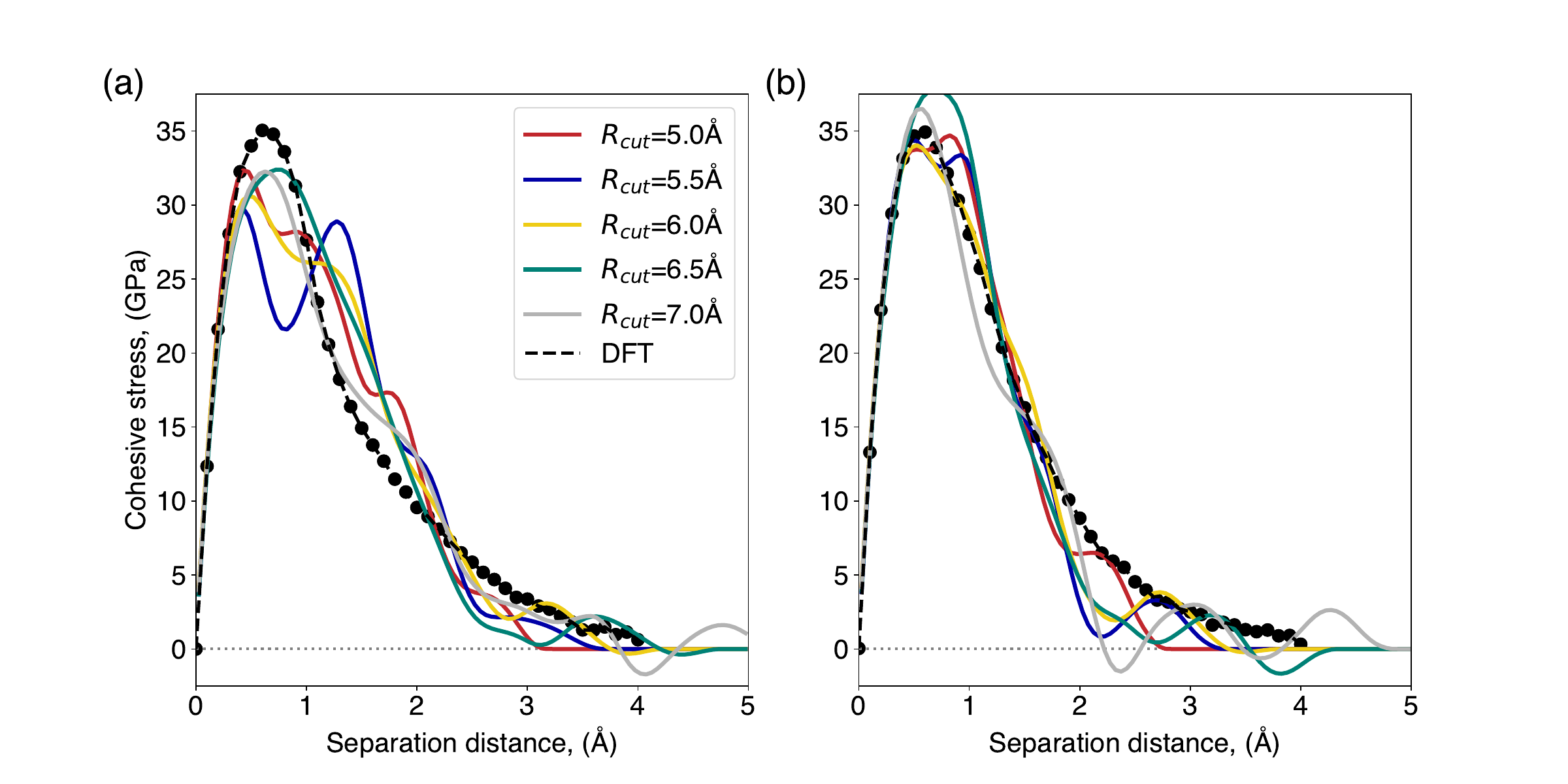}
	\caption{
        T-S profile predicted by PACE-FS with different cutoff radii by adding four time the separation data than Fig. \ref{fig:ts-pace}. 
		Cutoff ranges from $R_{\rm cut}$=5-7 Å with a interval of 0.5 Å.
		\textbf{(a)} \{100\}, and
		\textbf{(b)} \{110\} plane.
		Note that the potential is trained on the solid circle DFT data.
	}
	\label{fig:ts-data}
\end{figure}

Fig. \ref{fig:ts-data}shows the T-S curves of (100)  and (110) planes predicted by PACE-FS trained on four time the surface separation data.
Yet, the T-S curves fluctuate at the end of the separation.

\subsection{Model uncertainty quantification}

\begin{figure}[H]
	\centering
	\includegraphics[trim=0 20 0 0, width=14cm]{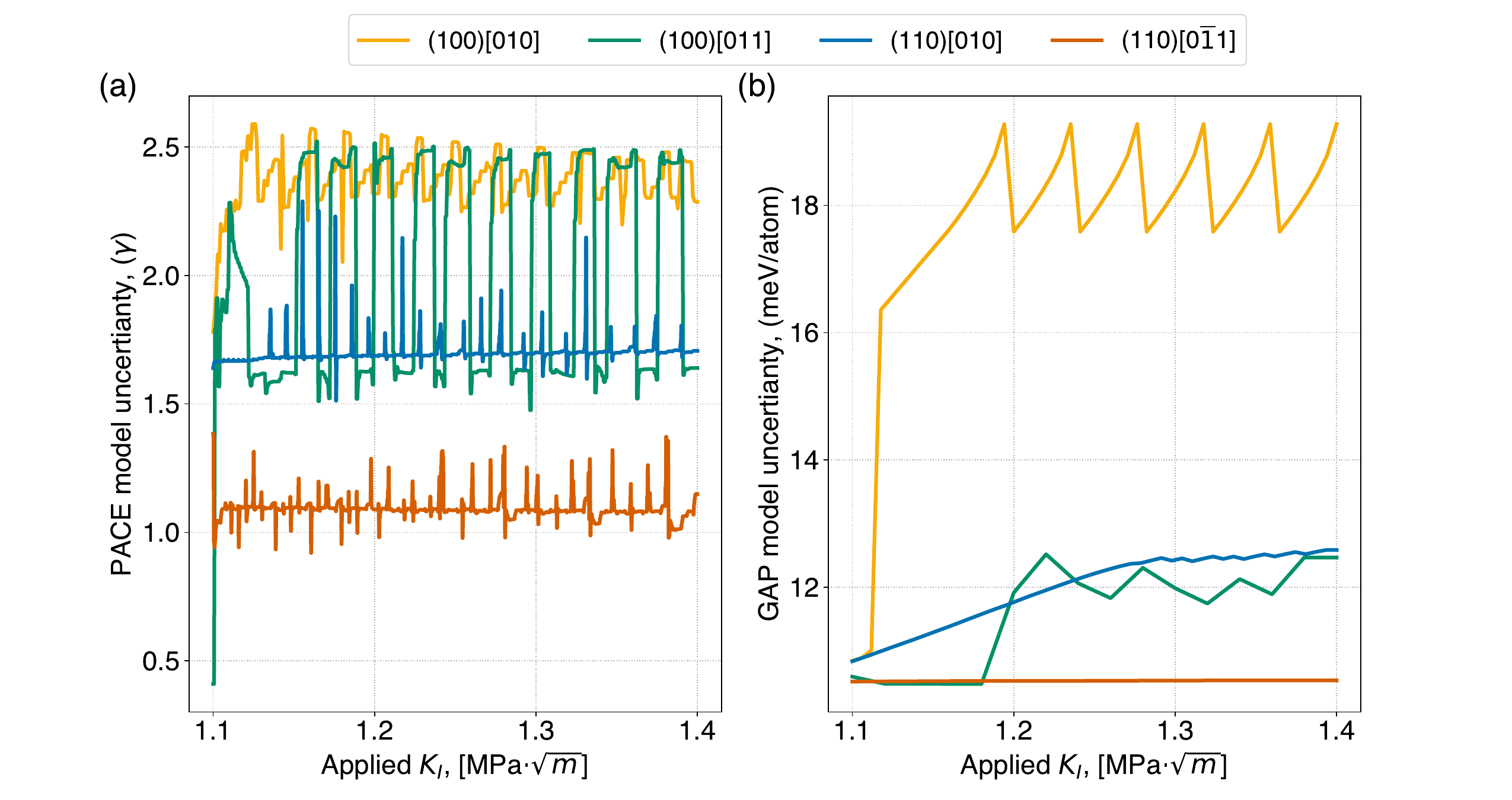}
	\caption{
		Maximum model uncertainty around the crack tip as a function of applied $K_{\rm I}$ predicted by
		\textbf{(a)} GAP-DB-I and
		\textbf{(b)} PACE-DB-I. 
		Four crack systems are considered (crack plane/crack front), i.e., (100)[010], (100)[011], (110)[001], and (110)[1$\bar{1}0$].
	}
	\label{fig:uncertainty_crack}
\end{figure}

We calculate the maximum model uncertainty for crack configurations as a function of applied $K_{\rm I}$ predicted by both GAP-DB-I and PACE-FS.
As shown in Fig. \ref{fig:uncertainty_crack}, the uncertainties fluctuate with the increase of applied $K_{\rm I}$.  
Maximum $\gamma$ predicted by PACE-FS indicates mild extrapolation, which is within the threshold ($\gamma < 3$).
GAP predicted uncertainty shows larger value for (100)[010] than other three crack systems.
However, the atomistic fracture mechanism agrees with the PACE-FS prediction.

\begin{figure}[H]
	\centering
	\includegraphics[trim=0 0 0 0,width=12cm]{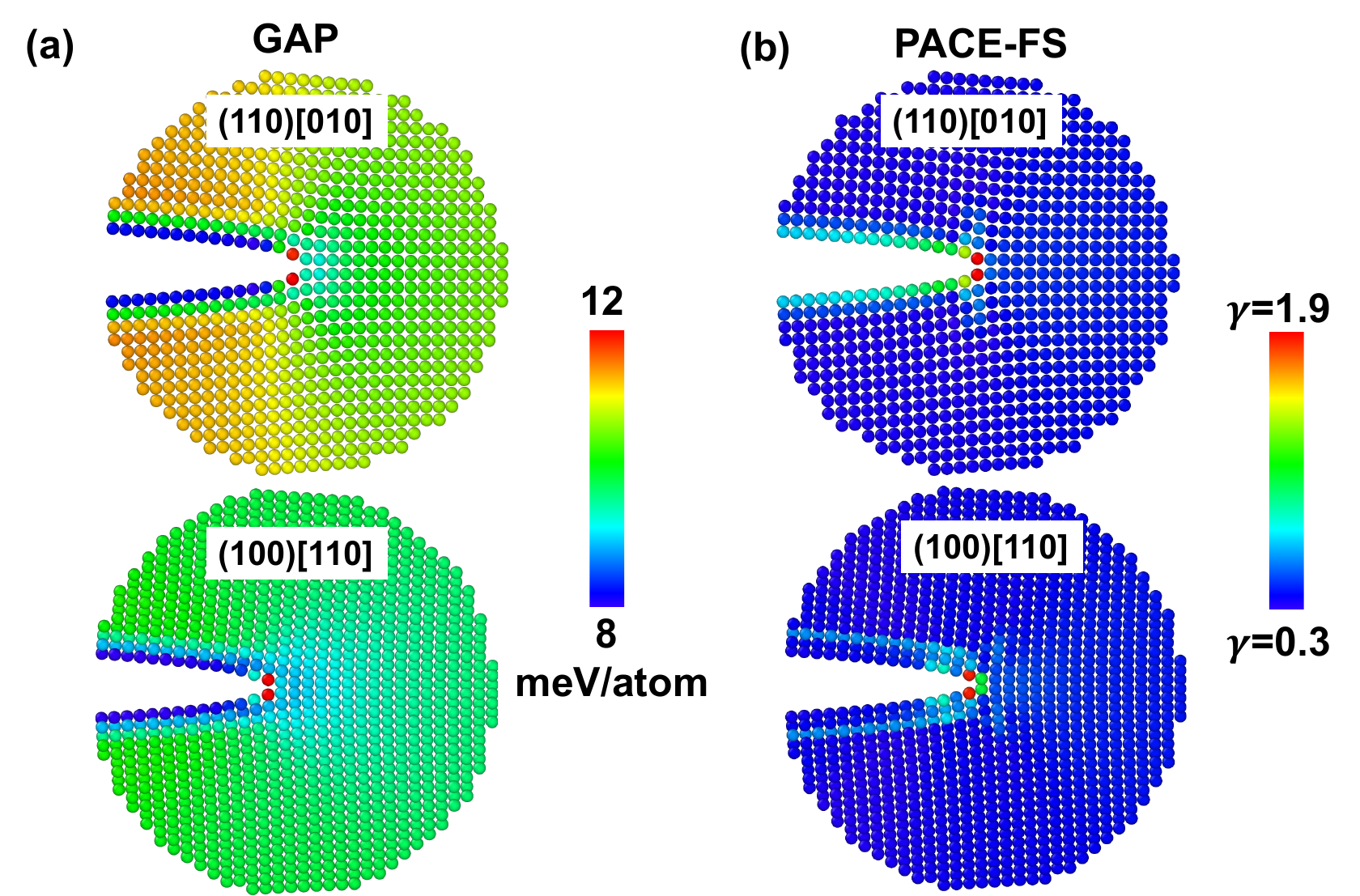}
	\caption{ Per-atom uncertainty for crack system (110)[010] and (100)[110] predicted by two model uncertainty quantification approaches. Atoms are colored according to the square root of the GP variance and extrapolation grade $\gamma$ in GAP and PACE-FS, respectively.  
        \textbf{(a)} GAP-DB-I and
		\textbf{(b)} PACE-FS.
        Atoms are colored according to the square root of the GP variance and extrapolation grade $\gamma$ in GAP and PACE-FS, respectively. 
	}
	\label{fig:crack_gamma_variance}
\end{figure}

Figure \ref{fig:crack_gamma_variance} shows a qualitative correlation of the position of the most uncertain atoms using GAP and PACE-FS model uncertainties.
The atoms with largest uncertainties are always located at the crack tip, which are identified by both approaches.

\subsection{Critical $K_{\rm Ic}$ predicted by ML-IAPs with $R_{\rm cut} = 5$ Å}

\begin{figure}[H]
	\centering
	\includegraphics[trim=0 20 0 0, width=14cm]{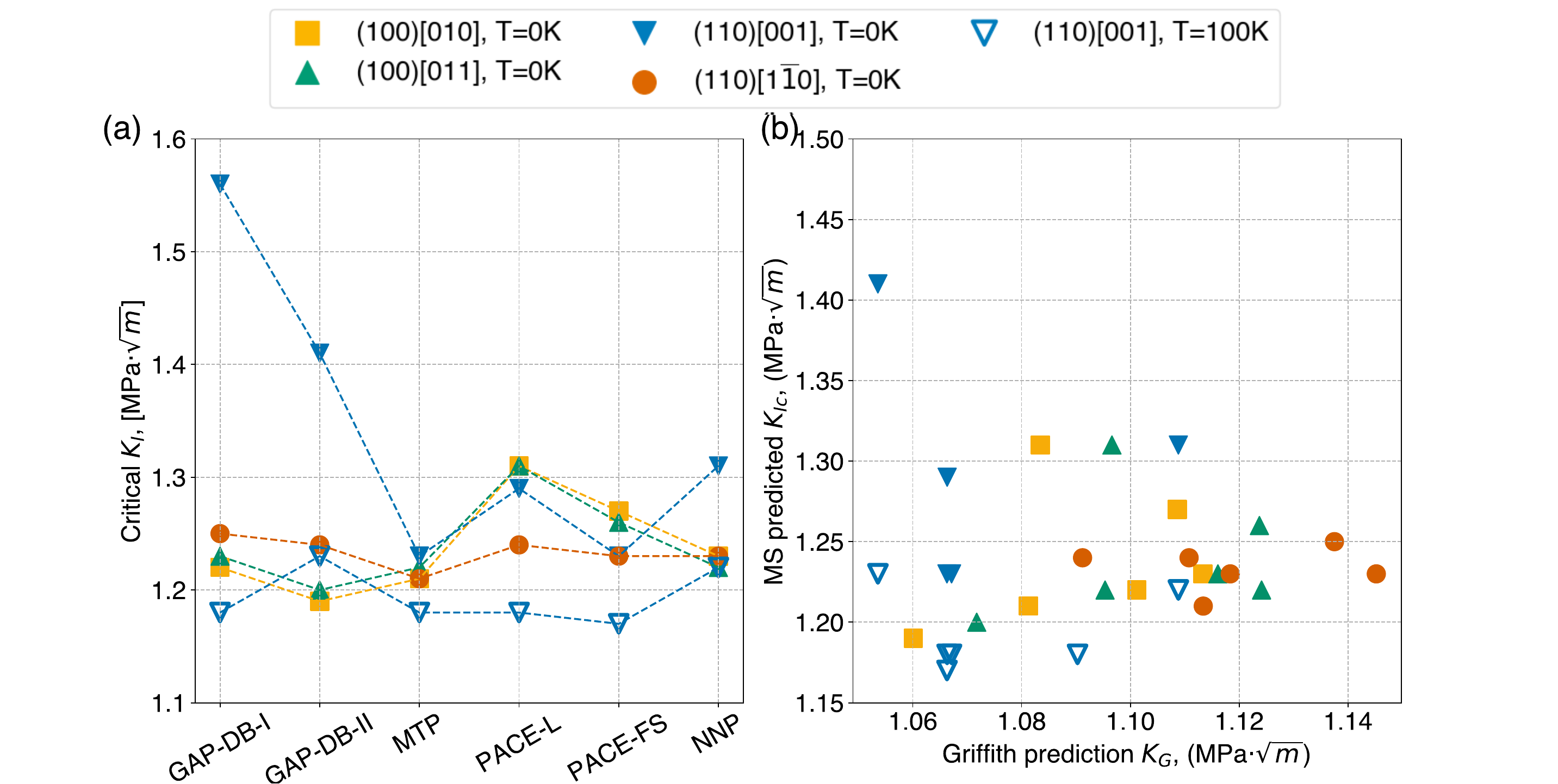}
	\caption{
		\textbf{(a)} Critical $K_{\rm Ic}$ predicted by different ML-IAPs ($R_{\rm cut} = 5$ Å) at T=0K for four crack systems and T=100K for crack (110)[010]. 
		Crack systems are indicated by crack plane/crack front. 
		\textbf{(b)} Critical $K_{\rm Ic}$ predicted by ML-IAPs versus Griffith criterion $K_{\rm G}$.
	}
	\label{fig:kgkic}
\end{figure}

ML-IAPs are trained with $R_{\rm cut} = 5$ Å and $R_{\rm cut} = 6.5$ Å to investigate the influence of the cutoff radius on fracture mechanism and predicted $K_{\rm Ic}$.
The results of both $R_{\rm cut} = 5$ Å and $R_{\rm cut} = 6.5$ Å reveal that the atomistic fracture mechanism is cleavage on the pre-cracked plane for \{100\} and \{110\} crack planes. 
Fig. \ref{fig:kgkic}a shows the $K_{\rm Ic}$ predicted by ML-IAPs trained with cutoff radius 5Å. 
For $R_{\rm cut}$ = 5 Å, the $K_{\rm Ic}$’s for crack system (100)[010] and (100)[011] predicted by any ML-IAP are rather close to each other, with a maximum difference of 0.02 $\rm MPa\sqrt{m}$. 
The fluctuation of predicted $K_{\rm Ic}$’s is expected due to the different extrapolation ability of different ML-IAPs. 
However, $K_{\rm Ic}$ for crack system (100)[010] is consistently larger than (100)[011] when $R_{\rm cut} = 6.5$ Å (\textit{Fig. 11 of the main manuscript}), which indicates that $R_{\rm cut} = 5$ Å is not enough to describe the bond decohesion of (100) plane. 
The short $R_{\rm cut}$  predicts the bond rupture of crack tip atoms when they can still physically interact with each other, leading to a smaller predicted $K_{\rm Ic}$. 

\bibliographystyle{unsrt}
\bibliography{supref}